\documentclass[aps,rmp,reprint,amsmath,amssymb,graphicx,longbibliography]{revtex4-1}
\usepackage{bm}

\usepackage[utf8]{inputenc}
\usepackage[a4paper, total={7in, 10.5in}]{geometry}
\usepackage{lipsum}
 
\usepackage{caption}
\usepackage{subcaption}
\usepackage{amsmath,array, makecell}
\usepackage{mathtools}
\usepackage{bibentry}
\usepackage{url}
\usepackage{color}
\usepackage[dvipsnames]{xcolor}
\usepackage{graphicx}
\usepackage{grffile}
\usepackage{natbib}  
\usepackage{comment}
\DeclareGraphicsExtensions{.pdf}
\usepackage{titlesec}
\titleformat*{\section}{\large\bfseries}
\usepackage{epstopdf}
\usepackage{float}
\usepackage{wrapfig}
\usepackage{accents}

\usepackage[hidelinks]{hyperref}
\hypersetup{colorlinks=true,
  linkcolor=blue,urlcolor=black,citecolor=blue}

\newcommand{\rg}[1]{\textcolor{magenta}{#1}}

\newcommand{\ajn}[1]{\textcolor{brown}{#1}}

\newcommand{\bu}{\mathbf{u}}
\usepackage[normalem]{ulem}
\usepackage{cellspace} 
\usepackage[normalem]{ulem}
\allowdisplaybreaks

\begin{document}

\preprint{APS/123-QED}

\title{Viscosity variation in fluid flows across scales}
\author{Arjun Sharma}
\affiliation{Center for Computational Research, Sandia National Laboratories, Albuquerque, USA}

\author{Ritabrata Thakur}
\affiliation{Department of Applied Mechanics, Indian Institute of Technology Delhi, New Delhi, India}

\author{Sharath Jose}
\affiliation{International Centre for Theoretical Sciences, Tata Institute of Fundamental Research, Bengaluru, India}

\author{Rama Govindarajan}
\altaffiliation{Corresponding author: rama@icts.res.in}
\affiliation{International Centre for Theoretical Sciences, Tata Institute of Fundamental Research, Bengaluru, India}

\date{\today}
\begin{abstract}
A plethora of natural and engineering flows exhibit spatial and temporal variation in viscosity. The variations occur in flows across a colossal range of length and time scales, from microbial motion to Earth-scale mantle convection, and give rise to new physical mechanisms that are absent in constant-viscosity fluid flows. This review surveys such phenomena across scales, and examines these mechanisms. In Stokes (zero Reynolds number) flows, viscosity gradients lead to translation-rotation coupling, and enable novel particle dynamics such as rotation in response to uniform forcing -- a feature that may be exploited by microorganisms. The dynamics of laminar shear flows across scales are transformed by viscosity variation, by modifications to the base flow profile, and by the breaking of symmetries. In high-Reynolds-number shear flow, viscosity stratification fundamentally alters the mathematical structure of the singular perturbation problem which describes the production of disturbance kinetic energy. Even a minor stratification within the layer of kinetic energy production can dramatically enhance or suppress standard instabilities, and create new instabilities. Shear flows are prone to transition to turbulence via linear mechanisms: here the linear stability operator is non-self-adjoint (nonnormal), resulting in algebraic perturbation growth. Viscosity variations introduce new nonnormal and  nonlinear pathways for perturbation energy growth, apart from altering the structure and dynamics of the perturbations by the broken symmetries. While laminar and fully developed turbulent flows have received enormous attention over the decades, the process by which laminar flows transition to turbulence is not well-understood except in a few canonical constant-viscosity flows, and there too, only partially. There are myriad routes to fully-developed turbulence, and viscosity-variations will likely occupy centre-stage as we learn more about them. In fully developed turbulence, viscosity variations influence the structure of wall-bounded flows, jets, and mixing layers. At even larger scales, the accounting correctly for stratification of eddy viscosity in the ocean in global circulation models can lead to better weather predictions. Moving to the planetary scale, viscosity variations of several orders of magnitude occur in Earth's mantle, and these play a central role in geological evolution and mantle convection. Flows laden with solid particles and larger objects, whether in dilute suspension or in mushy regimes, are ubiquitous across scales. Effective viscosity variations from inhomogeneous particle loading can produce effects similar to actual viscosity variations. Throughout this review, we emphasize scenarios where viscosity variation qualitatively, and not merely quantitatively, alters flow physics. Although some aspects of viscosity varying flows have been studied for decades, an understanding of the physics of many other aspects is just beginning, we believe, with big questions lying wide open. The review is written with graduate students in mind, and in every discussion we attempt to identify well-posed, tractable research questions. We highlight cross-scale, and therefore cross-disciplinary, connections, which are rarely made but will likely be insightful.

\end{abstract}

\maketitle
\tableofcontents{}

\section{Viscosity-Stratified Flows across Scales}

Practically no flow on Earth, or elsewhere, for that matter, occurs in a fluid of strictly constant viscosity. Variations in viscosity can arise at virtually any length scale, from hundreds of nanometres to planetary dimensions. This naturally raises foundational questions about viscosity variations: Are these effects not negligible? Do viscosity variations merely adjust quantitative predictions, or do they qualitatively alter the physics? Can physics at one scale inform another? For answers, read on.

\subsection{Variable viscosity in natural and engineered systems}\label{sec:VariableViscIntroA}
The viscosity of a fluid depends on local thermodynamic conditions, particularly temperature and pressure, which are rarely, if ever, constant. In multicomponent systems, the viscosity also depends on the local composition. Yet, in the overwhelming majority of theoretical and computational studies of fluid motion, viscosity is assumed constant. Classical texts on fluid dynamics (e.g., \citet{Batchelor_1967book,Bird_etal_2005book,White_2006book}) typically acknowledge the dependence of viscosity on temperature and composition before proceeding to assume it fixed. To be fair, this simplification is justified in a plethora of flows. But there also exists a diverse class of problems, spanning biology, geophysics, engineering, and daily life, where such an assumption fails. Even small viscosity variations can break symmetries, introduce singular perturbations, and couple rotational and translational motion in ways that overturn dominant mechanisms and radically change flow behavior. Viscosity gradients can even give rise to `inviscid' instabilities. This review explores such situations and aims to unify the conceptual, mathematical, and physical consequences of viscosity stratification across scales and disciplines. We also discuss eddy viscosity, a model for the effective diffusivity afforded by turbulence.

\begin{figure*}
\centering
\includegraphics[width=\textwidth]{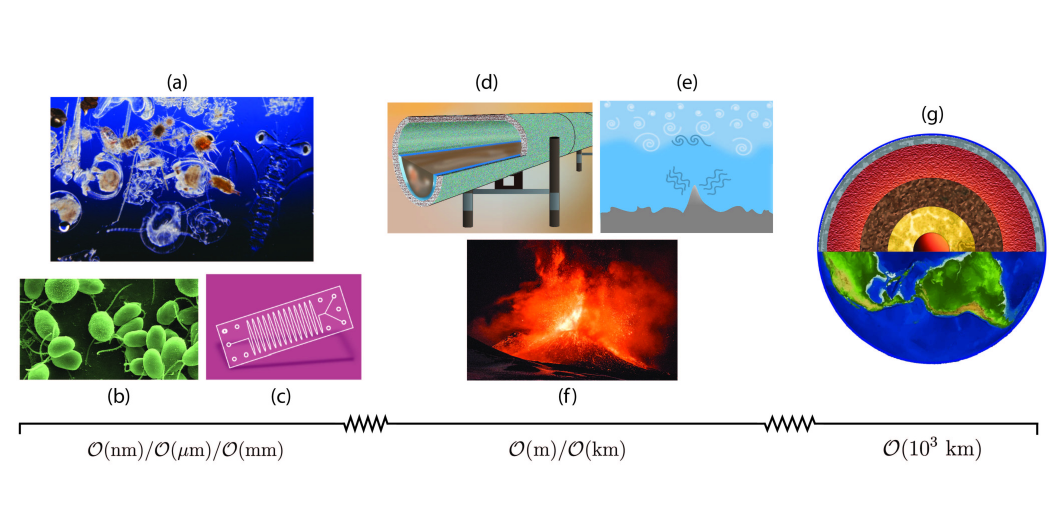}
\caption{Examples of flows where viscosity variations significantly affect dynamics, spanning about 15 orders of magnitude in length scale. (a) Mixed phytoplankton in the ocean (Peter Parks, imagequestmarine.com); (b) Scanning electron microscope image of \textit{Chlamydomonas reinhardtii}, $\sim 10 \mu$m per cell; (c) schematic of a microfluidic device; (d) industrial pipe flow; (e) ocean mixed layer and thermocline with internal waves and topographic forcing; (f) eruption at Mt. Etna (Tom Pfeiffer, volcanodiscovery.com); (g) Schematic of Earth’s crust, mantle and core. Figure compiled by Anugraha A.}
\label{fig:schematic}
\end{figure*}
Figure~\ref{fig:schematic} shows a range of systems in which viscosity variations impact flow dynamics. The associated length scales in these examples span about 15 orders of magnitude, from microorganisms to planetary interiors. Chemotactic swimming of microbes in aquatic environments often leads to aggregation, locally altering the effective viscosity of the suspension. These organisms also secrete mucus with viscosities dramatically higher than those of the ambient fluid, leading to steep viscosity gradients. In such cases, microbes may exhibit viscotaxis-- directed migration in response to viscosity gradients. Thus, active flows are often also viscosity-varying flows.

Industrial flows are constantly miniaturising into the micro- and even the nano- scale, with consequent demand for achieving enhanced mixing at small scale. Can controlling viscosity gradients in micro- and nano-fluidic devices enable enhanced mixing at low Reynolds numbers? We are not sure but think so. In larger-scale (of the order of metres to tens of metres) industrial flows (e.g., chemical reactors, pipelines, prilling towers), substantial temperature and composition gradients are ubiquitous. Evaporative cooling in
data centers is, as they say, a hot topic! Viscosity variations can be of some consequence here. Processes such as sublimation and solidification also naturally introduce spatially variable viscosity. An everyday reminder of spatio-temporal viscosity variation is in the kitchen, where familiar examples include the thickening of chocolate or syrup as it cools.

Geophysical flows, which can extend up to $O(10^4 \textrm{ km})$ in scale, offer another arena wherein viscosity stratification plays an essential role. Especially in Earth’s interior, but also in the atmosphere and oceans, temperature, composition and pressure all affect viscosity, resulting in non-trivial consequences for convection, turbulence, stratified mixing, and biological activity. The famous term ``Panta Rhei'' -- everything flows, attributed to the Greek philosopher Heraclitus, applies to Earth’s mantle: though solid on human timescales, the mantle flows as an extremely viscous fluid over geological epochs. Viscosity varies by several orders of magnitude across depth below the surface of Earth and other planets, strongly influencing mantle convection and plate tectonics. Lava flows exhibit more severe viscosity changes as they cool and crystallize.  While atmospheric gradients in molecular viscosity are generally weak, spatial variations in eddy viscosity need not be. Moreover in particulate flows such as dust storms, gradients in particle number density produce gradients in the effective viscosity. Temperature and salinity determine the viscosity in the ocean, and this can affect many things including marine life. While the upper reaches of the ocean are often turbulent, it is quieter deep below. As a result, sharp variations in eddy viscosity are possible; to make things more interesting, the level of turbulent activity changes with the time of the day and the season, and this has significant impact on heat redistribution. 

We note that viscosity variation is routinely incorporated in certain subfields — e.g., in non-Newtonian rheology and high-Mach number compressible flows (e.g., \citet{wan2022wall, zheng2024effect}); but these are outside the scope of this review. Except in a few contexts such as mantle flow, we focus for the most part on Newtonian fluids in the incompressible, continuum (scales of $\mu$m or larger) limit, at low Mach numbers and across a wide range of scales. In Newtonian fluids, stress is linearly related to the strain rate. In principle, a fourth-order viscosity tensor relates the stress tensor to the rate of strain tensor, but isotropy and frame indifference reduce this to two scalar coefficients. For incompressible flows, one of these, the bulk viscosity, is irrelevant \citep{Aris_1989book}, and the stress can be written in terms of a single spatially varying scalar field: the dynamic viscosity. We present general formulations for low-Reynolds-number flows involving passive or active bodies, as well as for flows with dominant shear. In the former, viscosity gradients can dramatically alter propulsion, force generation, and energy balance. Special materials to which isotropy does not apply can display features such as odd viscosity, which we touch upon. 

The associated time scales span many orders of magnitude as well, from microseconds to geological. Interesting behaviour is displayed at even shorter time scales, which are outside our purview. For example, liquids can display viscoelasticity, as water does when subjected to oscillatory shear or extensional flows in the 50 GHz–10 THz frequency range \citep{o2019viscoelasticity}. We move on to discuss the relevant non-dimensional parameters and their ranges of interest.

\subsection{The Non-dimensional Parameters}
\label{sec:nondim}

A natural way to classify flows is by the range of non-dimensional parameters they reside in. The {\em Reynolds number} is defined as
\begin{equation}
Re \equiv \frac{L U_0}{\nu_0},\label{eq:Reynolds}
\end{equation}
where $L$ and $U_0$ are characteristic length and velocity scales of the flow, and $\nu_0 = \mu_0 / \rho_0$ is a reference kinematic viscosity, defined in terms of reference dynamic viscosity $\mu_0$ and density $\rho_0$. High $Re$ indicates inertia-dominated flows such as turbulence, whereas very low $Re$ corresponds to viscous or creeping-flow regimes (Stokes flows). 

The relative magnitude of viscosity variation may be characterized by the ratio 
\begin{equation}
 \frac{\Delta \nu}{\nu_0},
  \label{eq:viscrat}
\end{equation} 
or equivalently by $m = (\nu_0 + \Delta \nu) / \nu_0$, with $\nu_0$ being chosen as the smallest viscosity in the flow. Note that for convenience we define this ratio in terms of kinematic viscosity, and in density-varying flows this will not be the same as the ratio of dynamic viscosities. We shall make use of the dynamic viscosity ratio too, where more convenient.
The variations typically arise from underlying scalar fields such that the spatiotemporal viscosity field may be generically expressed as 
\begin{equation}
\nu(\mathbf{x},t)=\nu(s_1(\mathbf{x},t),s_2(\mathbf{x},t),\cdots ,s_i(\mathbf{x},t),\cdots ,s_n(\mathbf{x},t)),\label{eq:viscscalardependence}
\end{equation} 
where $\mathbf{x}$ and $t$ denote the spatial location and time respectively, and each scalar $s_i, \ i \in [1,n]$ (e.g., temperature, solute concentration, salinity) evolves due to advection and diffusion. For each scalar, a {\em Péclet number} estimates the relative importance of its advection to its diffusion,
\begin{equation}
Pe_i \equiv \frac{L U_0}{\kappa_{s_i}}, \quad i \in [1,n], \label{eq:Péclet}
\end{equation}
where $\kappa_{s_i}$ is the molecular diffusivity of scalar $s_i$ (e.g., thermal diffusivity for temperature, or mass diffusivity for solute). A small Péclet number implies fast diffusion, while a large $Pe$ implies advection-dominated transport. The Péclet number may also be written as the product of the Reynolds number and a second non-dimensional group representing the ratio of momentum to scalar diffusivity. For thermal diffusion, the relevant number is the \textit{Prandtl number},
\begin{equation}
Pr \equiv \frac{\nu_0}{\kappa_T}, \label{eq:Prandtl}
\end{equation}
and for solute transport in multicomponent flows, it is the \textit{Schmidt number},
\begin{equation}
Sc_i \equiv \frac{\nu_0}{\kappa_{m_i}},
\end{equation}
where $\kappa_T$ is the thermal diffusivity and $\kappa_{m_i}$ is the mass diffusivity of species $i$. Thus, $Pe_i = Re \, Pr$ or $Re \, Sc_i$ as appropriate. The diffusivities $\kappa_{s_i}$ merit discussion. In gases, molecules diffuse rapidly, transporting momentum, heat, and mass at similar rates, resulting in Prandtl and Schmidt numbers typically of order one. This is because the velocity and length scales in gaseous diffusion are set by the Maxwell–Boltzmann distribution and the mean free path, respectively. In liquids, however, molecular transport behaves quite differently. Due to close molecular packing, mass diffusion is slow as molecules must wait for sufficiently large gaps between neighbours. In contrast, momentum is transferred more readily because shear forces can be communicated through collective rearrangements of neighboring molecules, leading to a coordinated large scale motion. As a result, liquids diffusing into each other, or solutes in solutions diffusing into liquid, often exhibit large Schmidt numbers: $Sc \sim \mathcal{O}(10^2 -10^4)$ for salt water diffusing into fresh water, sugar solution into fresh water, or glycerol in water. Thermal diffusion in non-metallic liquids lies between these two extremes. Individual molecules at higher temperatures vibrate more vigorously, transferring energy to neighbors via collisions, but the requirement of the physical contact implies that heat does not propagate as efficiently as momentum. This results in Prandtl numbers that are $\mathcal{O}(10)$ (e.g., $Pr \approx 7$ for water), except in metallic liquids, which have extremely high heat conductivity, i.e., extremely low Prandtl numbers ($Pr \approx 10^{-2}$ for mercury).
In this review, we assume constant scalar diffusivities for simplicity, but in general each of these quantities themselves may depend on temperature, composition, and pressure, i.e., 
\begin{equation}
\kappa_{s_i} = \kappa_{s_i}(s_1, s_2, \ldots, s_N, p).
\end{equation}
This functional dependence introduces nonlinearity, and further coupling between the scalar advection–diffusion equations and the Navier–Stokes equations via the viscosity field. Such couplings, between multiple scalars and between scalars and momentum, remain relatively unexplored and represent a rich avenue for future work. 

When viscosity variations co-occur with density variations, such as in heated or stratified multicomponent flows, buoyancy forces must also be included via the {\em Rayleigh number} for each species which contributes to density changes:
\begin{equation}
Ra_i \equiv \frac{g \, \alpha_i \Delta s_i \, L^3}{\rho_0 \, \nu_0 \, \kappa_{s_i}}, \quad i \in [1,n], \label{eq:Rayleigh}
\end{equation}
where it is assumed that each scalar concentration change contributes linearly to density change via the factor $\alpha_i$, and $g$ is the magnitude of gravitational acceleration. A positive Rayleigh number indicates that fluid density decreases in the direction along gravity, which is an inherently unstable arrangement. Finally, in particle-laden flows, the {\em Stokes number} or the ratio of the particle response time $\tau$ to a flow timescale, given by
\begin{equation}
St \equiv \frac{\tau U_0}{L},
\label{eq:stokesnumber}
\end{equation}
becomes relevant.

\begin{table*}
\caption{Typical length scales and non-dimensional numbers across systems discussed in this review. The Rayleigh number’s sign depends on the alignment of density stratification with gravity, and hence can take a negative value. Viscosity variation is measured as a ratio to a reference value.}
\begin{tabular}{| S{m{2cm}} | S{m{2.7cm}} | S{m{2.7cm}} | S{m{2.7cm}}  | S{m{3.4cm}} | Sl|}
\hline
& Microorganisms  & Microchannels  & Industrial/ Kitchen & Ocean/ Atmosphere   & Mantle  \\ [1.5ex] 
\hline
Length scales & $10^{-7}$ to $10^{-6}$ m &  $10^{-7}$ to $10^{-4}$ m &  $10^{-2}$  to  $10$ m  & $10^2$  to  $10^6$ m & $10^4$ to $10^6$  m\\ [1.5ex] 
\hline
$Re$  & $10^{-4}$ to $10^{-1}$ & $10^{-1}$ to $10$ & $10^3$ to $10^4$ & $10^{8}$  & $\approx 10^{-20}$ \\ [1.5ex] 
\hline
$Ra$  & $0$ &   $0$ & $-10^8$  to  $10^8$ & $-10^{13}$  to  $10^{22}$  & $10^{4}-10^{8}$ \\ [1.5ex] 
\hline
$Pr$ or $Sc$  &  $10^0$ to $10^4$ & $10^0$ to $10^4$  & $10^0$ to $10^4$ & $10^0$ to $10^2$ & $\approx 10^{25}$  \\ [1.5ex] 
\hline
$(\Delta \nu)/\nu_0$  &  $0$ to $10^2$ & $0$ to $10^3$ & $0$ to $10^4$ & $0$ to $10^{-1}$  &$\approx 10^{-18}$ \\ [1.5ex]
\hline
\end{tabular}
\end{table*}

Table~1 summarizes the range of length scales and non-dimensional parameters that will be encountered in this review. Paradoxically, the two extremes in length scale: the microscale motion of swimming organisms and the macroscale convection of Earth’s mantle, share the characteristic of extremely low Reynolds number and can both be modelled within the Stokesian regime. In the intermediate length scale range, industrial and everyday flows span the entire range of states from laminar to turbulent, through a variety of intermediate, or transitional, states. Microchannel flows tend to remain laminar, while oceanic and atmospheric flows have extremely high Reynolds numbers due to their vast spatial scales. The mantle has a high Rayleigh number, with the lower portions being hotter and lighter than the upper, and this leads to significant large-scale convection. Besides, vigorous smaller-scale convection is possible due to large spatial heterogeneities. All the flows can support a vast range of Prandtl or Schmidt numbers. When the Prandtl number is close to zero, temperature perturbations smooth themselves out instantaneously, simplifying the physics somewhat, but a mean temperature gradient can be maintained by imposing suitable boundary conditions. At large Schmidt numbers, a sharp interface across which there is a jump in viscosity can retain itself practically forever, resulting in instabilities and other singular behaviour.

\subsection{Dynamical effects of viscosity stratification and organization of the review}
\begin{figure*}
\centering
\includegraphics[width=0.8
\textwidth]{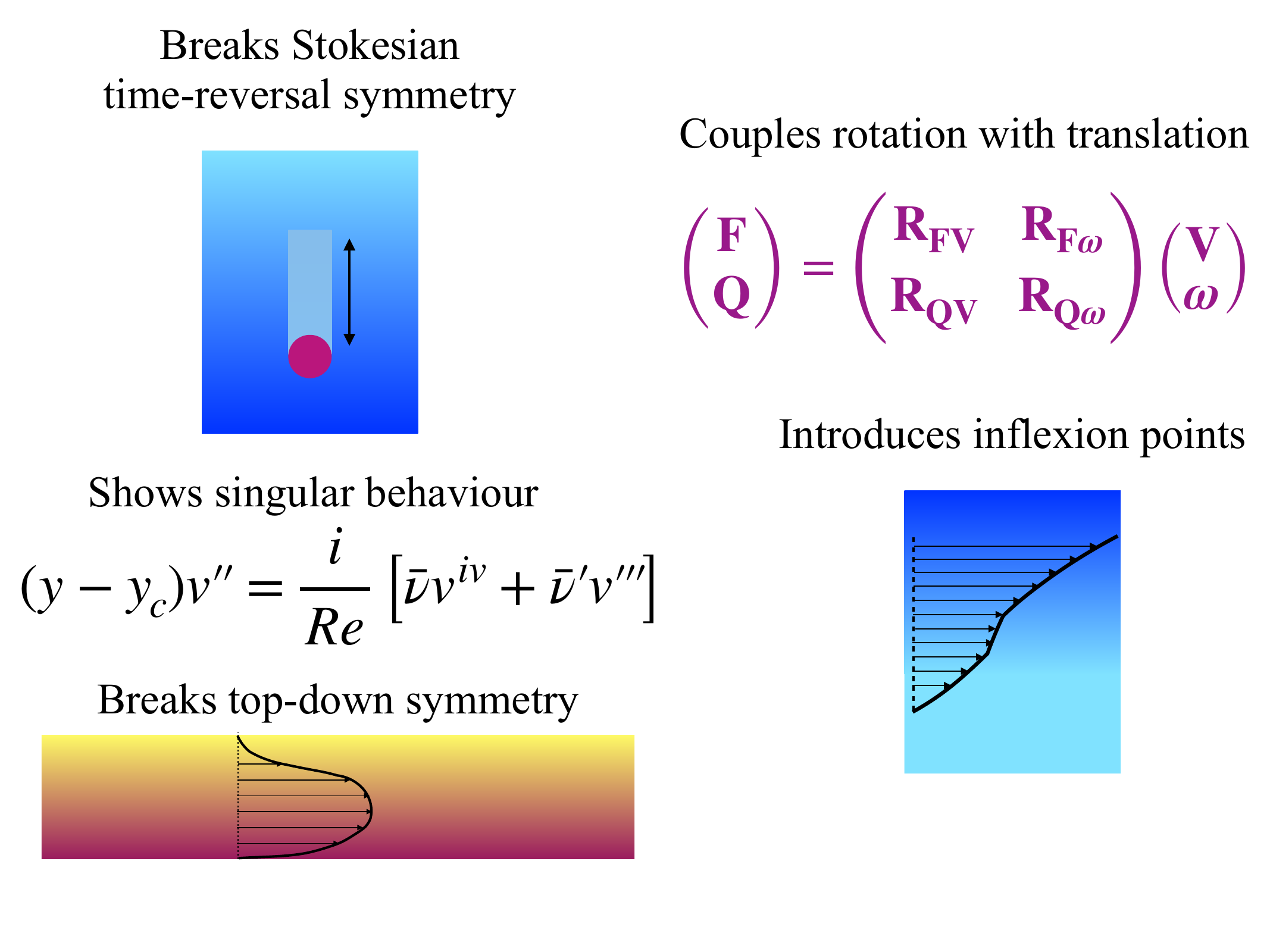}
\caption{Alterations in dynamics due to viscosity stratification (indicated by the colour gradient). Top left: a Stokesian particle in reciprocal motion cannot retrace its steps because the surrounding fluid has been disturbed on its onward motion, see section \ref{sec:PassiveinStratified}. Top right: Viscosity gradient makes a linearly forced object rotate and a body under external torque execute linear motion, as discussed in section \ref{sec:PassiveinStratified}. Middle left: at high Reynolds number, viscosity variation provides a singular perturbation, as discussed in section \ref{sec:singular}). Bottom left: at any length scale, the velocity profile in an asymmetrically heated channel or pipe is not top-down symmetric, see e.g., section \ref{sec:ShearFlowInstability}. Bottom right: a constant shear stress flow can develop inflexion points, see e.g. section \ref{couette_pipe}. All of these can produce intricate structures and vorticity patterns.}
\label{fig:what_visc_does}
\end{figure*}

Across length and time scales variations in viscosity affect flow systems in a fundamental way. Figure \ref{fig:what_visc_does} summarises the physics that we will discuss in the following sections. Before we go into each system we have chosen, it is useful to list some canonical alterations to the dynamics due to viscosity stratification. The mechanisms that are brought in are often common to several classes of viscosity-stratified systems. 

Stokes flow is time-reversal symmetric, which has important consequences for the dynamics. But viscosity variation can break this symmetry. Imagine a body falling under gravity through a very viscous fluid, covering a distance $z$ in time $t=z/v_t$ where $v_t$ is its terminal velocity based on local viscosity. Now in a constant viscosity fluid, if we reverse the sign of gravity, the body would retrace its dynamics. However, if the viscosity is stratified as shown in figure \ref{fig:what_visc_does}, with lower viscosity on top, the body would drag some low viscosity fluid on its way down, and its return journey upon reversing gravity would be faster on average. When we apply a linear external force to a non-spherical particle in the presence of a viscosity variation, it can make the particle rotate. So an object merely falling under gravity through a very viscous stratified fluid can execute interesting rotational motion on its way down. Conversely, when we apply an external torque, the viscosity variation can produce linear drift. 

In shear flows, such as the channel shown in figure \ref{fig:what_visc_does}, a gradient in viscosity breaks top-down symmetry, with consequences for the dynamics and for the stability of the system. Also even after the flow becomes fully turbulent, the fact that molecular viscosity is different at the two walls makes turbulence and turbulent structures top-down asymmetric. In particular the large structures near the two walls are different. Since shear stress is continuous across a flow, sharp viscosity variations introduce sharp shear changes, and viscosity variations can introduce inflexion points in velocity profiles where there were none with constant viscosity. This fundamentally alters the stability behaviour of the viscosity-varying flow which now lends itself to Kelvin--Helmholtz and other instabilities, which were absent in the case with constant viscosity. The Rayleigh-Fj{\o}rtoft theorem states that inflexional profiles with a vorticity maximum will be `inviscidly' unstable. This means that in the absence of viscous effects, i.e., upon setting the Reynolds number to infinity, such a mean flow profile would be unstable. Though there is no corresponding theorem for finite Reynolds number, the concept is extremely useful: a thumb rule is that inflexional profiles are strongly associated with low Reynolds number instability, and the term `inviscid instability' is applied even at finite Reynolds number. Thus, strangely, viscosity variations can create inviscid instabilities. We remark that although inflexional profiles do not `need' viscosity in the stability equation to go unstable, the profiles themselves were created by viscosity, and often by variations in viscosity. Finally viscosity variations can accentuate and modify the singular nature of high Reynolds number shear flows, making new instabilities and dramatic stabilisation possible. 

We end this section hoping that the reader is intrigued, and ready to be convinced about the appeal, the applicability, the breadth, the physics and the technological relevance of studying viscosity variations. This review is organized as follows. Section \ref{sec:equations} discusses how viscosity variations enter the governing equations, followed by a brief discussion on the origins of viscosity and some standard trends in viscosity as a function of the relevant physical quantities in section \ref{sec:ViscPropertyofState}. In section \ref{sec:ParticleDynamics}, we discuss dynamics of passive and active particles in the Stokesian limit in viscosity-varying flow and contrast this to the uniform viscosity situation. In section \ref{sec:shear_flows_var_visc}, we discuss how viscosity variation modulates the singular perturbation structure in high Reynolds number flows. Viscosity variations significantly alter stability characteristics, and we discuss modal and non-modal stability of viscosity-varying flows in multiple model systems in section \ref{sec:instab_due_to_visc_strat}. In section \ref{sec:turb}, we discuss implications of viscosity variations in fully developed turbulent flows, including those relevant to geophysical phenomena. In section \ref{sec:EarthFlows}, we discuss how earth-scale motions like mantle flows are impacted by viscosity variations. In each section, several areas of possible future research are discussed. Finally, in section \ref{sec:conclusions} we return to the commonalities across length scales and applications, and outline some general thoughts. 

\section{The equations for viscosity-varying flows}\label{sec:equations}
We present the governing equations in non-dimensional form, using the parameters introduced in the previous section. The mass and momentum conservation equations for an incompressible fluid are given by the continuity and Navier–Stokes equations:
\begin{align}
&\nabla \cdot \bu = 0, \label{eq:incomp} \\
&\frac{\partial \bu}{\partial t} + \bu \cdot \nabla \bu = \frac{1}{Re}\nabla \cdot \bm{\sigma} + \bm{f}_\text{buoyancy}, \label{eq:ns}
\end{align}
where boldface quantities are vectors or matrices, $\bu$ is the velocity field, $\bm{\sigma}$ is the fluid stress tensor, and $\bm{f}_\text{buoyancy}$ (discussed below) is the buoyancy force relevant in geophysical and other density-stratified flows.  The fluid stress tensor is given by
\begin{equation}
\bm{\sigma} = -Re\, p \bm{I} + \mu(\mathbf{x},t)\left(\nabla \bu + \nabla \bu^\mathsf{T} \right), \label{eq:StressDef}
\end{equation}
where $p$ is the pressure field over the hydrostatic value, which acts as a Lagrange multiplier to enforce incompressibility, and $\bm{I}$ is the identity tensor. Note that, except where explicitly specified otherwise, $\mu(\mathbf{x},t)$ and $\nu(\mathbf{x},t)$ have been nondimensionalised by their characteristic values $\mu_0$ and $\nu_0$ respectively. At zero Reynolds number (the Stokesian limit), we will, for convenience, absorb the factor of $Re$ into the definition of pressure (see equation \eqref{eq:MassandMomentum} in section~\ref{sec:ParticleDynamics}). Compared to the constant-viscosity formulation, where the stress divergence term in the momentum equation simplifies to $\nabla^2 \bu$, by virtue of incompressibility, in flows with spatially varying viscosity, we have
\begin{equation}
\nabla \cdot \bm{\sigma} =-Re \nabla p + \mu \nabla^2 \bu + (\nabla \bu + \nabla \bu^\mathsf{T}) \cdot \nabla \mu.\label{eq:StressDiverDecomp}
\end{equation}
Thus, in addition to the pressure gradient and a weighted (by variable viscosity) diffusion of the velocity field ($ \mu \nabla^2 \bu$), an additional term proportional to the viscosity gradient is introduced. This term is responsible for many of the distinct physical effects discussed in this review.

As discussed, viscosity may depend on multiple scalar fields such as temperature or solute concentration (equation \eqref{eq:viscscalardependence}). The transport of each of these scalars is governed by
\begin{equation}
\frac{\partial s_i}{\partial t} + \bu \cdot \nabla s_i = \frac{1}{Pe_i} \nabla^2 s_i, \qquad i \in [1,n]. \label{eq:species}
\end{equation}
In all the flows under discussion, the incompressibility condition \eqref{eq:incomp} is reasonable, and further, we make the Boussinesq approximation \citep{boussinesq1903theorie,spiegel1960boussinesq,kundu2024fluid}. By this approximation, density variations are neglected everywhere except in the buoyancy term of the momentum equation. The buoyancy force is
\begin{align}
\bm{f}_\text{buoyancy} 
= \sum_{i=1}^N \frac{1}{\rho_0} \left[\frac{\partial \rho}{\partial {s}_i}\right] \delta {s}_i {\mathbf{g}} \frac{L}{U_0^2} = \sum_{i=1}^N \frac{Ra_i}{Pe_i Re} \delta s_i\mathbf{\mathbf{\underaccent{\tilde}{z}}}, \label{eq:BuoyancyForcing}
\end{align}
where $\delta s_i = s_i - s_{0,i}$ is the deviation from a reference (hydrostatic for temperature) scalar profile, ${\mathbf{g}}$ is the acceleration due to gravity, $\mathbf{\underaccent{\tilde}{z}}$ is a unit vector in the gravity direction and $Ra_i$ is the Rayleigh number associated with scalar $s_i$. A linear relationship with a proportionality constant $\alpha_i$ is assumed for density variation due to every scalar, reconciling the definition in equation \eqref{eq:Rayleigh}. The Boussinesq approximation is not valid close to very strong vorticity, large density changes, and in some other situations, but those are outside our purview.

High Reynolds and Rayleigh numbers are associated with flow instability, transition to turbulence, and fully developed turbulence, while low values are indicative of steady laminar flow. How much is low and how much is high is extremely sensitive to geometry, background noise and other specifics. Consider first the high-$Re$ ($Re \gg 1$) regime. The viscosity ($\propto 1/Re$) multiplies the highest-order derivatives in the momentum equation \eqref{eq:ns}. A visual consideration of this equation gives the indication that in the singular limit $Re \to \infty$, we may neglect viscosity itself, as well as its gradients. But let us look again. If we drop the viscous terms, the differential order of the equation is reduced, i.e., viscous effects, no matter how small, cannot be neglected outright if all boundary conditions are to be satisfied. As a result, the viscosity field brings about a singular perturbation, and can exert substantial influence on the solution, especially in regions of strong gradients or near boundaries. We return to this point in detail in section~\ref{sec:singular}.

Large Rayleigh numbers, where $Ra_i \gg Pe_i Re$, imply strong buoyancy forcing (via equation \eqref{eq:BuoyancyForcing}), which can drive convection and transition to turbulence. Similarly, large Péclet numbers (in equation \eqref{eq:species}) indicate that scalar transport is advection-dominated, leading to scalar turbulence with sharp gradients and filamentation of scalar fields.

The interdependence of scalar fields $s_i$ and the velocity field $\bu$ in the momentum and scalar transport equations \eqref{eq:ns} and \eqref{eq:species} implies that these scalars are \textit{active} --- in the sense that they influence the flow field, rather than merely being passively advected. This notion of active scalars is distinct from the notion of \textit{active particles}, which we will discuss separately in section~\ref{sec:ParticleDynamics}. The feedback from scalar fields to momentum transport occurs not only via the buoyancy forcing term $\bm{f}_\text{buoyancy}$ (equation~\eqref{eq:BuoyancyForcing}), but also through the dependence of viscosity on these scalars, as indicated in equation~\eqref{eq:viscscalardependence}. In the next section, we explore from first principles how viscosity depends on temperature, concentration, and other scalar quantities.

\section{What determines viscosity?}
\label{sec:ViscPropertyofState}
Most of us would agree that honey is a highly viscous fluid, while we would regard air as having low viscosity. Yet, to a microorganism, air must feel like a highly viscous goo! The concept of a fluid’s “stickiness” has been discussed for centuries—by philosophers and naturalists, but the formal notion of \emph{viscosity} emerged much later \citep{tanner1998rheology}. Viscosity quantifies momentum transport by molecular processes; it smooths out velocity gradients and dissipates kinetic energy. It is a well-defined and measurable property, routinely characterized using viscometers. Viscosity is interesting not only because of its functional role in flows, but also due to its nontrivial dependence on the physical state of a fluid. Gases and liquids, for example, exhibit opposite trends in viscosity with increasing temperature, and the viscosity of a mixture can sometimes exceed that of its individual components. 

In his classic paper on low Reynolds number flows, \citet{purcell1977life} asked: why does the kinematic viscosity of liquids never fall below a certain minimum value? Some explanation from a condensed matter physics perspective is now available for this phenomenon, as summarised in the review of \citet{Trachenko_2021AP}. This minimum arises because, although increasing temperature reduces a liquid’s relaxation time (and hence its viscosity), this reduction is bounded below by the Debye vibration period, $\tau_D \approx r_m/c_s$, where $r_m$ is the intermolecular separation and $c_s$ is the speed of sound in the medium. Beyond this point, further temperature increases transform oscillatory (liquid-like) molecular motion into diffusive (gas-like) behavior, which again increases viscosity. This sets a floor at $\nu_{\text{min}} = {r_m^2}/{\tau_D} \approx r_m c_s$. Using further fundamental considerations, this can be related to Planck’s constant $h$, yielding $\nu_{\text{min}} = {h}/{4\pi \sqrt{m_e m_p A_w}},$ where $m_e$ and $m_p$ are the masses of the electron and proton, respectively, and $A_w$ is the atomic weight. The weak dependence on $A_w$ makes this minimum viscosity similar across liquids.

Beyond this `universal' floor, viscosity varies considerably across substances and conditions. In the rest of this section, we explore how viscosity depends on temperature, solute concentration, and pressure. We also consider the case of polar fluids where viscosity is no longer a scalar quantity.

\subsection{Viscosity variation with temperature}
\begin{figure*}
\centering
\includegraphics[width=1.0\linewidth]{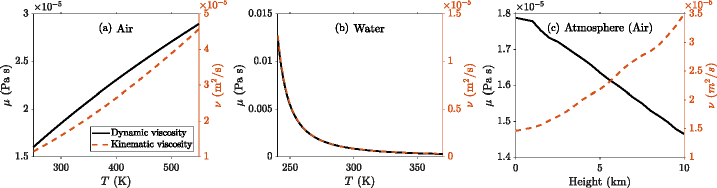}
\caption{Dynamic viscosity $\mu$ (dimensional, in Pa$\cdot$s; left axis; solid black curve) and kinematic viscosity $\nu$ (dimensional, in m$^2$/s; right axis; dashed orange curve) of (a) air and (b) water as functions of temperature $T$ at standard pressure. Data generated using the NIST REFPROP toolbox \citep{Huber_etal_2022IECR}. Data in the supercooled region in (b) is from \citet{dehaoui2015viscosity} and \citet{hare1986densities}. (c) Dynamic and kinematic viscosity of air as functions of altitude in Earth's atmosphere.}\label{fig:air_water_visc}
\end{figure*}

In the late nineteenth century, \citet{mallock1889iv,mallock1896iii} conducted some of the first systematic experiments demonstrating the dependence of water's viscosity on temperature. We have noted the remarkable feature of liquids and gases exhibiting opposite trends in viscosity (both dynamic and kinematic) with temperature ($T$). These dependencies are shown for air and water in  figures \ref{fig:air_water_visc}(a) and (b), respectively. We notice that the dynamic and kinematic viscosities of water follow the same trend. The is because density changes in liquid water are small across the temperature range shown. But some difference is seen in the case of $\mu$ and $\nu$ for air. The difference between liquids and gases has important implications in applications where fluid temperature is controlled through the bounding surfaces. For example, if wall \emph{heating} stabilizes a liquid flow, by creating a negative viscosity gradient as one approaches the wall, and thence a fuller velocity profile, then wall \emph{cooling} would be required to achieve the same effect in a gas. For gases, kinetic theory predicts a viscosity scaling of $\mu \sim \sqrt{T}$, while Sutherland’s law \citep{Sutherland_1893}, which is a refinement on this scaling, offers a better empirical fit across wider temperature ranges. Further improvements for real gases incorporate intermolecular interactions through the Boltzmann equation \citep{Chapman_Cowling_1970book}. For gas mixtures, Wilke’s model \citep{wilke1950viscosity} is widely used in engineering contexts (see e.g., \citet{berger2022synergistic}). Viscosity variations with altitude in Earth's atmosphere are worth a mention. Figure~\ref{fig:air_water_visc}(c) illustrates how dynamic viscosity decreases with altitude, while kinematic viscosity increases, due to the more rapid drop in density with height. 

For liquids, the simplest kinetic theory models invoke an activation energy required for molecules to escape the constraints of their neighbours \citep{Bird_etal_2005book}, leading to an Arrhenius form for viscosity: $\mu \sim \exp(B/T)$ (first proposed by \citet{andrade1930viscosity}), where $B$ is a liquid-dependent constant. While the Arrhenius law offers a convenient parametrization of the temperature dependence of liquid viscosity, deriving transport properties such as viscosity from microscopic principles is far more challenging for liquids than for crystalline solids or gases. To address this, \citet{rizk2022microscopic} used molecular dynamics simulations to study single-particle Lennard-Jones liquids -- simplified model systems that reproduce key structural and dynamical features of van der Waals or metallic fluids such as rare-gas liquids, liquid CO$_2$, and simple hydrocarbons. They found that, fortuitously, the
Arrhenius law holds over a wide range of temperatures and pressures. However, the origin of this behavior does not lie in thermal activation over energy barriers, since thermal energies are an order of magnitude lower than the nominal activation energy. Instead, the temperature dependence of viscosity in these systems is better captured by the Stokes–Einstein relation, which links viscosity to temperature and diffusivity. \citet{rizk2022microscopic} also proposed an alternative framework, the free volume model, where molecular motion is treated as a sequence of hard-sphere-like collisions. In this view, viscosity is governed by the probability of finding local free volume for molecular rearrangement, and the diffusion coefficient is derived from measurable microscopic quantities and mean free paths. For more structurally complex liquids, such as water, the microscopic origins of viscosity remain an open question. 

When solidification occurs in response to reducing temperatures, the behaviour in the liquid phase depends on whether a glass or a crystal is formed. The liquid phase shows a dramatic increase in viscosity as one approaches glass transition, whereas just before crystallisation, the liquid in equilibrium with the crystal has unremarkable viscosity properties. During crystallisation, small crystal structures are nucleated within the liquid background, which grow to cover the entire space. Once the process of crystallisation begins, we expect, and see, a sharp increase in apparent viscosity. But that is not all that can happen. In partially crystallizing fluids such as oil–paraffin mixtures, \citet{himo2021chaos} found that the apparent viscosity oscillates in time in some temperature range, and shows chaotic fluctuations with time in a lower temperature range. This highlights the complexity of ``mushy" systems, where solid and liquid phases coexist, and points to open questions regarding how the effective viscosity should be defined in such regimes. 
Supercooled liquids are obtained in the absence of impurities or nucleation sites by cooling below the freezing temperature. There are many models available for viscosity of supercooled liquids, see e.g., \citet{mauro2009viscosity}. A common feature during solidification is the occurrence of chemical changes, such as resulting from the ejection of solute into the liquid. For example, as Arctic sea-ice freezes, it ejects salt into the surrounding water, changing the ambient viscosity. Incidentally Arctic waters can be significantly supercooled, and, as seen in figure \ref{fig:air_water_visc}(b), viscosity can respond sensitively to temperature. Thus supercooling is not merely academic; it occurs naturally in important contexts.

\subsection{Variation with concentration}\label{sec:SoluteConcent}
Consider this simple experiment: put five large spoonfuls of sugar into 200~ml of water, stir well, and measure the viscosity. Now take 400 ml of water, add the same amount of sugar, and boil the mixture until it reduces to 200~ml. After cooling, measure the viscosity of the second solution too. Which viscosity is higher? Just the fact that the two solutions, of the same concentration at the same temperature, are of different viscosities, highlights a subtle but important point: viscosity can depend on the thermodynamic path taken to reach a given state, since the path can alter intermolecular interactions.

\begin{figure}
\centering
\includegraphics[width=0.50\textwidth]{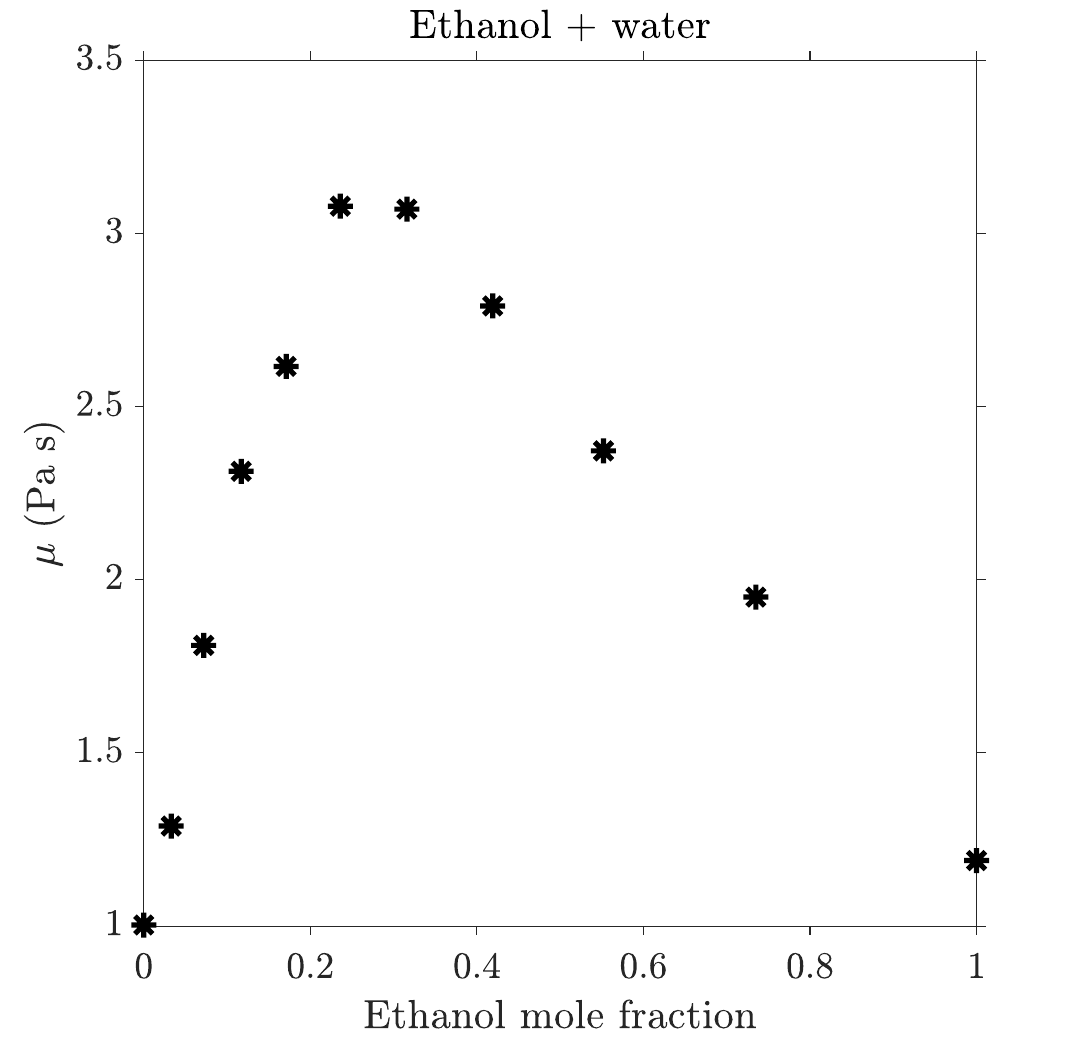}
\caption{Dynamic viscosity for an ethanol-water mixture at 293 K; experimental data from \citet{khattab2012density} was used for this figure.}	\label{fig:ethanol_water_visc}
\end{figure}

When a solute dissolves in a solvent, the viscosity typically increases. A rough empirical rule, proposed by \citet{arrhenius1887innere,arrhenius1917viscosity}, is that $\mu \propto \exp(c_s)$, where $c_s$ is the concentration of the solute (this viscosity-concentration relation is similar to the temperature-concentration relation for liquids discussed earlier). Mixtures of water and glycerol (commonly used to access a wide viscosity range), oil–oil combinations, and some ionic liquids deviate from Arrhenius’s concentration rule \citep{trejo2011viscosity,Shankar_Kumar_1994PRSA,niedermeyer2012mixtures,hayes2015structure}. A mixture of water and ethanol, so common at the dinner table, is an interesting example  that shows non-monotonic variation in viscosity with concentration of ethanol as shown in figure \ref{fig:ethanol_water_visc} (also see experiments mentioned in \citet{yusa1977viscosity,gonccalves2010pvt}). Such variation is common to a wide range of alcohol-water mixtures and also to chemically non-reacting solutes. This behavior is difficult to model and remains an active area of research. Various predictive methods for viscosity, such as the principle of corresponding states \citep{Ely_Hanley_1981IECF}, succeed in specific regimes, but comprehensive models are still lacking \citep{Thol_Richter_2021IJT}.

In the case of particulate suspensions, a different set of models applies. For semi-dilute suspensions with particle volume fraction $\phi$, a classical expression is
\begin{equation}
\frac{\mu_{\text{susp}}}{\mu_{\text{ps}}} = 1 + 2.5 \phi + 6.2 \phi^2, \label{eq:SuspRheology}
\end{equation}
where $\mu_{\text{susp}}$ and $\mu_{\text{ps}}$ denote the viscosities of the suspension and the pure solvent, respectively. The linear term represents Einstein’s viscosity law \citep{einstein1906new,einstein1911_correction}, valid in the dilute limit. The quadratic correction, derived by \citet{batchelor1977effect}, accounts for Brownian motion and hydrodynamic interactions between suspended particles. \citet{arrhenius1917viscosity} compared early experimental measurements with Einstein’s law and found good agreement under idealized conditions, for example, with suspensions of gamboge particles (0.3–4~$\mu$m) in a glycerol-water mixture of viscosity approximately $15$ times that of pure water. However, in most other cases, deviations from the ideal theory giving equation \eqref{eq:SuspRheology} were observed, often due to effects such as flocculation, non-sphericity, or enhanced Brownian interactions. Over the decades, empirical improvements over \eqref{eq:SuspRheology} have been proferred, to be applicable over the range of particle number density. One example is the correlation of \citet{krieger1959mechanism}, where viscosity diverges as particles approach close-packing. For an excellent summary of the rheology of dense suspensions in Newtonian and non-Newtonian fluids, the reader is directed to \citet{guazzelli2018rheology}.

\subsection{Variation with pressure}

Except at extremely high pressures, the viscosity of liquids is relatively insensitive to pressure. For instance, at pressures around 400–500 atmospheres, such as those found at the bottom of the ocean, the dynamic viscosity of fresh water remains practically unchanged from its value at atmospheric pressure at the same temperature. The pioneering studies of \citet{bridgman1925viscosity, Bridgman_1926PAAS} extended the investigation of liquid viscosities to pressures exceeding thousands of atmospheres, conditions relevant to planetary and stellar interiors, as well as to the development of high-performance materials \citep{Mcmillan_2005NM}. 

In most liquids, viscosity increases approximately exponentially with pressure due to reduced molecular mobility under extreme compression \citep{Bridgman_1926PAAS, dymond1981transport}. This effect becomes significant in technological applications such as lubrication, where mechanical components operate under high-pressure conditions \citep{Bair_2007book,Szeri_2011book}. In such regimes, constitutive models that relate stress and strain rate implicitly have been employed to capture the pressure dependence of viscosity \citep{Rajagopal_2006JFM}. Bridgman observed that at low temperatures (0 and 10.3$^\circ$C), the viscosity of water initially decreases as pressure is increased, before eventually rising in response further pressure increase. The initial decrease is attributed to the disruption of the tetrahedral hydrogen-bond network, which enhances molecular mobility \citep{debenedetti2003supercooled}. At lower temperatures, the tetrahedral network is more pronounced at normal pressures, leading to stronger pressure dependence and the appearance of a clear viscosity minimum \citep{bett1965effect, debenedetti1996metastable, huber2009new, singh2017pressure}. For gases too, viscosity increases with pressure, though the rate and functional form of the increase depend on both the gas and the temperature, as shown by early experimental work on CO$_2$ and N$_2$ \citep{phillips1912viscosity, michels1931measurement}. 

Thermodynamic and transport properties of a wide range of fluids have been systematically compiled in engineering handbooks and databases \citep{Poling_etal_2001book,Bejan_Kraus_2003book,Tropea_etal_2007book,stephan2013viscosity,viswanath2007viscosity}. A notable resource is the NIST REFPROP software, which provides accurate thermophysical properties of pure fluids and mixtures under a wide range of operating conditions \citep{Huber_etal_2022IECR}. Numerous empirical correlations for pressure-dependent viscosity exist and continue to be developed. However, a unified theory of viscosity variation, valid across broad ranges of pressure, temperature, and composition in all classes of fluids, remains an open challenge.

\subsection{Generalised Newtonian fluids}

Fluids with polar molecules may necessitate the consideration of the internal distribution of moments, so the stress tensor may not be symmetric \citep{Stokes_1984book}. Even if the stress tensor is symmetric, the underlying viscosity tensor may be anti-symmetric. [The two are not mutually exclusive. For detailed analyses of the viscosity tensor based on symmetry arguments, see \citet{machado2023hamiltonian} in two dimensions (2D) and \citet{khain2022stokes} in three dimensions (3D).] This is the case in polyatomic molecules \citep{korving1967influence}, plasmas in magnetic fields \citep{landau1987fluid}, and more generally in active matter, where microscopic internal rotational degrees of freedom can manifest themselves at the hydrodynamic scale \citep{markovich2021odd}. 

Anti-symmetric viscosity coefficients are termed odd viscosity, and historically the term odd, or Hall, viscosity dates back to the work of \citet{avron1995viscosity} and \citet{avron1998odd} in 2D quantum fluids. Several additional terms appear now in the momentum balance equation (see \citet{Fruchart_etal_2023ARCMP} for a recent review). In 2D, the effects of odd viscosity always manifest themselves at boundaries, modifying free surface dispersions and chiral edge modes \citep{ganeshan2017odd}. The only bulk effect is to modify the pressure. However, in 3D, the bulk flow is generally modified, whereby the `odd' Stokeslet produces a spiral flow pattern \citep{khain2022stokes}. Contemporary research on odd viscosity aims to clarify its effect on lift and drag forces acting on immersed bodies \citep{lier2023lift}, and to further develop a theory of odd turbulence \citep{chen2024odd}. Both directions promise to flesh out the many intricate properties associated with parity-broken flows. We will discuss a stability example in section \ref{sec:jets_films}. 

Having introduced the governing equations for flows with spatially varying viscosity in the previous section, and the fundamental mechanisms by which such variations arise in real fluids, we are now ready to examine their consequences across a variety of fluid mechanical problems. These problems are not only physically rich, but also of significant technological relevance. We begin by exploring the effect of viscosity stratification on the motion of solid particles embedded in flow.

\section{Particulate matter in viscosity stratified flows}\label{sec:ParticleDynamics}

Viscosity stratification and particle loading frequently coexist in natural and industrial flows. Such coexistence is prominent in systems like inks, biological fluids (e.g., mucus, blood), and seawater, where viscosity gradients arise from variations in plasma content, red blood cell concentration \citep{nader2019blood}, salinity \citep{sharqawy2010thermophysical}, or suspended biomass such as phytoplankton \citep{guadayol2021microrheology}. The interplay between viscosity stratification, particle properties (e.g., shape, porosity, activity), and fluid characteristics (e.g., density, solute concentration, non-Newtonian rheology) can generate complex behaviors relevant across a range of applications. These include tuning rheology in direct ink writing \citep{ye2021study}, removing oil droplets from seawater \citep{gao2013seawater}, designing biomedical delivery systems, and understanding phytoplankton sedimentation patterns \citep{chajwa2024hidden}. Phytoplankton alone account for nearly 30\% of anthropogenic carbon sequestration via sedimentation into the deep ocean \citep{gruber2019oceanic}, highlighting the importance of understanding whether viscosity -- or eddy viscosity -- gradients influence their settling dynamics—knowledge that could inform artificial carbon sequestration strategies \citep{jones2022climate}. To isolate the underlying mechanisms, researchers often examine the effects of individual particle or fluid features in controlled settings.

The dynamics of particles in both Newtonian and non-Newtonian fluids have been extensively studied, yet the influence of viscosity stratification, even in non-Brownian particle dynamics, has only recently gained attention. Stratification, especially when coupled with particle geometry, can significantly affect both translational and rotational motion. Particles in these systems span a wide range of geometries: from spheroidal microorganisms such as E. coli \citep{bai2006dielectric} and Paramecium \citep{keller1977porous}, to irregularly shaped microplastics, which are often modeled as ellipsoids \citep{kooi2021characterizing, wang2021settling}. Spherical particles are commonly used in fluidized beds to enhance heat transfer \citep{feng2014direct}, while fibrous particles serve to reinforce mechanical strength in composite materials \citep{mortazavian2015effects}.

Particle–particle interactions mediated by hydrodynamics are critical in real-world systems. But first, a foundational understanding is most effectively developed by analyzing the flow around a single, isolated particle. In some cases, such as heated particles in thermally sensitive fluids, viscosity gradients are generated by the particles themselves. In others, stratification arises from ambient environmental gradients. In both scenarios, the resulting particle dynamics differ qualitatively from those in constant-viscosity fluids. To date, all studies of particle motion in viscosity-stratified fluids, including our review, have adopted the zero-inertia approximation. This assumption is appropriate for small particles or highly viscous media. However, we anticipate that incorporating particle inertia, along with associated complexities such as cross-streamline migration \citep{segre1962behaviour, saffman1965lift}, will qualitatively alter particle behavior. Exploring how inertia interacts with viscosity stratification represents an exciting direction for future work.

An important distinction in particle dynamics is that between \emph{passive} and \emph{active} particles, defined in terms of their mechanical behavior. Passive particles, such as microplastics or colloids, do not self-propel but respond to ambient flows and gradients. Despite their passivity, they can exhibit rich and nontrivial dynamics in viscosity-stratified fluids. In contrast, active particles, such as motile microorganisms, convert internal or environmental energy into directed motion, often interacting with, and exploiting, viscosity gradients through mechanisms such as viscotaxis \citep{schweitzer2003brownian}. We first review the dynamics of passive particles (section~\ref{sec:Passive}), followed by mechanically passive but biologically active systems such as phytoplankton (section~\ref{sec:ecology}), and then motile active particles (section~\ref{sec:Active}).

\subsection{Passive particles}\label{sec:Passive}
We saw that the stress tensor for a Newtonian fluid with spatially varying viscosity is given by equation \eqref{eq:StressDef}. At zero fluid inertia, i.e., in the steady Stokes limit, ($Re = 0$), the governing equation \eqref{eq:ns} for momentum conservation becomes
\begin{equation}
\nabla \cdot \bm{\sigma} = 0, \label{eq:MassandMomentum}
\end{equation}
while the incompressibility condition (mass conservation) remains unchanged, as given in equation \eqref{eq:incomp}. Note that the Reynold number is absorbed into the nondimensionalisation of pressure is the expression for stress in equation \eqref{eq:MassandMomentum}.

These equations are closed by appropriate conditions at the domain boundaries, including no-slip conditions on the particle surface. The hydrodynamic force $\mathbf{F}$ and torque $\mathbf{Q}$ on a particle with surface $\mathbf{r}_p$ are given by
\begin{equation}
\mathbf{F} = \int_{\mathbf{r}_p} \mathrm{dS} \, (\mathbf{n} \cdot \boldsymbol{\sigma}), \quad
\mathbf{Q} = \int_{\mathbf{r}_p} \mathrm{dS} \, \mathbf{x} \times (\mathbf{n} \cdot \boldsymbol{\sigma}),
\end{equation}
where $\mathbf{x}$ is the position vector, measured from the centroid of the particle in this case (and from some other appropriate origin elsewhere in this paper), $\mathbf{n}$ is the outward normal from a surface element $\mathrm{dS}$. 
The Stokesian dynamics of a particle of arbitrary shape can be described using the resistivity formulation:
\begin{equation}
\begin{pmatrix}
\mathbf{F} \\ \mathbf{Q}
\end{pmatrix}
= \mathbf{R}
\begin{pmatrix}
\mathbf{V} \\ \boldsymbol{\omega}
\end{pmatrix}, \quad
\mathbf{R} =
\begin{pmatrix}
\mathbf{R}_{\mathbf{F}\mathbf{V}} & \mathbf{R}_{\mathbf{F}\boldsymbol{\omega}} \\
\mathbf{R}_{\mathbf{Q}\mathbf{V}} & \mathbf{R}_{\mathbf{Q}\boldsymbol{\omega}}
\end{pmatrix}.
\label{eq:ResistivityFormulation}
\end{equation}
where the $6 \times 6$ grand resistivity tensor $\mathbf{R}$ relates hydrodynamic forces and torques to the particle’s translational ($\mathbf{V}$) and rotational ($\boldsymbol{\omega}$) velocities \citep{kim2013microhydrodynamics}. Its inverse, $\mathbf{M} = \mathbf{R}^{-1}$, is the mobility tensor. 

We revisit the well-established case of motion in a uniform viscosity fluid before addressing the effects of viscosity variations.

\subsubsection{Uniform-viscosity fluid}

For a sphere of radius $a$, both $\mathbf{R}$ and $\mathbf{M}$ in equation \eqref{eq:ResistivityFormulation} are diagonal when viscosity is uniform, with $\mathbf{R}_{\mathbf{F}\mathbf{V}} = 6\pi a\,\bm{I}$, the classical Stokes drag \citep{stokes1851effect}, and $\mathbf{R}_{\mathbf{Q}\boldsymbol{\omega}} = 8\pi a^3\,\bm{I}$ for torque. These expressions are non-dimensional, with the sphere radius being the length scale, so $a=1$. Still, to highlight the size dependence on force and torque, we include $a$ in the expressions for the elements of the resistance tensor.

For ellipsoids in uniform-viscosity fluids, the resistivity tensor remains diagonal when expressed in the particle's principal axis frame, resulting in a decoupling between translational and rotational motions. The differing diagonal components reflect the anisotropic drag and torque imposed by the particle’s shape. This decoupling breaks down for arbitrary particle shapes, which have off-diagonal entries in the resistivity tensor, see e.g., \citet{joshi2025sedimentation}.  Interestingly, viscosity variation can behave similarly to shape variation, as we shall see, to give rise to novel particle dynamics. 

The resistivity expressions for spheroids in uniform-viscosity fluids were derived by \citet{jeffery1922motion}: 
\begin{equation}
\mathbf{R}_{\mathbf{F}\mathbf{V}}=6\pi a \begin{bmatrix}
f_1 & 0 & 0 \\
0 & f_2 & 0 \\
0 & 0 & f_2 \\
\end{bmatrix},\hspace{0.02in}\mathbf{R}_{\mathbf{Q}\boldsymbol{\omega}}=8\pi a^3 \begin{bmatrix}
q_1 & 0 & 0 \\
0 & q_2 & 0 \\
0 & 0 & q_2 \\
\end{bmatrix}.
\end{equation}
For a prolate spheroid of aspect ratio $\mathcal{A}$, the coefficients are 
\begin{align}
\begin{split}
f_1 &= \frac{8}{3\xi_0\left(\xi_0 + (3 - \xi_0^2)\coth^{-1}(\xi_0)\right)},  \\
f_2 &= \frac{4}{3\xi_0\left(-\xi_0 + (1 + \xi_0^2)\coth^{-1}(\xi_0)\right)}, \\
q_1 &= \frac{8(1 - 2\xi_0^2)}{9\xi_0^3\left(\xi_0 - (1 + \xi_0^2)\coth^{-1}(\xi_0)\right)}, \\
q_2 &= \frac{8(\xi_0^2 - 1)}{3\xi_0^3\left(\xi_0 - (\xi_0^2 - 1)\coth^{-1}(\xi_0)\right)},
\end{split}
\end{align}
where $\xi_0 \equiv \mathcal{A}/\sqrt{\mathcal{A}^2 - 1}$. All coefficients reduce to unity in the spherical limit. Analogous expressions are available for oblate spheroids. Analytical extensions of these expressions to weakly stratified viscosity fields have been developed by \citet{gong2024active} and \citet{sharma2025sedimentation}, as discussed below.

\subsubsection{Viscosity-stratified fluid}\label{sec:PassiveinStratified}

Under spatially varying viscosity $\mu(\mathbf{x}) = 1 + \hat{\mu}(\mathbf{x})$, where $\hat{\mu}(\mathbf{x})$ is the fluctuation viscosity non-dimensionalized with the viscosity scale, $\mu_0$ (as mentioned in section \ref{sec:equations}), the velocity and pressure fields can be decomposed as
\begin{align}
\begin{split}
\mathbf{u} &= \mathbf{u}^{\text{Stokes}} + \mathbf{u}^{\text{Stratified}}, \\
p &= \left(1 + \hat{\mu}(\mathbf{x})\right)p^{\text{Stokes}} + p^{\text{Stratified}}.
\end{split}
\end{align}
This decomposition allows the governing equations \eqref{eq:incomp} and \eqref{eq:MassandMomentum} to be split into two parts. The first corresponds to the classical Stokes problem in a uniform-viscosity fluid:
\begin{equation}
\nabla \cdot \mathbf{u}^{\text{Stokes}} = 0, \quad 
\nabla \cdot \boldsymbol{\sigma}^{\text{Stokes}} = 0,
\label{eq:StokesProblem}
\end{equation}
with 
\[
\boldsymbol{\sigma}^{\text{Stokes}} = -p^{\text{Stokes}}\boldsymbol{I} + \nabla \mathbf{u}^{\text{Stokes}} + (\nabla \mathbf{u}^{\text{Stokes}})^\mathsf{T},
\]
subject to the imposed flow and particle motion at the boundaries. The second set of equations captures the perturbative effect of viscosity stratification: 
\begin{equation}
\nabla \cdot \mathbf{u}^{\text{Stratified}} = 0, \quad 
\nabla \cdot \boldsymbol{\sigma}^{\text{Stratified}} + \nabla \mu \cdot \boldsymbol{\sigma}^{\text{Stokes}} = 0,
\label{eq:StratificationProblem}
\end{equation}
where
\begin{align}
\begin{split}
\boldsymbol{\sigma}&^{\text{Stratified}} = -p^{\text{Stratified}} \boldsymbol{I} \\& + (1 + \hat{\mu}(\mathbf{x})) \left(\nabla \mathbf{u}^{\text{Stratified}} + (\nabla \mathbf{u}^{\text{Stratified}})^\mathsf{T}\right).
\end{split}
\end{align}
Zero boundary conditions are imposed on $\mathbf{u}^{\text{Stratified}}$ to be consistent with the boundary conditions prescribed already for the unperturbed flow. The total fluid stress is then given by 
\begin{equation}
\boldsymbol{\sigma} = \boldsymbol{\sigma}^{\text{Stokes}} + \hat{\mu}(\mathbf{x})\boldsymbol{\sigma}^{\text{Stokes}} + \boldsymbol{\sigma}^{\text{Stratified}},
\label{eq:NetStress}
\end{equation}
where the Stokes stress acting in a spatially varying viscosity field is $\hat{\mu}(\mathbf{x})\boldsymbol{\sigma}^\text{Stokes}$, and the stress $\boldsymbol{\sigma}^{\text{Stratified}}$ arises from stratification-induced velocity and pressure fields. The Stokes equations \eqref{eq:StokesProblem} exhibit time-reversal symmetry — that is, they remain invariant under the transformation $t \rightarrow -t$. This symmetry has important consequences for the locomotion of microorganisms, which we revisit in section~\ref{sec:Active}. \citet{esparza2021rate} demonstrated that this symmetry is preserved even when viscosity varies smoothly in space. However, it may break down when viscosity at a given location varies in time, for instance due to the advection of a scalar field that modulates viscosity, as described by equations \eqref{eq:viscscalardependence} and \eqref{eq:species}, and indicated by the schematic in figure \ref{fig:what_visc_does}.

The decomposition introduced above is central to understanding particle dynamics in viscosity-stratified fluids. In the following description, we classify recent developments into those for uniform flows (including sedimentation) and linear flows (such as simple shear and extension). For clarity and interpretability, we will frequently reference the stratification-induced pressure term $\hat{\mu}(\mathbf{x}) p^{\text{Stokes}}$ — a component of $\hat{\mu}(\mathbf{x})\boldsymbol{\sigma}^{\text{Stokes}}$ — to illustrate key mechanisms in viscosity-stratified fluids. The influence of other stratification-induced terms has been found to yield similar effects \citep{sharma2025sedimentation}.

\paragraph{Uniform flows and sedimenting particles} 
A heated particle in a temperature-sensitive fluid will induce local viscosity gradients purely through its presence. Under the assumption of weak viscosity variation, \citet{oppenheimer2016motion} employed regular perturbation theory and the reciprocal theorem to quantify how these gradients influence the motion of a heated sphere. At small Péclet number and slow time variation, equation \eqref{eq:species} reduces to
\begin{align}
\begin{split}
&\nabla^2 T = 0, \quad \text{with} \quad 
T(\mathbf{x}_\text{particle}) = T_\text{prescribed}, \\
&T \rightarrow T_\infty \text{ as } |\mathbf{x}| \rightarrow \infty.
\label{eq:TemperatureTransport}
\end{split}
\end{align}
When the surface temperature is asymmetric — for instance, between hemispheres — the off-diagonal components of the resistivity matrix ($\mathbf{R}_{\mathbf{F}\boldsymbol{\omega}}$ and $\mathbf{R}_{\mathbf{Q}\mathbf{V}}$) become non-zero, indicating translation–rotation coupling induced by the anisotropic viscosity field.
A multipole expansion of the temperature field in equation~\eqref{eq:TemperatureTransport} reveals that monopole and quadrupole components modify the diagonal entries of the resistivity matrix ($\mathbf{R}_{\mathbf{F}\mathbf{V}}$ and $\mathbf{R}_{\mathbf{Q}\boldsymbol{\omega}}$), while the dipole component generates translation–rotation coupling. Notably, isotropic heating (monopole) reduces the overall drag, lowering the diagonal entries of $\mathbf{R}_{\mathbf{F}\mathbf{V}}$ below the constant-viscosity drag coefficient of $6\pi\mu_\infty a$ (for an equivalent isothermal viscosity $\mu_\infty$ far from the particle). 

These theoretical results are supported by experiments on the diffusion of heated gold nanoparticles in water \citep{rings2010hot}. Under constant temperature and viscosity, such particles would display Brownian motion, with a diffusion coefficient $k_B T/(6\pi \mu_\infty a)$, $k_B$ being the Boltzmann constant [The fluctuation-dissipation provides this relationship via the Einstein relation, see below equation (2) in \citet{einstein1905molekularkinetischen}.] \citet{rings2010hot} obtained an enhanced diffusion coefficient for the heated particles which could not be explained by the increased temperature alone. The super-linear increase in diffusivity can be better explained by accounting for the fact that viscosity varies in the radial coordinate away from the particle surface, as later calculated by \citet{oppenheimer2016motion}. In a complementary study, \citet{datt2019active} examined the effect of externally imposed spatial viscosity gradients on passive and active spherical particles. They considered weak, linear stratification of the form $\mu(\mathbf{x}) = 1 + \mathbf{x} \cdot \nabla \mu$, with $|\mathbf{x} \cdot \nabla \mu| \ll 1$,  and derived explicit expressions for the stratification-induced hydrodynamic forces and torques. Remarkably, even perfectly spherical particles exhibit translation–rotation coupling when the viscosity gradient has a component perpendicular to the velocity. The resulting resistivity matrix $\mathbf{R}$, though no longer diagonal, remains symmetric and positive definite [like in the uniform-viscosity case \citep{kim2013microhydrodynamics}], as established by \citet{oppenheimer2016motion}.
For a sphere rotating with angular velocity $\boldsymbol{\omega}$, stratification produces an additional linear drag $\mathbf{F}^{\text{Stratified}} = 2\pi a^3 (\nabla \mu \times \boldsymbol{\omega})$. Conversely, when a stationary sphere is placed in a uniform flow with velocity $\mathbf{U}$, there is a stratification-induced torque $\mathbf{Q}^{\text{Stratified}} = -2\pi a^3 (\nabla \mu \times \mathbf{U})$. These arise from differential viscous stresses across the particle due to the imposed viscosity gradient. The mechanism can be physically interpreted by examining equation~\eqref{eq:NetStress}. One contribution to the force comes from the Stokes pressure field $p^{\text{Stokes}}$ in a nonuniform medium, yielding a stress contribution proportional to $\hat{\mu}(\mathbf{x}) p^{\text{Stokes}}$. The remaining contributions include the stratification-induced pressure field $p^\text{Stratified}$, the viscous stress due to the Stokes flow in a nonuniform medium, $\hat{\mu}(\mathbf{x})[\nabla\mathbf{u}^\text{Stokes} + (\nabla\mathbf{u}^\text{Stokes})^\mathsf{T}]$, and the viscous stress from the stratification-induced flow, $\nabla\mathbf{u}^\text{Stratified} + (\nabla\mathbf{u}^\text{Stratified})^\mathsf{T}$. Thus a freely sedimenting sphere of excess density $\Delta \rho$ in a fluid with a viscosity gradient $\nabla \mu$ will experience a torque that causes it to rotate at an angular velocity
\begin{equation}
\boldsymbol{\omega}^{\text{Stratified}} = \frac{2 a^2 \Delta \rho}{9} (\mathbf{g} \times \nabla \mu).
\end{equation}
This represents a fundamental departure from the classical picture of Stokes sedimentation, where spherical particles settle without rotating.

These insights open avenues for controlling particle behavior in stratified environments. For instance, \citet{laumann2019focusing} employed viscosity variations in microflows to control the trajectories of soft particles. \citet{esparza2021dynamics} experimentally investigated a helical swimmer traversing viscosity gradients, finding that its swimming speed and direction depend sensitively on the gradient’s orientation relative to the swimmer’s body. Such mechanisms, by which viscosity variations qualitatively change the dynamics, have significant implications for biomedical engineering, including targeted drug delivery, microfluidic sorting, and the design of micro- and nano-scale robotic systems \citep{nelson2010microrobots, li2017micro, palagi2018bioinspired}. \citet{ziegler2022hydrodynamic} analyzed the dynamics of two spheres of radius \( a \), separated by a distance \( d \) [which is the magnitude of the center-to-center vector \( \mathbf{d} \)], in the presence of both externally imposed linear viscosity gradients and viscosity variations induced by thermal effects (the particles being heated or cooled). They derived first-order corrections for weak gradients and small size-to-separation ratios (\( a/d \)). Due to the imposed viscosity gradient, they found that the sedimenting particles undergo rotation relative to each other, an effect absent in constant-viscosity. For thermally induced viscosity variations, their results were consistent with those of \citet{oppenheimer2016motion}. But while \citet{oppenheimer2016motion} found no rotation–translation coupling for a single uniformly heated sphere, \citet{ziegler2022hydrodynamic} demonstrated that such coupling does arise in a two-particle system, appearing at order \( \mathcal{O}(d^{-2}) \). 

\begin{figure*}
\centering 
\includegraphics[width=0.75\textwidth]{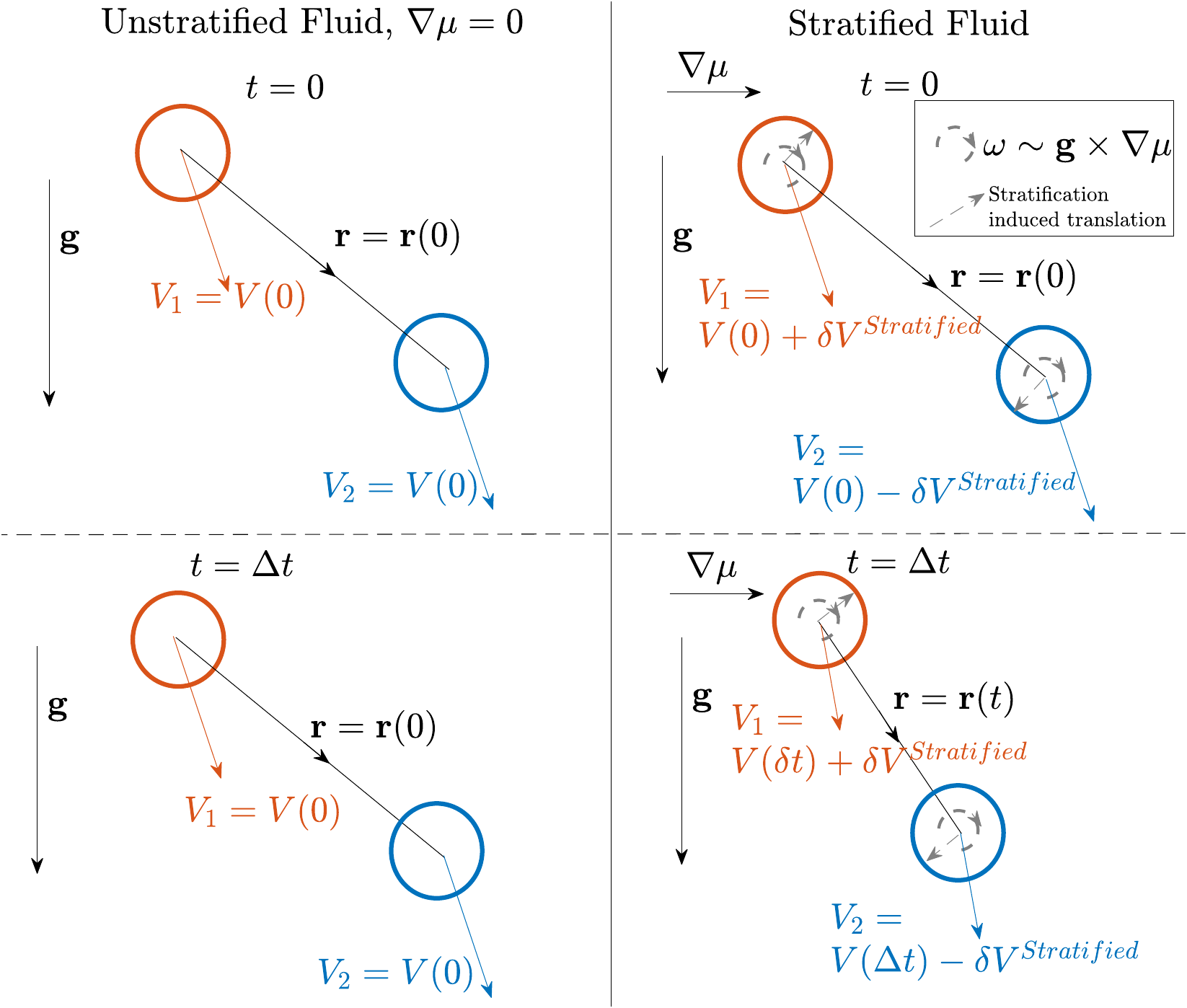} 
\caption{In a quiescent fluid with uniform viscosity, two identical spheres (left panel) sediment at the same speed $V$, given by equation~\eqref{eq:2SphereVel}, maintaining constant relative distance and orientation. In a stratified fluid (right panel), however, each sphere rotates with angular velocity $\boldsymbol{\omega} = -\frac{2 a^2 \Delta \rho}{9} \mathbf{g} \times \nabla \mu$. This rotation generates a perturbation in the surrounding fluid velocity, $\pm \delta \mathbf{U}^{\text{Stratified}} = \pm \boldsymbol{\omega} \times \mathbf{d}$ (grey arrows), inducing relative motion between the spheres.\label{fig:selfrotating}}
\end{figure*}

In constant viscosity, using the method of reflections \citep{guazzelli2011physical}, the sedimentation velocity of the two identical spheres is given by
\begin{eqnarray}
\begin{split}
&\mathbf{V}_1 = \mathbf{V}_2 = \mathbf{V}\\ 
&= {a^2 \Delta \rho} \left( \frac{2}{9} \mathbf{g} + \frac{a}{6} \left( \frac{\mathbf{g}}{d} + \frac{(\mathbf{g} \cdot \mathbf{r}) \, \mathbf{d}}{d^3} \right) \right)+\mathcal{O}((a/d)^{3}),
\end{split}    \label{eq:2SphereVel}
\end{eqnarray}
where we use the notation $\mathbf{V}$ for a particle's velocity to distinguish from $\mathbf{U}$ for fluid velocity.
Because the particles have identical velocities, their relative positions remain fixed during sedimentation, as illustrated in the left panel of figure~\ref{fig:selfrotating}, and the pair settles faster than an isolated sphere \citep{kim2013microhydrodynamics}. In contrast, in fluid with varying viscosity, each sphere experiences a torque that induces particle rotation, which in turn generates a flow field that perturbs the motion of the neighboring particle. Specifically, the rotation, $\boldsymbol{\omega}$, of one sphere produces an additional velocity \( \pm \delta \mathbf{U}^{\text{Stratified}} =\boldsymbol{\omega}\times \mathbf{d} \) at the location of the other, breaking symmetry and resulting in relative motion. This effect is illustrated schematically in the right panel of figure~\ref{fig:selfrotating}, where both the direction and magnitude of the separation vector \( \mathbf{d} \) evolve over time. Such relative motion has important consequences for the stability of particle suspensions. Viscosity-induced interactions may drive particles toward one another, potentially leading to clustering or aggregation — instabilities relevant in industrial contexts such as slurry transport, sedimentation in fixed-bed reactors, and food processing applications.

\begin{figure*}
\centering 
\subfloat{\includegraphics[width=0.5\textwidth]{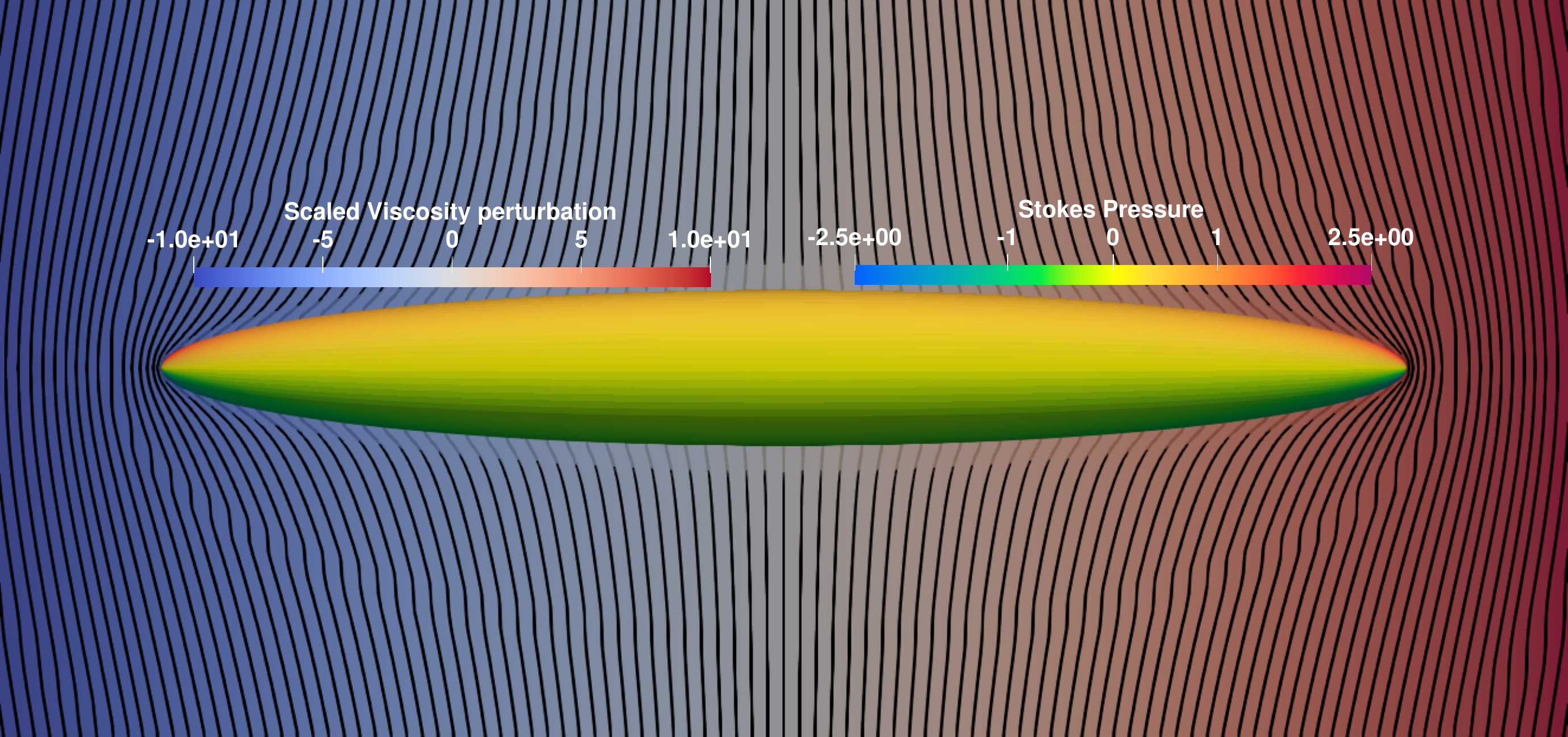}}
\caption{Surface Stokes pressure \( p^{\text{Stokes}} \), shown in colour, on a prolate spheroid with aspect ratio 8 held stationary in uniform downward flow. Fluid viscosity increases from left to right.  The spheroid develops pressure maxima on the top surface and minima on the bottom at both ends. Because the two ends are exposed to different viscosities, an asymmetry arises in the pressure distribution, resulting in a net clockwise torque. Similar visualizations were used by \citet{sharma2025sedimentation}.}
\label{fig:AR8}
\end{figure*}

These mechanisms naturally extend to non-spherical particles — a relevant consideration since real-world particles in industrial, geophysical and biological contexts often exhibit more complex shapes: fibers, discs, rings and many more. As discussed earlier, these geometries can give rise to richer dynamics due to their anisotropic hydrodynamic resistance, and a spherical or spheroidal shape, though studied most often, may be wide of the mark in predicting the dynamics. Using resistive force theory, \citet{kamal2023resistive} computed the forces and torques on slender fibers and rings suspended in linearly stratified fluids and subjected to both uniform and linear background flows. Their analysis assumes large aspect ratios as well as a close alignment of the direction of viscosity variation with the particle’s major axis. Like in the spherical case, consistent with the stratification-induced torque mechanisms we have described, they found that a viscosity gradient perpendicular to the flow direction generates a torque on both fibers and rings, inducing rotation. Additionally, while pressure does not contribute to the torque on a sphere due to radial symmetry, it contributes significantly to it for spheroids \citep{sharma2025sedimentation}. To illustrate this, consider a prolate spheroid with aspect ratio 8 held fixed in a uniform flow orthogonal to the viscosity gradient. Figure~\ref{fig:AR8} shows the surface Stokes pressure distribution \( p^{\text{Stokes}} \). As this pressure is symmetric, it generates no torque in a constant-viscosity fluid. However, when weighted by a spatially varying viscosity, the term \( \hat{\mu}(\mathbf{x}) p^{\text{Stokes}} \) being asymmetric, generates a net torque.

\begin{figure*}
\centering
\includegraphics[width=0.9\textwidth]{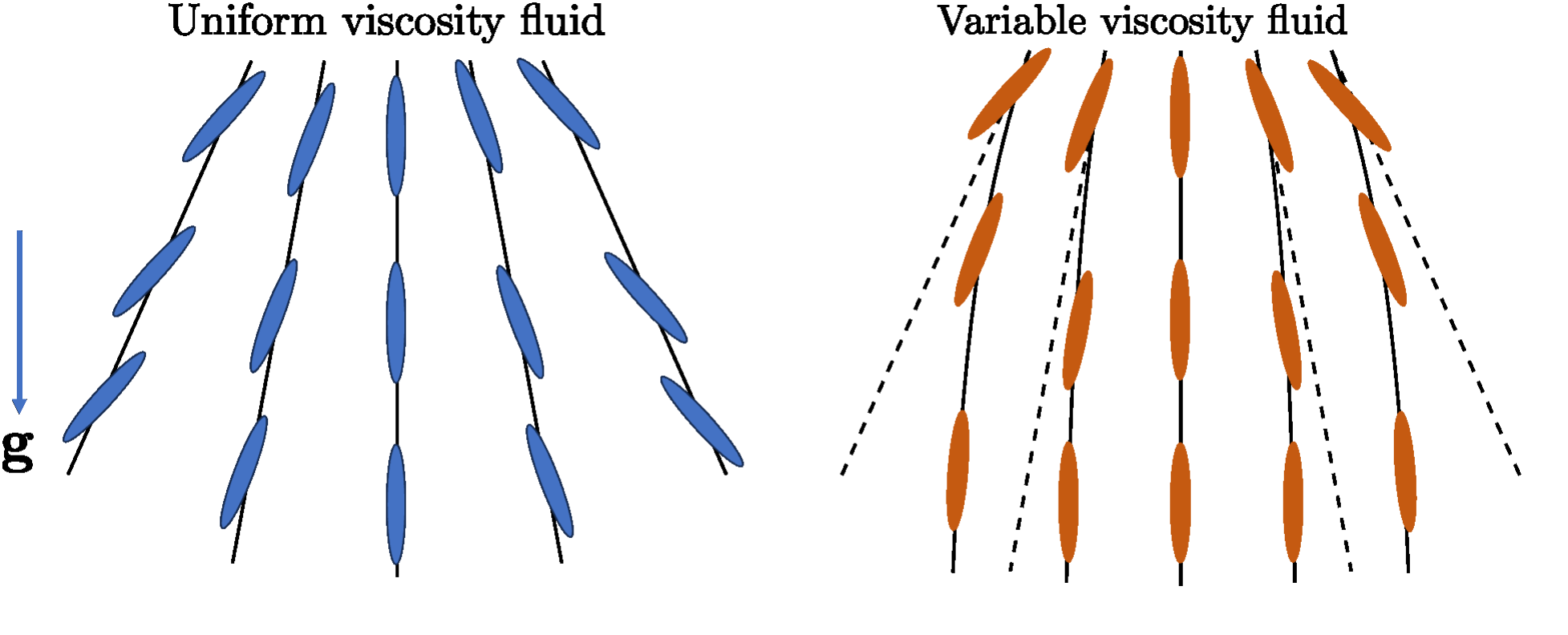}
\caption{Effect of viscosity stratification on the sedimentation of a prolate spheroid. Left: In a uniform-viscosity fluid, the particle sediments at a constant, orientation-dependent angle. Right: In a viscosity-stratified fluid, stratification-induced torque reorients the particle towards a steady-state orientation, leading to a sedimentation angle and speed persistent across a mixture of particles that were initially randomly oriented. The dashed lines show the sedimentation path the particle would have taken in uniform viscosity.}
\label{fig:SedimentationSchematic}
\end{figure*}

While stratification induces torque at the leading order on a sedimenting sphere or spheroid, sedimentation speed or direction are typically altered only at higher orders. However, for anisotropic particles such as a spheroid, the induced rotation at first order can already modify translational motion. This is because, ignoring stratification effects, the instantaneous velocity of the spheroid depends on its orientation. Now, as stratification drives rotation, the evolving orientation alters the sedimentation trajectory, even under weak gradients. Recent theoretical studies by \citep{anand2023sedimentation,gong2024active,sharma2025sedimentation} have shown that viscosity gradients can produce a variety of rotational behaviors in sedimenting spheroids. \citet{sharma2025sedimentation} further developed a phase diagram characterizing the orientation dynamics of spheroids as a function of aspect ratio and the angle between the viscosity gradient and gravity. The distinct dynamical regimes which were found include: monotonic alignment or spiralling toward orientations either parallel or perpendicular to the stratification–gravity plane; tumbling confined to the stratification–gravity plane; and continuous, non-uniform rotation in closed orbits akin to Jeffery orbits in simple shear flow. Notably, the sensitivity of rotational behavior to the stratification direction is greatest for aspect ratios between 0.55 and 2.0—a range relevant to many naturally occurring and anthropogenic anisotropic particles, including microplastics, and various microorganisms such as bacteria \citep{kaya2009characterization, kooi2021characterizing}. In some regimes, particles approach  the same orientation as each other, as illustrated in Figure~\ref{fig:SedimentationSchematic}. The significant trajectory changes induced by stratification are highlighted in the figure. 	
These insights suggest new strategies for controlling particle motion using viscosity gradients. The ability to reorient particles independently of their initial orientation has potential applications in reactor design and particle filtration. In fixed-bed reactors or microfluidic systems, for instance, viscosity stratification can be engineered to align particles, minimizing wall collisions and enhancing transport efficiency. Beyond industrial relevance, such mechanisms may also benefit environmental monitoring and biomedical technologies involving anisotropic colloids or engineered microswimmers.

\begin{figure*}
\centering 
\subfloat{\includegraphics[width=0.49\textwidth]{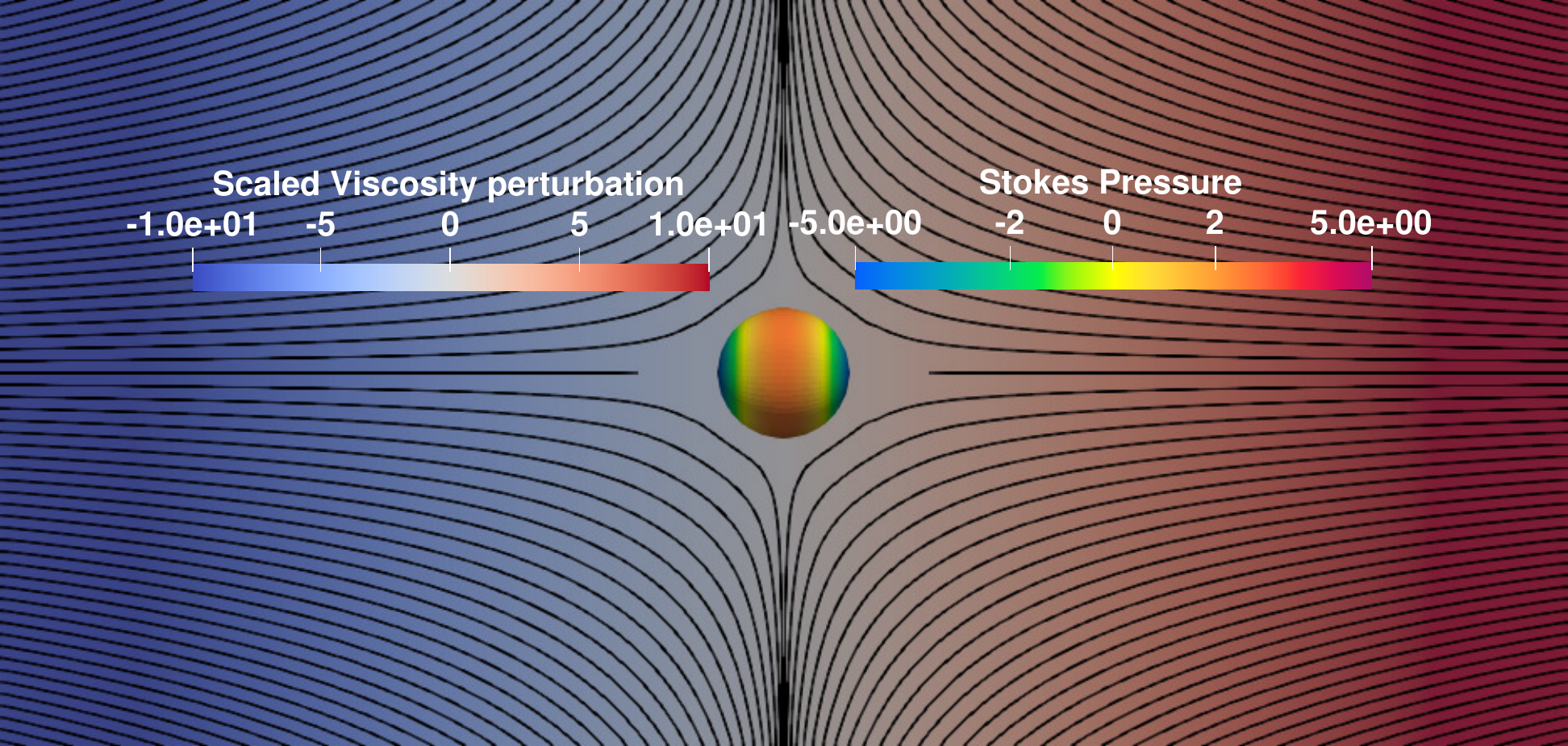}}\hfill
\subfloat{\includegraphics[width=0.49\textwidth]{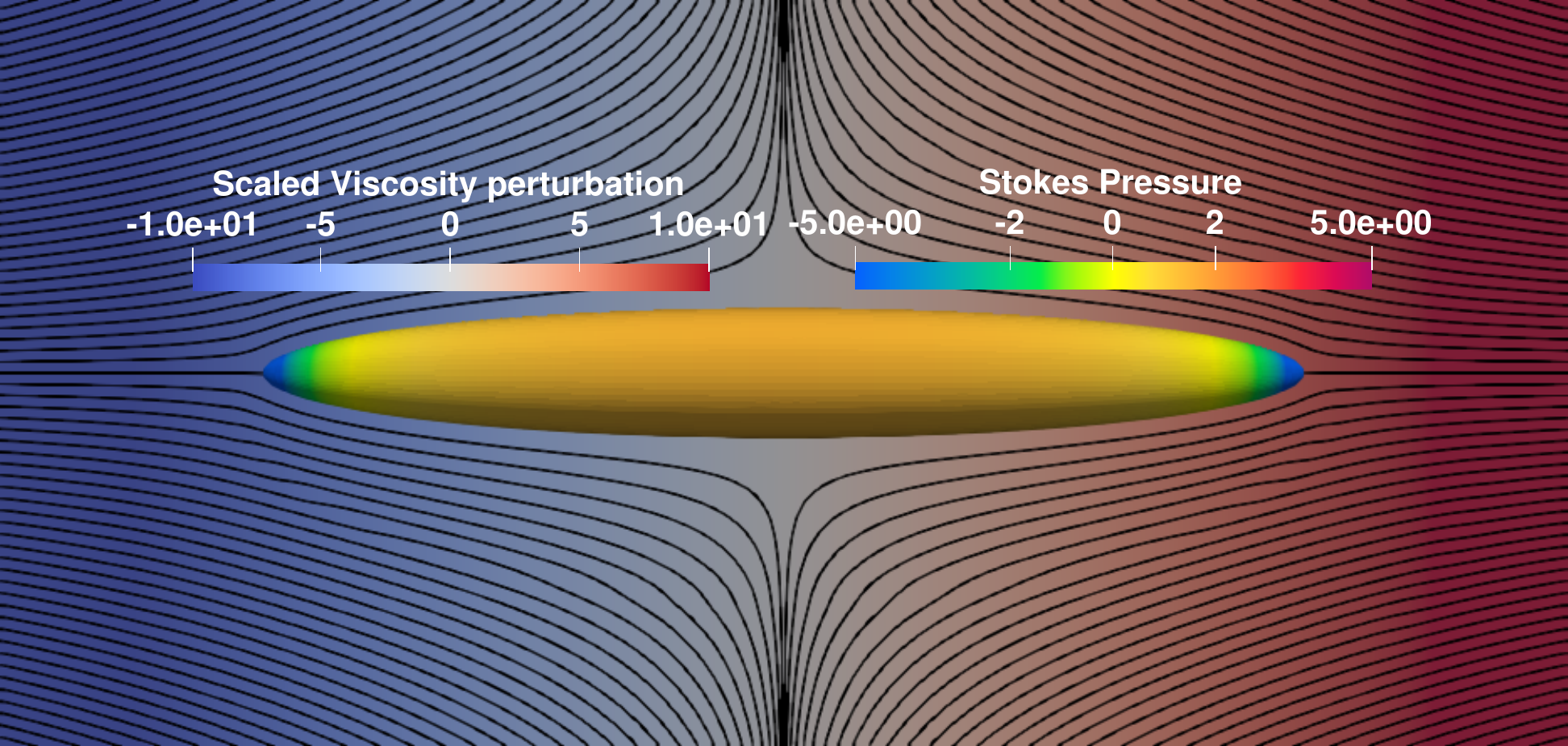}}
\caption{Same caption as in figure~\ref{fig:AR8}, but with an imposed uniaxial extensional flow and a sphere in the left panel and aspect ratio 8 spheroid in the right panel. The extensional axis is horizontal, and the compressional plane is aligned vertically. In this case, the low-pressure regions at the horizontal ends of the particle lie in zones of different viscosity. This asymmetry generates a net force on the particle, in contrast to the force-free situation in a fluid of uniform viscosity.}
\label{fig:PressureinShearandExtflow}
\end{figure*}

\paragraph{Linear flows}
In industrial, geophysical and biological contexts, particles encounter flow fields beyond uniform or sedimentation-dominated regimes. Near solid boundaries, flow can be approximated as being in simple shear, while extensional flows arise at pore inlets or outlets. When viscosity varies spatially, it can induce lateral forces that drive particles across streamlines — an effect absent in constant-viscosity fluids. Theoretical predictions by \citet{kamal2023resistive} first demonstrated this effect for slender fibers and rings placed in linearly stratified fluids under rotational or extensional flow. In fact the symmetry of classical Stokes flow is broken even when the particle is fixed in space. More generally, \citet{sharma2025sedimentation} explain the behavior of spheroids of arbitrary aspect ratio subjected to linear flows in the presence of viscosity gradients.
As an illustration, consider a sphere suspended in a fluid in uniaxial extensional flow with velocity field \( \mathbf{U} = [-x/2,\ -y/2,\ z]^\mathsf{T} \), where viscosity increases along the extensional axis \( z \). Unlike in constant viscosity fluid, an asymmetric force due to the pressure arises — through the term \( \hat{\mu}(\mathbf{x}) p^{\text{Stokes}} \) in equation~\eqref{eq:NetStress}.
This mechanism is visualized in figure~\ref{fig:PressureinShearandExtflow}, which shows the surface pressure distribution for a sphere and an aspect ratio 8 spheroid, both in extensional flow. When viscosity increases along \( z \), one hemisphere lies in a region of lower viscosity and lower pressure, while the opposite hemisphere experiences the same pressure in a higher-viscosity region. As these pressures are weighted by $\hat{\mu}(\mathbf{x})$, the imbalance yields a net force along the viscosity gradient. Thus, even in symmetric flows, stratification can induce transverse migration of particles.

A similar effect occurs in simple shear flow. When viscosity varies in the flow direction, a sphere will drift along the velocity gradient due to the stratification-induced force. Since this force scales with particle size, viscosity stratification offers a pathway for size-selective sorting in microchannels. The dynamics become even richer for non-spherical particles, where anisotropic resistance and particle shape alignment lead to three-dimensional migration.

The theoretical studies of particles in viscosity-stratified fluids discussed thus far have all focused on the regime of weak viscosity gradients. Despite this limitation, the novel physical mechanisms uncovered already motivate a range of experimental investigations and potential applications, as highlighted throughout the preceding sections. However, further theoretical and computational developments -- particularly those that go beyond the weak-gradient approximation — are likely to reveal qualitatively new particle behaviors.
One such conjecture was proposed by \citet{sharma2025sedimentation}. As a sedimenting sphere rotates due to stratification-induced torque, it experiences a relative angular velocity with respect to the surrounding fluid. Also, a particle rotating in a linearly stratified fluid can experience a stratification-induced force. Thus, the rotation of a sedimenting sphere may lead to a sideways drift in stratified fluids. Such lateral forces arise only at second order in the viscosity gradient and therefore lie outside the formal regime of validity of the weak-gradient assumption. However, in experiments or simulations that allow for stronger viscosity variations, this effect could be manifested as curved sedimentation paths. Lastly, before moving to active particles, we mention that the only study to date (to our knowledge) that considers interactions between more than one passive particle in a viscosity-stratified fluid is that of \citet{ziegler2022hydrodynamic}. The schematic in figure~\ref{fig:selfrotating} suggests rich interaction mechanisms that remain largely unexplored. These include scenarios beyond the dilute or well-separated particle limit, which may be probed through immersed boundary simulations or dense-suspension experiments. We conclude this subsection by saying that viscosity variation in the context of particulate flows is an open question which has the potential to become a powerful subject with richness in theory and in applications.

\subsection{Phytoplanktonic systems: mechanically passive but biologically active}\label{sec:ecology}
\begin{figure*}
\centering 
\subfloat{\includegraphics[width=0.48\textwidth]{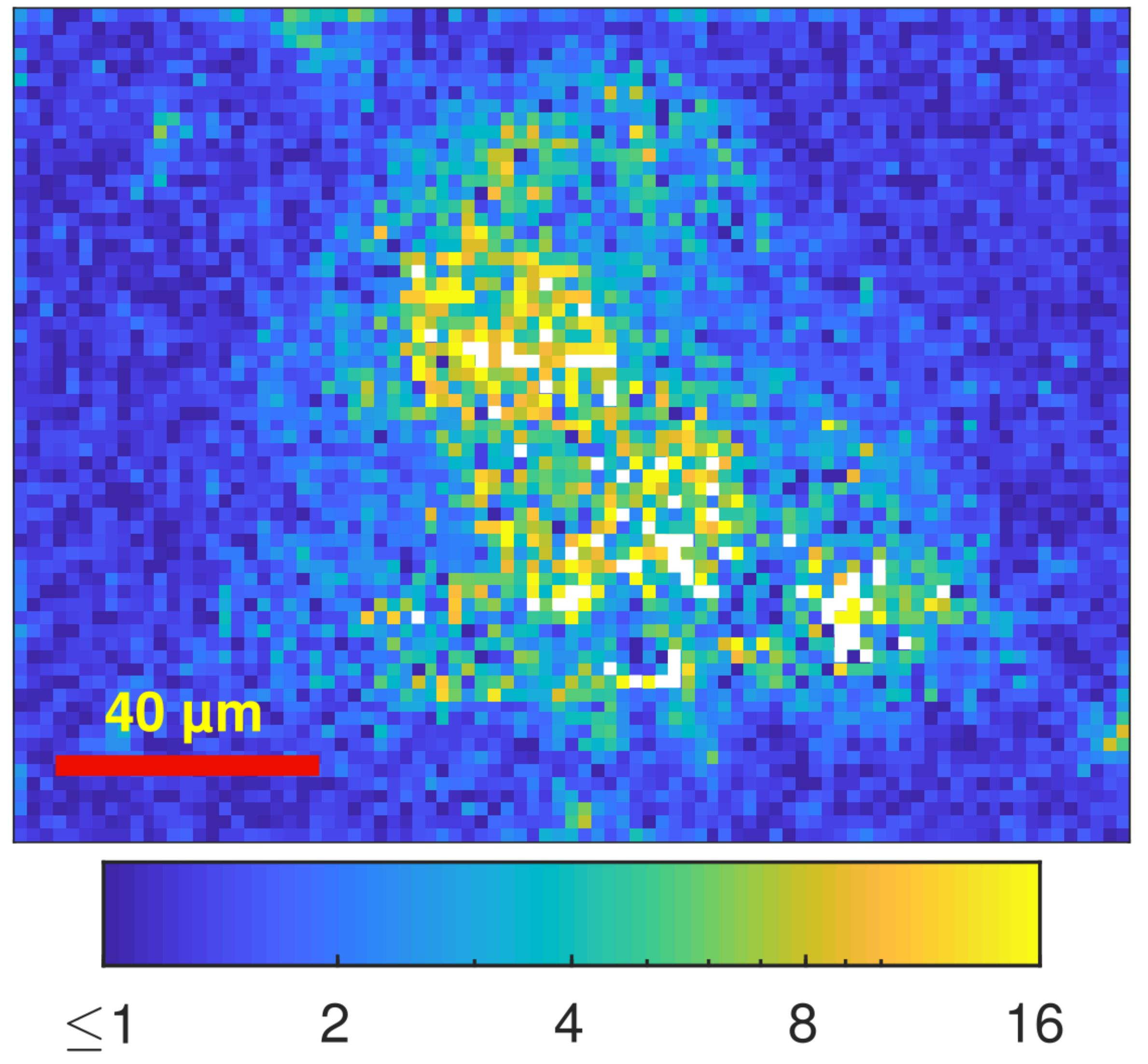}}
\subfloat{\includegraphics[width=0.5\textwidth]{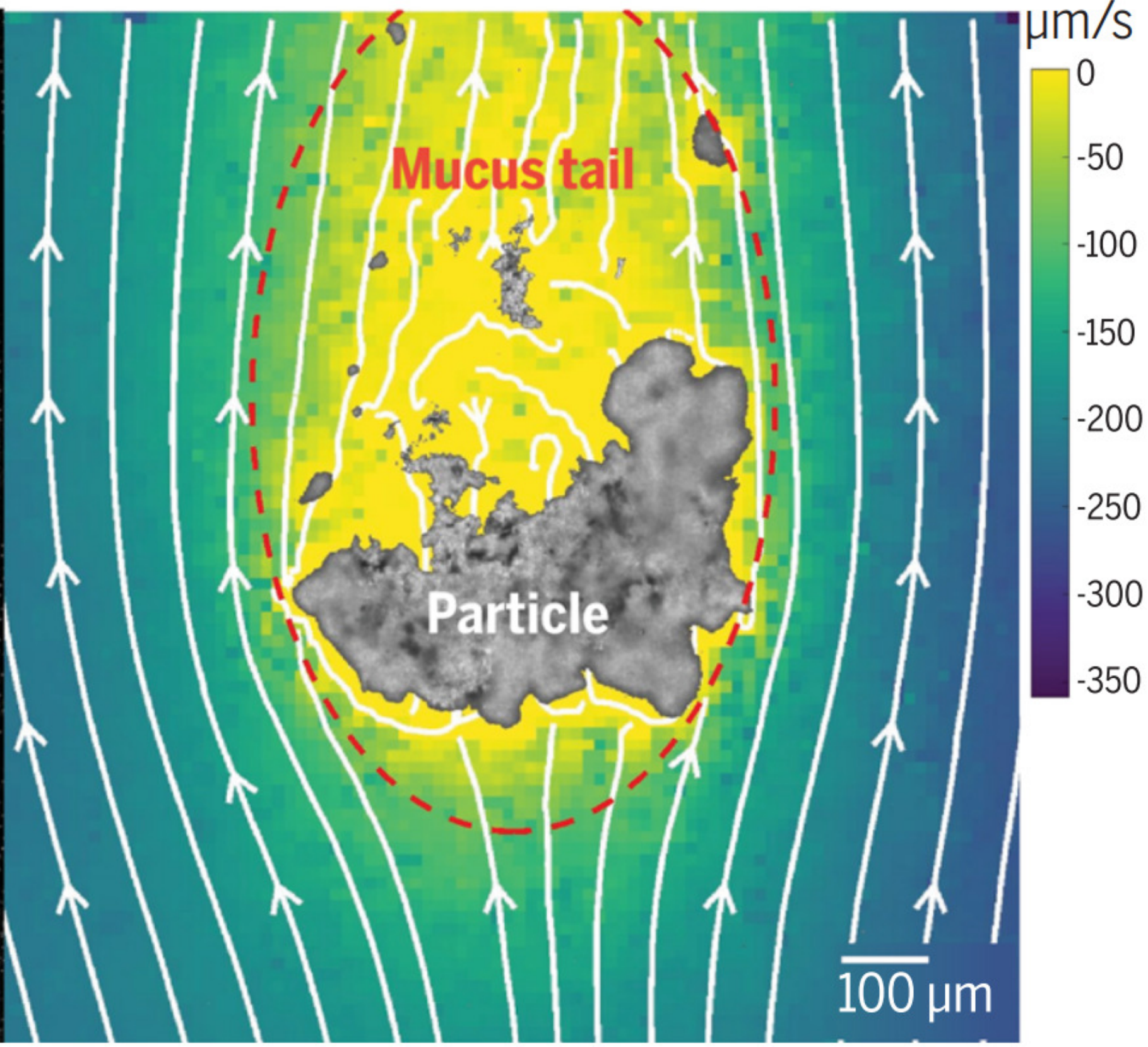}}
\caption{(Left) Viscosity in and around an aggregate of \textit{Chaetoceros affinis}, a diatom. The colour is the ratio of local dynamic viscosity to the dynamic viscosity of seawater in 2 \textrm{$\mu$}m $\times$ 2 $\mu$m bins. Adapted from \citet{guadayol2021microrheology}. (Right) Color shows the velocity of a sinking marine snow particle with a mucus tail. The particle was collected during an algal bloom in the Gulf of Maine by \citet{chajwa2024hidden}. Adapted with permission from the American Association for the Advancement of Science.}\label{fig:gudayol_image}
\end{figure*}

Planktonic systems play a key role in carbon sequestration, as mentioned above. Microscale heterogeneity in viscosity is a hallmark of many planktonic environments, with important implications for microbial behavior. And their interactions with the stratified environments are central to oceanic ecological dynamics. Besides molecular viscosity, which individual plankton perceive in their neighborhood, collections of plankton will be affected by eddy diffusivity in the turbulent upper ocean. We will discuss this in section \ref{sec:EddyViscosity}.
One reason for viscosity stratification is that phytoplankton release some of their cellular contents, which are polymeric substances of very high viscosity, into the neighbourhood. \citet{guadayol2021microrheology} experimentally showed that  viscosity in the immediate vicinity of phytoplankton such as \textit{Chaetoceros affinis} (see figure~\ref{fig:gudayol_image}) can be elevated by several orders of magnitude, giving rise to strong local viscosity gradients. Aspects of the dynamics which arise from these microorganisms acting as `viscosity sources' can be studied in analogous situations in non-living systems. We have already discussed heated particles, which may be thought of as `viscosity sinks'. A solid object which can emit solute of a different viscosity would be another system in this class. Though it was not about viscosity variation, the numerical exploration of \citet{zhu2023self} of a phoretic disc is of relevance to note. By emitting solute which modified the surface slip velocity, this disc exhibited a variety of behaviors—including steady translation, orbiting, periodic motion, and chaos. 

Besides their own dynamics, these microscale viscosity enhancements by plankton directly affect the dynamics of neighbouring small-inertia microorganisms, and shape the spatial distribution and interaction patterns of marine microbes surrounding the particle releasing the contents. For instance, high-viscosity zones reduce bacterial motility leading to increased bacterial concentrations near phytoplankton cells. \citet{seymour2017zooming} and \citet{guadayol2021microrheology} hypothesized that this mechanism promotes symbiotic bacterial growth, potentially benefiting both partners through nutrient exchange and protection. Idealized simulations by \citet{inman2022hidden} suggest that increased local viscosity can slow nutrient uptake by microorganisms. However, quantifying this effect precisely at both cellular and aggregate scales remains an open challenge. Observational studies by \citet{seuront2006biologically} revealed that extracellular polymer secretion during springtime phytoplankton blooms significantly increases the apparent viscosity of seawater. This may represent an ecological adaptation: by suppressing turbulent mixing, phytoplankton could create more stable local environments favorable for colony formation \citep{smayda2002turbulence}.

Recent work by \citet{chajwa2024hidden} highlights the complex dynamics of marine snow, which are aggregates of organic matter and microorganisms that contribute to carbon sequestration in the ocean \citep{boyd2019multi, alldredge1988characteristics}. These sinking particles form mucus-rich tails that increase their effective size and reduce their sedimentation speed (see figure~\ref{fig:gudayol_image}). Understanding the full dynamics of marine snow, including lateral motion, interactions with internal waves, and encounters with zooplankton, can improve numerical estimates of the global carbon budget and biogeochemical fluxes \citep{burd2010assessing, giering2014reconciliation}. To improve our understanding we will require more detailed laboratory and field-scale measurements to better parametrise these processes in ocean biogeochemical models \citep{rohr2023zooplankton}. Bridging the gap between microscopic dynamics and their emergent macroscopic effects, such as energy cascades and large-scale circulation in active turbulence, remains a major research frontier \citep{wensink2012meso, dunkel2013fluid, bechinger2016active, deng2022temporal, alert2022active, waigh2023heterogeneous}.


In the next section, we shift focus to the dynamics of individual active particles and microorganisms in viscosity-stratified environments.

\subsection{Active particles/ microbes}\label{sec:Active}
While \citet{rafai2010effective} found that suspensions of motile \textit{Chlamydomonas reinhardtii} exhibited increased effective viscosity (similar to passive particles), \citet{sokolov2009reduction} observed the opposite effect for \textit{Bacillus subtilis}, with active swimming leading to viscosity reduction. Remarkably, \citet{lopez2015turning} reported a superfluid-like regime with effectively zero viscosity in moderately concentrated suspensions of motile \textit{Escherichia coli}. These contrasting behaviours underscore the sensitivity of active suspension rheology to organism type, motility strategy, and concentration.

It is becoming increasingly evident that viscosity stratification significantly influences the locomotion of microorganisms such as bacteria and algae. These organisms, often referred to as active particles, propel themselves by consuming internal energy and respond to environmental stimuli \citep{hatwalne2004rheology, ramaswamy2010mechanics}. In this section, we first review key experimental observations of their behavior in viscosity-stratified environments and then discuss theoretical frameworks developed to explain these phenomena.

Early studies by \citet{kaiser1975enhanced} found that the pathogenic bacterium \textit{Leptospira interrogans} swims more efficiently in fluids with higher viscosity. Building on this, \citet{petrino1978viscotaxis} observed that these bacteria actively migrate toward regions of higher viscosity when suspended in a viscosity-stratified fluid—a behavior termed positive viscotaxis. The helical body of this animal, coiled around an imaginary centerline, is thought to enhance its swimming efficiency in viscous environments. One proposed explanation, as noted by \citet{berg1979movement}, is that the helical body acts as an efficient corkscrew: in low-viscosity fluids such as water, the organism slips significantly and requires several revolutions to advance by one pitch length, whereas in more viscous media it propels itself more effectively with reduced slippage. \citet{petrino1978viscotaxis} further hypothesized that viscotaxis may aid the survival of free-living Leptospira in moist soils and mud, and raised concerns about its potential role in facilitating mucosal membrane penetration during infection. Interestingly, the morphology of Leptospira changes with ambient viscosity \citep{takabe2017viscosity}. The bacterium's centerline may terminate in either a hook (H) or sinusoidal (S) shape at each end, and at certain viscosities, it transitions from a swimming mode (with asymmetric SH configuration) to a rotational mode (with symmetric HH or SS configurations). Spiroplasma are another organism that display viscotaxis, and similar to Leptospira, swim faster in more viscous environments \citep{daniels1980aspects}. This enhanced propulsion in high-viscosity media, facilitated by their helical body morphology and undulatory motion, contrasts with the behavior of most flagellated bacteria, which experience reduced motility beyond a certain viscosity, as shown in the experiments of \citet{schneider1974effect}. Further experimental studies of morphology and flow interaction in stratified media may provide deeper insight into the diverse viscotactic strategies of microorganisms.

To mechanistically explain the influence of body shape on viscotactic behavior in fluids with linearly variation in viscosity, \citet{liebchen2018viscotaxis} modelled the shapes by configurations of rigidly connected spheres. These ranged from a simple dumbbell (two identical spheres) to more complex assemblies of three or four spheres of differing radii \( \mathbf{a}_i,\ i \in [1,N] \), arranged in triangular (\(N=3\)) or square (\(N=4\)) geometries. Each sphere was assumed to experience local hydrodynamic drag given by Stokes' law, \( 6\pi \mu(\mathbf{r}) \mathbf{a}_i \mathbf{u}(\mathbf{r}) \), where \( \mu(\mathbf{r}) \) is the viscosity at the center of the sphere located at \( \mathbf{r} \). Their analysis showed that uniaxial swimmers — such as one gets from the dumbbell model — do not exhibit viscotaxis, as the net hydrodynamic force (the vector sum of forces on each sphere) remains aligned with the swimmer’s axis of symmetry. In contrast, swimmers with triangular or quadrilateral arrangements experience a net torque due to misalignment between the net force and their instantaneous swimming direction, causing them to reorient in response to the viscosity gradient. The model predicted positive viscotaxis, consistent with experimental observations. These shape-induced asymmetries thus allow the swimmer to ``sense" the viscosity gradient. This simple framework provides a qualitative explanation for how shape asymmetry enables viscotactic behavior in living organisms, offering mechanistic insight into observations such as those by \citet{takabe2017viscosity}, where changes in swimming direction were found to accompany morphological transitions in \textit{Leptospira} bacteria.

This framework also echoes the discussion in section~\ref{sec:Passive} on translation–rotation coupling in non-spherical particles. There, anisotropy in shape led to off-diagonal components in the resistivity tensor, producing coupling between translational and rotational dynamics. Similarly, in the viscotaxis model, geometrical asymmetry creates an effective coupling between propulsion and reorientation. Although this minimal model successfully captures several features of viscotactic alignment, it omits certain finer effects, such as additional rotation–translation couplings due to viscosity variation along the surface of each sphere. In more realistic models that resolve the full surface of the swimmer, small asymmetries in surface stresses due to spatial viscosity gradients can produce rotation in uniaxial swimmers as well.

\begin{figure*}
\centering
\includegraphics[width=0.6\textwidth]{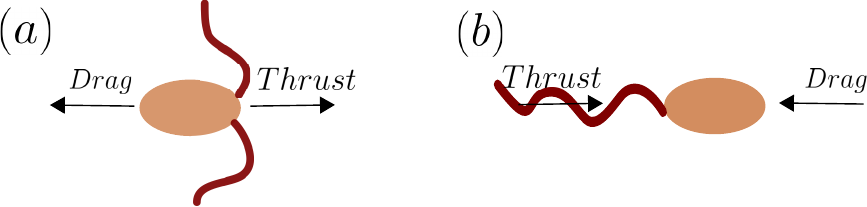}
\caption{Two types of squirmers moving from left to right. A squirmer is a simplified representation of a swimming microorganism, modeled as a particle with a prescribed surface slip velocity. (a) Puller: thrust is generated at the front. (b) Pusher: thrust is generated at the rear.}
\label{fig:pusher_puller}
\end{figure*}

In contrast to \textit{Leptospira}, \citet{sherman1982viscosity} showed that \textit{Escherichia coli}, which propels itself using a single flagellar bundle, exhibits negative viscotaxis, swimming down the viscosity gradient. More recently, \citet{stehnach2021viscophobic} documented more complex behavior in the bi-flagellated microalga \textit{Chlamydomonas reinhardtii}, which displays competing responses to viscosity gradients. First, the organism undergoes viscophobic turning, hypothesized to result from differential drag on its two flagella in stratified viscosity fields. This drag imbalance produces asymmetric thrust and causes the organism to reorient. Second, the alga slows down in high-viscosity regions. Under strong viscosity gradients, the turning response dominates, leading thereafter to motion along the direction they are pointing in, and to accumulation in low-viscosity zones. Under weak gradients, however, the slowdown effect prevails, causing higher concentrations in high-viscosity regions. The turning mechanism is absent in \textit{Escherichia coli} (E. coli, discussed below), which swims using peritrichous flagella that cluster into a single bundle \citep{chattopadhyay2006swimming}. A key distinction between these two swimmers is the swimming gait — the mode of propulsion used to achieve self-locomotion. 

Bacteria such as E. coli swim via a run-and-tumble mechanism, where periods of forward, ballistic motion (runs) are punctuated by stochastic reorientations or “tumbles.” During the run phase, multiple helical flagella bundle together and rotate like a screw, pushing fluid backward to propel the organism forward. These organisms that generate thrust from the rear are classified as pushers. In contrast, the green alga \textit{Chlamydomonas reinhardtii} employs a breaststroke-like gait, pulling fluid from the front with its pair of anterior flagella. Such organisms are termed pullers. Figure~\ref{fig:pusher_puller} schematically illustrates these two classes. Other swimming gaits too are observed in nature: spermatozoa use undulatory motion, sending travelling waves along their flagella; \textit{Vorticella} utilize jet propulsion; amoebae move by crawling via surface adhesion; and ciliates like \textit{Paramecium} swim using coordinated ciliary beating distributed across their surfaces.
To unify the modeling of such diverse gaits, the \textit{squirmer model} offers a minimal yet powerful framework \citep{lighthill1952squirming, blake1971spherical}. In this model, a spherical swimmer is assigned a prescribed tangential slip velocity on its surface, rather than imposing a no-slip condition. The slip velocity is expanded in terms of Legendre polynomials in the polar angle \( \theta \), measured from the swimmer’s axis of symmetry. Retaining just the first two modes, the tangential surface velocity may be written as
\begin{equation}
\mathbf{u}_\text{squirmer} = B_1 \sin \theta + B_2 \sin \theta \cos \theta,
\end{equation}
where \( B_1 \) determines the swimming speed and captures the source dipole occurring due to the source and sink flow around the swimmer, and \( B_2 \) encodes the stresslet (force dipole) strength occurring due to the equal and opposite forces exerted by their head and tail. Pushers correspond to \( B_2/B_1 > 0 \), pullers to \( B_2/B_1 < 0 \), and neutral squirmers (passive swimmers) to \( B_2 = 0 \). The velocity field in the surrounding fluid due to \( B_2 \), decays as \( B_2 r^{-2} \), whereas that due to the first mode decays more rapidly as \( B_1 r^{-3} \) in the far-field. Here, $r$ is the distance from the center of the squirmer. 

Using this model, \citet{datt2019active} showed that spherical squirmers in fluids with weak linear viscosity gradients exhibit negative viscotaxis—they rotate toward regions of lower viscosity. Extending this framework to prolate spheroids using spheroidal squirmer model, \citet{gong2024active} found that viscotaxis remains negative for elongated squirmers. However, the magnitude of the viscotactic response decreases with increasing aspect ratio. As illustrated in figure~\ref{fig:pusher_puller}, the tail and head exert equal and opposite forces in order to satisfy the force-free condition required in inertialess (Stokes) flow. \citet{gong2024active} hypothesized that for slender swimmers, the viscotactic response is dominated by the head, with the tail primarily serving as a thrust generator. They proposed that when the tail exerts an external force ($\mathbf{F}_\text{ext}$) on the head along the swimming direction, it could produce a \textit{positive} viscotactic response with angular velocity \( \mathbf{F}_\text{ext} \times \nabla \mu \), counteracting the \textit{negative} viscotaxis due to the imposed swimming velocity \( \mathbf{U}_\text{ext} \), causing a rotation velocity equal to \( -\mathbf{U}_\text{ext} \times \nabla \mu \). This idea reconciles the model with the experimental observations of \citet{takabe2017viscosity} showing positive viscotaxis.
In follow-up work, \citet{gong2024swimming} investigated the relative swimming efficiency of spheroidal pushers and pullers in stratified fluids. They showed that pushers—which generate thrust from the rear—are more efficient when swimming down the viscosity gradient, whereas pullers are less so. In contrast, when swimming up the viscosity gradients the pullers are more efficient and pushers less efficient than neutral swimmers.  This outcome is intuitive: placing the thrust-generating rear end in a region of higher viscosity improves propulsion, while the drag-experiencing front end benefits from lower viscosity.

The preceding discussion assumes that viscosity gradients are externally imposed and unaffected by the swimmer’s motion. However, as with passive particles \citep{oppenheimer2016motion, ziegler2022hydrodynamic}, active swimmers can also generate local viscosity variations by altering scalar fields such as temperature or solute concentration. \citet{shaik2021hydrodynamics} examined this scenario and found that although the type of boundary condition imposed on these scalar fields does not affect the direction of viscotaxis, it significantly influences both the swimming speed and the rate at which steady-state orientation is achieved.
For instance, under a no-flux boundary condition of the scalar that directly affects viscosity (e.g., for temperature as governed by equation~\eqref{eq:TemperatureTransport}), pushers tend to swim faster, pullers swim slower, and passive particles remain unaffected. These changes arise from local modifications to the viscosity field, primarily due to the term $\hat{\mu}(\mathbf{x})\boldsymbol{\sigma}^{\text{Stokes}}$ in equation~\eqref{eq:NetStress}. In contrast, when Dirichlet boundary conditions are applied — maintaining the value of the scalar (such as temperature) at the swimmer’s surface as constant and different from that of the bulk fluid — the flow is altered non-locally. This is reflected by \( \boldsymbol{\sigma}^{\text{Stratified}} \) in equation~\eqref{eq:NetStress}, which modifies swimming behavior in a manner that depends sensitively on the squirming mode ratio \( B_2/B_1 \) and the swimmer's temperature relative to the ambient fluid. These findings indicate that both local and non-local feedbacks from swimmer-induced viscosity variations can significantly alter motility characteristics. Moreover, the qualitative trends observed for spherical squirmers extend to elongated, spheroidal swimmers as well, as shown by \citet{gong2024active}, suggesting robustness of these effects across a range of swimmer geometries.

In section \ref{sec:PassiveinStratified} we briefly commented on the time-reversal symmetry of the Stokes equations governing low-Reynolds-number flows. This symmetry imposes a fundamental constraint on microscale locomotion - known as the \textit{scallop theorem}, first described by \citet{purcell1977life}.  According to this theorem, a reciprocal motion — one that retraces the same sequence of shapes forward and backward in time — cannot result in net propulsion in a fluid with uniform, constant viscosity. For example, a scallop opening and closing its shell in a time-reversible manner will not translate, because the flow induced during the forward stroke is exactly undone during the backward stroke. Even if the opening and closing occur at different rates, the net displacement remains zero. This principle explains why microorganisms employ non-reciprocal gaits, such as the helical rotation of flagella in \textit{E.~coli} or the breaststroke-like beating of flagella in \textit{Chlamydomonas}, to achieve propulsion in unstratified fluids. Interestingly, \citet{esparza2021rate} showed that the scallop theorem remains valid even when the viscosity is a smoothly varying function of space. This result is somewhat counterintuitive, given that different parts of a reciprocal swimmer can experience different local viscosities. However, the authors hypothesized that the advection of viscosity fields — such as including advection of temperature in equation~\eqref{eq:TemperatureTransport}, which becomes relevant at finite Péclet numbers or across sharp viscosity gradients — could break time-reversal symmetry and thereby invalidate the scallop theorem. Future numerical simulations exploring such advection-coupled viscosity fields may help clarify the extent to which time-reversibility can be broken by stratified environments.

The theoretical investigations of microswimmer locomotion in viscosity-stratified fluids discussed above reveal that spatial variations in viscosity can significantly influence both the swimming direction and efficiency, depending on the organism’s morphology and swimming gait. Thus far, our discussion has focused on smoothly varying (continuous) viscosity fields. However, in biological environments, microorganisms frequently encounter sharp interfaces — such as cellular membranes or mucus layers — where the viscosity changes abruptly. Note that here, and elsewhere in this review, our `interfaces' are \emph{miscible} expect when otherwise specified. The viscosity variation across these interfaces is sharp but continuous. In effect miscible interfaces are thin `mixed layers' within which the concentration varies from one fluid's to the other's. These step-like variations allow for additional swimming behaviors, further enriching the dynamical landscape of microswimmer motion in viscosity stratified fluids.
In their theoretical and numerical investigation, \citet{gidituri2022reorientation} examined how squirmers interact with sharp viscosity interfaces. In the absence of the source dipole (\( B_1 = 0 \)), they found that torque generated by the force dipole alone causes pushers (\( B_2 < 0 \)) to align parallel to the interface, while pullers (\( B_2 > 0 \)) tend to align perpendicular. When \( B_1 \ne 0 \), the source dipole introduces an additional torque that drives reorientation toward the lower-viscosity region, consistent with experimental observations by \citet{coppola2021green} on the green alga \textit{Chlamydomonas reinhardtii}. The interplay between these two torques leads to a rich set of behaviors. For instance, pushers with weak dipole strength (\( B_2/B_1 \approx 0 \)) reorient nearly perpendicular to the viscosity interface, while stronger pushers (\( |B_2/B_1| \gg 1 \)) settle at finite inclination angles and swim along the interface. Conversely, strong pullers consistently align normal to the interface, pointing either into or away from the lower-viscosity fluid depending on their initial orientation. These findings underscore how swimming behavior at viscosity discontinuities depends sensitively on the swimmer’s mode of propulsion.

In the experiments of \citet{coppola2021green} on the green alga \textit{Chlamydomonas reinhardtii}, pullers aligned normal to a sharp viscosity interface were observed to swim toward it. The interface separated methyl cellulose (a highly viscous fluid) from water (a low-viscosity medium). When swimming from the high-viscosity to the low-viscosity side, the cells showed only a slight change in orientation after crossing the interface, deviating modestly from their initially normal alignment. Remarkably, however, when approaching from the low-viscosity side, the swimmers exhibited a pronounced reluctance to enter the high-viscosity region. Instead, they frequently scattered back into the low-viscosity fluid. This counterintuitive behavior contrasts with classical expectations: in uniform media, organisms tend to accumulate in regions where they move more slowly. Yet in this case, despite reduced swimming speeds in the high-viscosity phase, \textit{Chlamydomonas} did not preferentially accumulate there. These observations highlight the profound impact of viscosity interfaces on microbial distribution.
A theoretical explanation was provided by \citet{gong2023active}, who modeled squirmers approaching viscosity interfaces. They showed that both sharp, and even smoothly diffuse, viscosity transitions can lead to a phenomenon analogous to the total internal reflection of light. Specifically, when the swimmer's angle of incidence relative to the interface normal exceeds a critical value, the swimmer reorients and scatters back into the high-viscosity region. This critical angle is inversely proportional to the viscosity ratio between the two phases. For neutral squirmers, the theoretical critical angle for a viscosity ratio of 2 is approximately \( \pi/5 \), in qualitative agreement with the experimentally observed angle of about \( \pi/3 \) \citep{coppola2021green}. Pushers and pullers exhibit similar behavior, with critical angles slightly larger and smaller, respectively, than for neutral squirmers. These results underscore the effectiveness of the squirmer model in capturing the  essential features of microswimmer-interface interactions.

\citet{esparza2021dynamics} conducted a combined theoretical and experimental study on a different class of swimmer — an artificial microswimmer with a helical tail, mimicking the propulsion mechanism of \textit{E. coli}, and traveling perpendicular to a viscosity interface. The swimmer's head had an aspect ratio of about  3.5. They considered four different modes of approach to the interface, paying greater attention to a head-first (pusher-like) and a tail-first (puller-like) approach. Their results revealed that when approaching the interface from the low-viscosity side, a head-first configuration led to a reduction in swimming speed due to increased drag on the head. Conversely, a tail-first orientation resulted in increased swimming speed, as the propulsive unit experienced higher viscosity, thereby enhancing propulsion efficiency. These observations are consistent with theoretical predictions from the spheroidal squirmer model presented by \citet{gong2024swimming}, as well as with resistive force theory calculations by \citet{esparza2021dynamics}. Both models predict that the response of a swimmer to crossing a viscosity interface depends on the direction of travel: for each orientation, a swimmer going from low to high viscosity should exhibit behavior opposite to the one going from high to low described above. However, the experiments showed that when the swimmer travelled from low to high viscosity it slowed down regardless of whether it approached head- or tail-first. The authors hypothesized that this discrepancy may be due to the entrainment of high-viscosity fluid around the swimmer. Future experiments or direct numerical simulations may help clarify the role of such effects.

These findings not only explain microbial navigation but also open possibilities for engineering particles that exploit viscosity gradients for functional outcomes. Using adjoint optimization techniques and novel shape parametrization formulation, \citet{eggl2020shape,eggl2022mixing} found the optimal shapes required to obtain maximum two dimensional mixing of a binary fluid using two stirrers initialized with circular shapes. The optimal shapes they obtained were non-intuitive and led to enhanced mixing. Similar novel shapes may be expected when optimising for various desired objectives in a stratified fluid. For example, the particle shapes required to obtain maximum and minimum sedimentation velocity in an inertia-less fluid are a sphere and a slender fibre, respectively. However, this is unlikely to be true when viscosity is stratified, as a fore-aft asymmetric particle with more surface area in the region of smaller viscosity is likely to experience less skin friction drag than a symmetric particle. Under the constraint of fixed particle volume, shapes with a fore-aft asymmetry along the viscosity gradient direction could be envisioned to have different drag than a shape that has this symmetry. In the case of active particles, \citet{piro2024energetic} study the energetics of different body shapes in swimming microorganisms and find that for different optimisation criteria, different body shapes are preferred. Such optimized particle or organism shapes will be useful, for example, in obtaining the slowest sedimentation speed for maximal mass transport in fluidized bed reactors or for the fastest swimming velocity in drug-delivery applications using smart active particles \citep{tsang2020roads}. 

In this subsection, we discussed the motion of active particles in viscosity-stratified environments. Similar to the effects of solute concentration gradients discussed earlier in section~\ref{sec:SoluteConcent}, suspensions of such active particles can themselves generate effective viscosity stratification, even without ejecting cellular or other material from their bodies. The active stresses in these systems depend on factors such as microbial concentration and orientation, and can lead to non-Newtonian rheological behavior, including shear thinning or thickening \citep{hatwalne2004rheology,rafai2010effective,saintillan2010dilute}. Spatial variations in microbial activity or alignment can thus induce viscosity gradients within the suspension. These gradients, in turn, can influence the motion of other passive or active particles, enabling new avenues for controlling transport via swimmer activity, gate, concentration or orientation fields. Such strategies may find use in applications ranging from targeted drug delivery to the design of smart materials with tunable rheology. These effects are analogous to viscosity stratification in passive non-Newtonian fluids—an important topic that we do not address in detail here, as it lies outside the scope of this review (see section~\ref{sec:VariableViscIntroA}).

We now turn from particle-laden flows to particle-free shear flows, wherein spatial variations in viscosity can generate novel hydrodynamic instabilities that, depending on the application, may either need to be suppressed or harnessed. The next three sections are closely interconnected. They address, in sequence, the mathematical structure of fluid instabilities, their manifestation in canonical flow configurations, and the eventual turbulent state arrived at as a consequence of these instabilities followed by more complicated processes. At one point, in section~\ref{sec:ParticleShear}, we revisit the role of suspended particles in the context of large-scale viscosity-stratified shear flows.

\section{Shear flows with variable viscosity}\label{sec:shear_flows_var_visc}
In section~\ref{sec:equations}, we briefly discussed the implications of neglecting the $1/Re$ term in the momentum conservation equation~\eqref{eq:ns} in the singular limit as $Re \rightarrow \infty$. We now examine this more carefully in the context of high Reynolds number shear flows, that often present us with singular perturbation problems. 
While a lot is understood about this in constant viscosity flows, it is not common knowledge that viscosity variations, arising from the third  term in equation \eqref{eq:StressDiverDecomp}, can appear as singular terms at the lowest order. The purpose of this section is to emphasise this aspect and discuss its consequences. Unless otherwise specified, in this section we assume density to be constant across the flow. It is instructive to start our discussion with a toy singular-perturbation problem. Further details can be obtained in classical texts, such as in chapter 4 of \citet{leal2007advanced}.

\subsection{A toy singular perturbation problem and its relevance}\label{toy_prob}

\begin{figure}
\centering 
\includegraphics[width=0.45\textwidth]{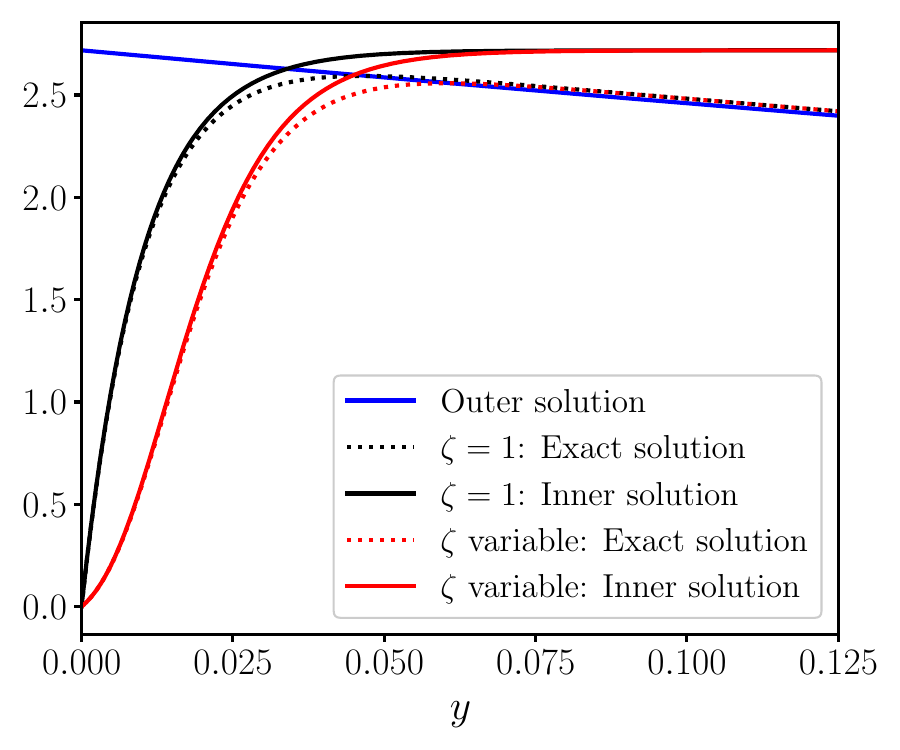}
\caption{Approximate and exact solutions of the modified Friedrich problem (equation \eqref{eq:friedrich}) with $\zeta=\left(1+0.2 \exp[-y/\epsilon_0]\right)$ and $\epsilon_0=10^{-2}$. Also shown are solutions for constant $\epsilon$ ($\zeta=1$).}
\label{fig:singular}
\end{figure}

Among the simplest singular perturbation problems is Friedrich's problem, which we have modified as given below:
\begin{equation}
(\epsilon f')'+ f' + f = 0, \textrm{ with } f(0)=0,~f(1)=1.
\label{eq:friedrich}
\end{equation}
Primes in this review denote a derivative with respect to the independent variable. In this case, this is the spatial variable $y$, and $\epsilon(y) = \epsilon_0 \zeta(y) \ll 1$ is our proxy for viscosity, or in non-dimensional terms, the constant $\epsilon_0$ is the inverse of the Reynolds number. In the classical Friedrich problem, $\zeta=1$ and the first term is presented as $\epsilon_0 f''$. We have allowed $\epsilon$ to vary, analogous to viscosity stratification within the momentum equation \eqref{eq:ns}. Equation \eqref{eq:friedrich}, with $\zeta=1$, is analytically solvable, and its solution is shown by the black dashed line in figure \ref{fig:singular} for a choice of $\epsilon_0=10^{-2}$. Most singular perturbations in fluid dynamics are not analytically solvable but lend themselves to construction of hierarchies of solutions up to the order of accuracy desired. The variable $\zeta$ case is in general not easy to solve analytically, but here too the complete solution is easily obtained numerically, and shown by the red dashed line in figure \ref{fig:singular}, for a specific choice of $\zeta(y)$. We shall construct the lowest-order solution to equation \eqref{eq:friedrich}, to acquaint the reader with the basic ideas behind solving singular perturbation problems. 

Given that $\epsilon$ is extremely small, it might be tempting to neglect the first term, obtain a far simpler equation, and hope to get a good approximation to the correct answer. That this approach will lead us into trouble becomes immediately evident: the order of the equation reduces by 1, and we cannot satisfy both boundary conditions. Thus, however small $\epsilon$ is, our solution is going to be quite wrong at least near one of the boundaries. Such a problem contains thus a thin ``boundary layer" near one of the boundaries where the term containing $\epsilon$ may not be dropped, and the remaining region, where it may be. This remaining region is known as the outer region (relative to the boundary layer), where the problem posed by
\begin{equation}
f_{outer}' + f_{outer} = 0,
\label{outer}
\end{equation}
is valid. We must make a choice, dictated by the nature of the problem, about where to apply the boundary condition in the outer problem. Here, it is applied at $y=1$, i.e. the right boundary where $f_{outer}(1)=1$ using the original boundary condition in equation \eqref{eq:friedrich}. The outer solution is an exponential, shown by the blue line in figure \ref{fig:singular}. 
As $\epsilon$ decreases, the outer solution is valid over a larger and larger fraction of the domain, i.e., the boundary layer becomes thinner and thinner. 

The solution within the boundary layer is described by the inner equation. We first derive the inner solution for $\zeta=1$ and then ask what happens if $\epsilon$ varies. To define the inner problem near $y=0$, we proceed as follows. We know that the term containing $\epsilon_0$ is of significant magnitude in this region. This can only be possible if the second derivative becomes large here. We may therefore define an inner variable $Y \equiv y/\delta$, where $\delta \ll 1$ will be chosen to ensure that derivatives with respect to $Y$ are $O(1)$. We expand the unknown variable in powers of $\delta$ as
\begin{equation}
f_{inner} = \sum_{i=0}^\infty f_i \delta^i.
\label{expan_inner}
\end{equation}
Substituting the expansion \eqref{expan_inner} into equation \eqref{eq:friedrich}, and retaining only the lowest-order terms, we get, in the inner layer,
\begin{equation}
\frac{\epsilon_0}{\delta}  \frac{ d^2 f_0}{dY^2} + \frac{df_0}{dY} = 0.
\label{inner}
\end{equation}
Without loss of generality, we choose $\delta=\epsilon_0$.
The `outer' boundary condition for the inner solution is obtained by requiring the matching of the inner and the outer solutions:
\begin{equation}
f_{inner}(Y \to \infty) = f_{outer}(y \to 0),
\label{matching}
\end{equation}
while the inner boundary condition at $y=0$ is satisfied exactly. This determines the constants of integration, and we obtain the inner solution for $\zeta=1$, shown by the black solid line in figure \ref{fig:singular}. Equations \eqref{outer} and \eqref{inner} constitute the zeroeth order (here, the lowest order) outer and inner equations, respectively, for $\zeta=1$. We see that neither does too well in the crossover region seen at $y \sim 0.04$ in figure \ref{fig:singular}, where both the dashed black (inner) and solid blue (outer) curves over-predict the exact solution given by dashed black curve. There are many methods to achieve good matching between the two solutions in the crossover region. In addition, it is possible to improve these solutions by writing a hierarchy of higher-order equations, and we leave this to the reader as an exercise. 

Let us now take up the case where the small quantity $\epsilon$ is not constant everywhere, but is given by
\begin{equation}
\epsilon=\epsilon_0 {\zeta}(y),
\end{equation}
where $\zeta$ is of $O(1)$, and the system satisfies the modified Friedrich problem of equation \eqref{eq:friedrich}.
While the outer equation remains unchanged, a different lowest-order equation will emerge for the inner layer that contains all terms that can contribute to the dominant balance. For this, we need to know the magnitude of $d{\zeta}/dY$, i.e., the variation of $\epsilon$. If it is $O(1)$, the lowest-order inner layer equation, with $\epsilon_0=\delta$ is
\begin{equation}
{\zeta} \frac{ d^2 f_0}{dY^2} + \left[1 + \frac{d {\zeta}}{dY}\right] \frac{df_0}{dY} = 0,
\label{inner2}
\end{equation}
from which it is evident that variations in $\epsilon$ contribute at the lowest order. The inner solution for the variable small-parameter case is shown by the solid red figure \ref{fig:singular}. It is clear that a twenty percent overall change (as specified for the modified Friedrich problem) in the small parameter can change the character of the inner solution completely. 

The analogy we wish to draw here is that at high Reynolds numbers, the Navier--Stokes (with or without viscosity stratification) poses a singular perturbation problem. If we neglect the viscous terms completely, we cannot satisfy all the boundary conditions, and in particular the no-slip boundary condition at solid walls. On the other hand, if we keep viscosity but neglect {\em variations} in viscosity, we may be able to satisfy all boundary conditions, but our solution could be quite wrong, even at the lowest order of approximation. As the Reynolds or P\'eclet number becomes larger (e.g. viscosity or diffusivity becomes smaller) we obtain a progressively thinner boundary layer near the walls, where velocity or viscosity gradients grow higher. Thus, the wall layer is an inner layer. There is another `inner' layer we shall encounter in stability problems, namely the critical layer, where the perturbation phase speed is close to the base velocity of the flow. 
While boundary layers are frequently encountered in fluid mechanics, they are not unique to fluid mechanics. Indeed, they are encountered in any situation where the highest derivative in the governing equation is multiplied by a small factor.

\subsection{Shear flows as singular perturbation problems}

\begin{figure}
\centering
\includegraphics[width=0.45\textwidth]{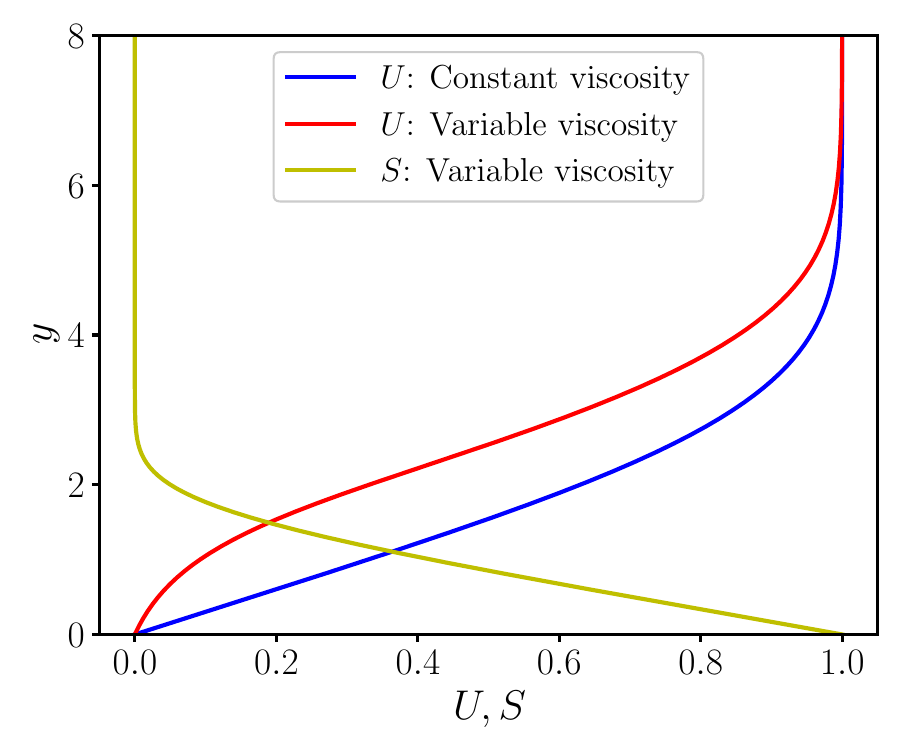}
\caption{Boundary layer velocity profile on a flat plate with constant viscosity (blue) and varying viscosity (red). The concentration profile for the latter is shown in yellow; the viscosity depends on the concentration as $\bar\nu = \exp{(2S)}$. Note that the $y$-axis is scaled by the boundary layer thickness, which is much thinner than the typical streamwise length scale.}   \label{fig:heated_bl}
\end{figure}

\subsubsection{Steady laminar shear flows}
\label{sec:base_flow}
As discussed in section \ref{sec:equations}, when the Reynolds and Péclet numbers are extremely high, a naive glance at equations \eqref{eq:ns} and \eqref{eq:species} suggests that we neglect their right hand side. But taking a hint from the toy problem above, we know that this could be a highly erroneous thing to do, because our system could constitute a singular perturbation problem. This becomes evident when we notice that the highest spatial derivative is multiplied by the reciprocal of the Reynolds number. So, if we neglect viscosity, or equivalently, set $Re \to \infty$, the equation reduces to one of lower order, and all the boundary conditions cannot be satisfied; the framework of inviscid flows has no way of satisfying the no-slip condition at the wall. 

It follows that viscous forces must enter the dominant balance somewhere, however large the Reynolds number. There are at least two ways this can happen. One is that the geometry of the flow results in the nonlinear terms in the Navier--Stokes equations being zero or small, so viscous forces must balance the pressure gradient, and are large everywhere. Plane Poiseuille flow is a canonical example. The other is that in a thin region close to the wall, there is a dominant balance between viscous and nonlinear terms. The boundary layer, created by uniform flow at velocity $U_\infty$ past a flat plate, is a textbook example of a flow profile that emerges from a singular perturbation; the constant-viscosity solution is shown by the blue line in figure \ref{fig:heated_bl}. 
In fact, besides boundary layers, practically all shear flows are created by viscosity, typically by the need to satisfy a no-slip boundary condition somewhere. In other words, it is very hard to create a shear flow in a lab without using solid surfaces of any kind. Mixing layers in the atmosphere are a counterexample, but here viscous effects create the region of shear and make it thicker with time. So obviously, viscous effects are important for defining the shear flow at any Reynolds number. 

While we may not neglect viscosity, can we not get away with disregarding its variations? 
Our purpose below is to show that even at the highest Reynolds numbers, variations in viscosity may not be neglected, even at the lowest order of approximation, in the following contexts: (i) when the variations are large, and (ii) when changes in viscosity are small, but occur over small length scales such that the gradients are significant. We discuss the velocity profiles in two canonical flows: the Blasius boundary layer and plane channel flow, both modified by a base viscosity stratification in the coordinate $y$. These flows are well-described by a parallel flow approximation, so $U=U(y)$. 

The heated boundary layer, or the `inner' layer, is created by a free stream at velocity $U_\infty$ flowing past a flat plate, at which the temperature (or  concentration) is maintained to be different from that of the free stream. The steady laminar profiles within this thin layer are described by the similarity equations \citep{Miller_etal_2018PRF}
\begin{eqnarray}
\bar\nu f''' + \bar\nu' f''  + \lambda f f'' & = & 0, \\
s'' + Sc \lambda f s' & = & 0 \textrm{ and } \bar\nu=\bar\nu(s), 
\label{blasius++}
\end{eqnarray}
where the primes refer to differentiation by the nondimensional wall-normal variable $y \equiv y_d/\delta$, $\delta(x_d)$ is a characteristic thickness of the velocity boundary layer given by $d\delta/dx_d=\lambda/Re$, $\lambda$ being an $O(1)$ constant. The species concentration is $s$, while $f$ is the streamfunction nondimensionalised by $U_\infty$ and $\delta$, giving $U=f'$. The Reynolds number $Re=U_\infty \delta(x_d)/\nu_\infty \gg 1$ does not appear in the above equation since we chose the similarity variable $\delta$ as length scale. But this brings in a feature that the Reynolds number is a function of distance $x_d$ from the leading edge. The subscripts $d$ and $\infty$ stand for a dimensional quantity and the freestream, respectively. Note that the boundary layer thickness $\delta$ has the same meaning as $\delta$ in the toy problem in section \ref{toy_prob}. Equation \eqref{blasius++} yields the Blasius boundary layer profile when viscosity is constant, i.e., $\bar\nu'=0$. Evidently, viscosity multiplies the highest derivative $f'''$ and, if viscosity differences are $O(1)$ relative to the freestream viscosity $\nu_\infty$, we have the second term $\bar\nu' f''$ of the same order of magnitude as the first. 

From our discussion in section \ref{sec:nondim}, we have gleaned that two liquids, or a solute in solution into the same of another liquid, typically diffuse very slowly into each other, i.e., Schmidt numbers are $O(10^2-10^4)$. Thus, rather sharp viscosity variations close to the wall can be sustained for extremely long times. The concentration boundary layer next to the wall is now extremely thin, and $\bar\nu' f''$ is very large within it. In other words, we have a concentration boundary layer that is much thinner than the momentum boundary layer, and the new behaviour that emerges from this effect in laminar shear flows is nothing short of amazing, with singular behaviour being arguably the most remarkable. Depending on the sign of $\bar\nu'$, new instabilities can result, and so can enormous stabilization. 

A sample velocity profile with the inclusion of a viscosity contrast is shown in figure \ref{fig:heated_bl} by the red line. For purposes of visualisation, a viscosity contrast of about $7$ has been used between the wall and the freestream, and a modest Schmidt number of $20$ has been prescribed. 
Evidently, the flow is highly modified, and, to a practised eye, it will be amply clear (upon noting the introduction of inflexion in the velocity profile due to the concentration gradient), that the modified velocity profile will go unstable at a far lower Reynolds number than the constant viscosity one. Thus, viscosity variations, and not just viscosity, can be important in the singular perturbation problem, and classical boundary-layer theory needs to be modified to account for them. 

Plane Poiseuille flow of a viscosity-varying fluid will satisfy the steady Navier--Stokes equation for one-dimensional flow:
\begin{equation}
G \equiv \frac{dP}{dx} = \bar\nu U'' + \bar\nu'U',
\label{plane_poise}
\end{equation}
where a prime now refers to a derivative with respect to the wall-normal direction $y$, when the flow is driven by a constant streamwise pressure gradient $G$. When the two walls of the channel are maintained at different solute concentrations or temperatures, then the importance of viscosity variations will depend on the size of the viscosity difference across the channel. On the other hand, in the two-fluid flow, slow diffusion ensures that the mixed layer is of thickness $q \ll 1$ so even though the viscosity change is small, its gradient $\bar\nu'$ could well determine the balance in equation \eqref{plane_poise}. 

\subsubsection{The departure from a steady laminar state}
\label{sec:ShearFlowInstability}
We have seen that viscosity variations can fundamentally alter the nature of singular perturbation problems in steady laminar flows. We have also hinted that such variations can profoundly impact the stability of these flows. In this section, we focus on the stability of a specific subset within the broad class of multiphase shear flows—namely, miscible two-fluid systems (such as glycerol and water for high viscosity contrast or water and salt water for mild contrast) and suspensions of particulate matter. Viscosity-stratified shear flows have been the subject of extensive investigation over the past several decades, and a number of comprehensive reviews are available \citep{Joseph_Renardy_1993book1,Boomkamp_Miesen_1996IJMF,joseph1997core,Mohammadi_Smits_2016JFE}. However, neither these reviews nor standard textbooks emphasize the singular nature of the associated stability problems. We therefore take a pedagogical approach in this section. We begin with a broad overview, followed by detailed discussion in the subsequent subsections.
\begin{figure*}
\centering
\includegraphics[width=0.9\linewidth]
{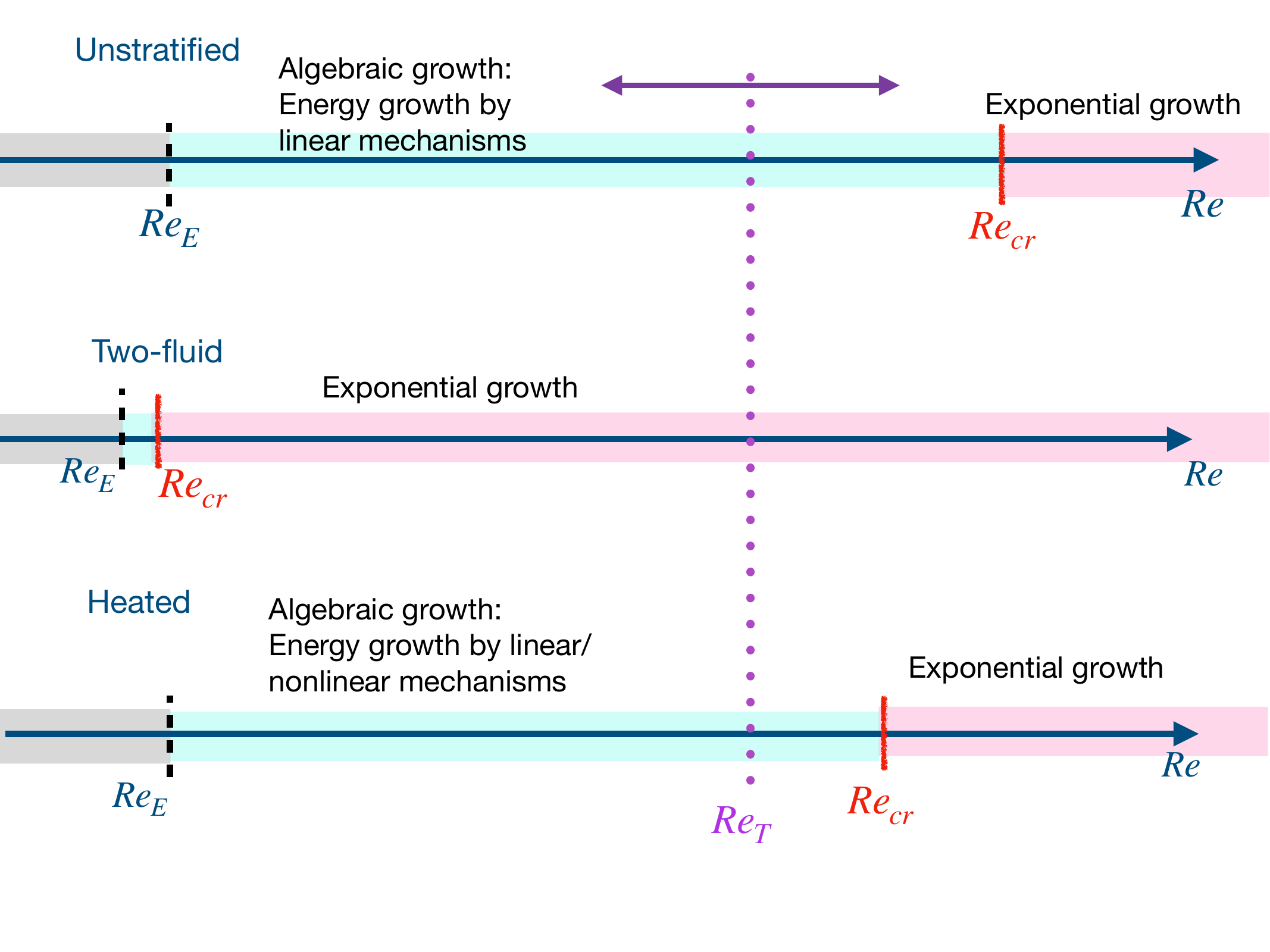}
\vskip-0.5in
\caption{In unstratified flows (top panel), beyond $Re_E$, stable but non-orthogonal linear eigenmodes can optimally combine to produce transient algebraic `nonmodal' growth, potentially triggering nonlinear effects, and thereby, turbulence. At $Re_{cr}$, at least one mode becomes linearly unstable and grows exponentially. This schematic contrasts scenarios in two-fluid (middle panel) and heated (bottom panel) flows with the constant-viscosity case. In constant viscosity, algebraic perturbation growth occurs via linear mechanisms. In two fluid flow, destabilization across a thin stratified layer may lead to exponential growth at lower $Re$, while heated flows may transition in a manner similar, at first sight, to constant viscosity flows, but the fastest growing perturbations could more likely be those that evolve in a nonlinear fashion. The onset of fully developed turbulence may occur across a range of $Re$ depending on the situation, and warrants further study.}
\label{fig:subcritical}
\end{figure*}

The process of transition to turbulence in viscosity-stratified shear flows is poorly understood. The schematic shown in figure \ref{fig:subcritical} gives a broad idea of the initial steps of the process, but since not enough is known about the process in viscosity-varying flows, future work may well lead to changes and additions in this schematic. It is standard to begin by linearising the system of equations and looking for the first departure from a laminar steady state. We define two Reynolds numbers relevant to the linear problem as shown in figure \ref{fig:subcritical}: $Re_E$, below which any general linear perturbation will monotonically decay, and $Re_{cr} \ge Re_E$, above which at least one linear eigenmode will grow exponentially. If the linear stability operator, $\mathcal{L}$ (as discussed in section \ref{sec:NonNormalVVF}), is self-adjoint, i.e., $\mathcal{L}\mathcal{L}^T(\cdot) = \mathcal{L}^T\mathcal{L}(\cdot)$, then $Re_E=Re_{cr}$, and a supercritical bifurcation would take place at $Re_{cr}$, i.e., the basin of attraction of the laminar state vanishes. Past this Reynolds number, the flow must go towards a new state in response to the smallest perturbation, so long as it contains a component of the unstable eigenmode. Note that depending on the flow situation, the relevant nondimensional number could be different from the Reynolds number. For Rayleigh--B\'enard convection, $\mathcal{L}$ is self-adjoint, so we directly have a supercritical bifurcation and convection rolls appear at $Ra_E=Ra_{cr}$. However, in shear flows, which are our focus in this section, the linear stability operator is not self-adjoint, i.e., $\mathcal{L}\mathcal{L}^T(\cdot) \ne \mathcal{L}^T\mathcal{L}(\cdot)$, so in general $Re_E < Re_{cr}$. 

We may now have a {\em subcritical} bifurcation to a new state at a Reynolds number $Re_{sub}$ lying between $Re_E$ and $Re_{cr}$, depending on the nature and amplitude of the perturbation. In shear flows, beyond  $Re_{sub}$, some initial perturbations may grow transiently and then decay, and thus take the system back to the laminar state. But others will show transient algebraic growth sufficient to trigger nonlinearities, and take the system to a new attracting state: in fact the state space may contain several other attractors, which could be time-varying states such as limit cycles, a fully or partially turbulent state, or nonlinearly saturated periodic states. Such transient algebraic growth occurs by a linear mechanism with several eigenmodes participating, giving it the name \emph{nonmodal} growth. For more comprehensive treatments of disturbance growth due to non-normality of the linear stability operator, we refer the reader to \citet{trefethen1993hydrodynamic,Schmid_2007ARFM,luchini2014adjoint,kerswell2018nonlinear}. 

 Thus a host of standard shear flows, such as channel and pipe flows, go turbulent at Reynolds numbers well below $Re_{cr}$. Pipe and plane Couette flows should, according to traditional stability analysis, never go to turbulence regardless of how high the $Re$, since eigenvalues of $\mathcal{L}$ for these systems are always stable, but we know they do display turbulence, as is evident for example in the rapid and unsteady flow emerging out of a garden hose. It is undergraduate textbook material that the Reynolds number for fully developed turbulence, $Re_T \sim 2000$ in a pipe, although depending on experimental conditions (such as ensuring constant radius, very smooth surfaces, and isolating the pipe from the lab floor to reduce noise transmission), laminar flow can be maintained up to far higher $Re$. Even in channel (plane-Poiseuille) flow  $Re_T \sim 1500$ in practice, even though the eigenvalues of $\mathcal{L}$ first become positive at $Re_{cr}=5772.2$. This attainment of a turbulence `bypassing' the traditional route of modal instabilities has evoked much interest \citep{grossmann2000onset,avila2011onset,barkley2016theoretical,manneville2016transition,chantry2017universal,avila2023transition}. 
 
 In contrast, past a supercritical bifurcation, turbulence is attained with $Re_T > Re_{cr}$, and the two can differ by a large amount. There is an elaborate route to turbulence, occurring over a range of $Re$, which is very different for different flows. As an example of a supercritical route, consider the boundary layer in the flow past a flat plate where the system is kept very quiet by eliminating background noise to the extent possible. The Reynolds number, as we mentioned, is an increasing function of distance from the leading edge, and one can see the entire route to turbulence in one snapshot taken from above. For the Blasius boundary layer profile, we have growing waves corresponding to the highest growing eigenvalue once instability sets in. By Squire's theorem, which we will discuss, the first instability is two dimensional, and viewed from above we will see spanwise invariant lines of disturbance maxima and minima. At some Reynolds number $Re_{3D} > Re_{cr}$, secondary instability sets in, and the straight lines become wavy (i.e., three-dimensional). These then undergo further nonlinear instability, with perturbations of a range of wavenumbers appearing. At some distance downstream, we have the onset of turbulence at $Re_O > Re_{3D}$, where the first turbulent spot appears. These spots then grow, spread and merge, until the flow asymptotically reaches fully developed turbulence. By setting a criterion to define this point, we have $Re_T~(> Re_O)$. How will this entire process proceed in a heated boundary layer, or one into which a solute is dissolving all along the wall? Only future work will tell. Jets, wakes and shear layers too display $Re_T > Re_{cr}$ while their $Re_{cr} \sim O(10)$, far lower than that of the channel flow.

Since the route to turbulence varies qualitatively between flows, the minimum Reynolds number $Re_T$ for fully developed turbulence is shown as a range in figure~\ref{fig:subcritical}. We also note that in many cases (like boundary layer flow), there is a range of Reynolds numbers, $Re_O < Re < Re_T$, where the flow is neither fully laminar nor fully turbulent, but `transitional'. The fraction of time that the flow is turbulent increases from zero at $Re_O$, and tends to $1$ at $Re_T$. The route to turbulence could be quite different in viscosity-stratified flows, and two likely scenarios are illustrated in Figure~\ref{fig:subcritical}. In flows with sharp viscosity gradients, such as two-fluid systems, an instability distinct from the classical Tollmien--Schlichting mode can emerge at low Reynolds numbers. This so-called overlap (discussed below) can trigger exponential growth at $Re$ values where the constant-viscosity counterpart would yield only algebraic growth. Whether this leads to a lower transition Reynolds number $Re_T$ remains currently unknown.

In cases with gentler viscosity gradients, such as in heated flows, the transition may resemble that of constant-viscosity flows, aside from quantitative differences, or it may involve fundamentally new mechanisms. To illustrate the latter, consider a streamwise-invariant, pressure-driven channel flow. Taking the dot product of the Navier--Stokes equation~\eqref{eq:ns} with the velocity field and averaging over the homogeneous streamwise and spanwise directions, one obtains a modified Reynolds--Orr energy equation that accounts for viscosity variations:
\begin{eqnarray}
\frac{\partial \langle e\rangle}{\partial t}=-\bigg\langle \hat u_i \hat u_j\frac{\partial U_i}{\partial x_j}\bigg\rangle -\frac{1}{Re}\Bigg[\bigg\langle (\bar{\mu}+\hat\mu)\frac{\partial \hat u_i}{\partial x_j} \frac{\partial \hat u_i}{\partial x_j}\bigg\rangle - \nonumber \\ {\bigg\langle \hat\mu\frac{\partial \hat u_i}{\partial x_j} \frac{\partial U_i}{\partial x_j}} {+\frac{\partial \hat \mu }{\partial x_j}\frac{\partial (\hat u_i \hat u_j+\hat u_i U_j)}{\partial x_i}}+\frac{\partial \bar{\mu} }{\partial x_j}\frac{\partial (\hat u_i \hat u_j)}{\partial x_i}\bigg\rangle\Bigg],\label{eq:ReyOrrEnergy}
\end{eqnarray}
where the angle brackets imply volume integration (averaging over $x$ and $z$). In the above, $\langle e\rangle = \tfrac{1}{2} \langle \hat u_i \hat u_i \rangle$ is the net fluctuation kinetic energy, and the flow is decomposed into a mean component $(U_i, \bar{\mu})$ and fluctuations $(\hat u_i, \hat \mu)$.

For constant viscosity, this reduces to the classical Reynolds--Orr equation \citep{Schmid_Henningson_2001book}: ${\partial \langle e\rangle}/{\partial t}=-\langle \hat u_i \hat u_j{\partial U_i}/{\partial x_j}\rangle -\frac{1}{Re}\langle{\partial \hat u_i}/{\partial x_j} {\partial \hat u_i}/{\partial x_j}\rangle$. The first term represents energy exchange with the base shear flow — it is the shear production term which also appears in the base flow energy budget (i.e., that of the Reynolds averaged equation) but with opposite sign. In other words, the base flow loses this energy to the fluctuations. The second term represents viscous dissipation of perturbation energy. Notably, non-linear terms such as $\partial/\partial x_j (\hat u_i \hat u_i \hat u_j)$ appear in divergence form in the unaveraged perturbation energy equation and therefore integrate to zero over the domain under appropriate boundary conditions. They merely redistribute energy spatially and do not contribute to net growth. In variable-viscosity flows, the modified Reynolds–Orr equation [equation~\eqref{eq:ReyOrrEnergy}] reveals four additional mechanisms for perturbation energy growth, arising from mean and fluctuating viscosity gradients (the second line of the equation). 
This enables novel, shear-independent pathways to amplification, potentially influencing the transition to turbulence.

In the following subsections, we discuss modal instabilities leading to exponential perturbation growth, followed by a discussion on nonmodal or algebraic perturbation growth. 

\subsubsection{Singular effects on flow stability due to viscosity variation: the overlap mechanism}
\label{sec:singular}

We have seen that singular perturbation can arise in high-Reynolds number shear flows, but we reiterate that such a mathematical structure is far more widespread, in fact extending to situations well beyond fluid mechanics. Even at low Reynolds numbers, (i) when species diffusivities are so low that the Péclet number is high, we can encounter singularity-induced physics; (ii) when the Rayleigh number is high, buoyancy effects balance singular viscous effects near the wall. The combination of low Reynolds number, high Péclet number and moderate to high Rayleigh number is encountered in Earth's mantle and outer core. Coupled with variations in geometry, which themselves can bring in singular perturbations, and the effects of rotation, there are many problems to be studied. We describe parallel shear flows at high Reynolds number, to give the reader the basic approach, so that they may apply it to the flow of their interest. 

Singular perturbation methods have long been used for constant-viscosity flows, since the seminal work of \citet{lin1946stability}, who showed that there are two singular layers, where the ``inner" solution in the singular perturbation formulation, such as described for a toy problem in section \ref{toy_prob}, has to be applied: the critical layer and the wall layer. In these layers, viscosity stratification has the potential for singular contributions. For every linear eigenmode, the critical layer is identified by the location where its phase speed matches the base flow speed. It is well-known that perturbation kinetic energy is primarily produced within the critical layer, and from there it gets transported to elsewhere in the shear flow. It gets dissipated primarily within the wall layer. So, viscous effects operating in the critical layer are responsible for driving instabilities, while near-wall viscous effects work to suppress them. If the critical layer produces more perturbation kinetic energy than the wall layer can dissipate, we have an instability. Thus, viscosity plays a dual role: it can create or destroy instabilities, and compete with itself to do both!

It was realised by \citet{ranganathan2001stabilization} and \citet{Govindarajan_2004IJMF} that viscosity variations can modulate the critical layer singularity at the lowest order. The necessary condition is that we must have some overlap of the viscosity-stratified layer with the critical layer. The singular nature of the resulting system means that major stabilisation or destabilisation, to the tune of orders of magnitude in the critical Reynolds number, can result from just a minor overall change in viscosity. Apart from this, a new mode of instability, termed the overlap instability, can emerge. Species diffusivity, especially when low, is an influential player in the dynamics.
By isolating the dominant physics within the critical layer, we can isolate what drives the instability, find out whether viscosity variation plays a part in it, and if so, through which term it acts. This is the aim of this exercise. We follow the same approach as described in \citet{Govindarajan_2004IJMF}.

We split all flow quantities into their mean and a perturbation, i.e.,  $S_t = S(y) + \hat s(x,y,z,t)$, $\nu_t = \bar\nu(y) + \hat\nu(x,y,z,t)$ and $\bu_t = U(y)\mathbf{\hat{i}} + \hat\bu(x,y,z,t)$, where $\hat\bu=\hat u \tilde {\mathbf i} +\hat v \tilde{\mathbf j} + \hat w \tilde{\mathbf k}$. 
 Considering the viscosity to be a function only of concentration $s$, and using a Taylor expansion about the mean state, the perturbation viscosity is $\hat s[d\bar\nu/dS]$. Further, the perturbed quantities are all written in the normal mode form, e.g., 
\begin{equation}
\hat\bu = \bu(y)\exp\left[i(k_x x + k_z z - k_x c t)\right] + \ {\rm complex \ conjugate},
\label{normal_mode}    
\end{equation}
where $c$, the phase speed in the streamwise direction, will be an important quantity in later discussions. Substituting these in equations \eqref{eq:incomp}-\eqref{eq:ns} where we set gravity to zero, and linearising (i.e., neglecting nonlinear terms in the perturbation quantities), we obtain equations for the perturbations. We may eliminate pressure by taking the curl of the momentum equation and, after some algebra, write the system as 
\begin{align}
&(U-c)\mathcal{D}_{2-}{v} = U''{v} - \frac{i}{k_x Re} \left(\bar{\nu} \mathcal{D}_{2-}^2  +2\bar{\nu}'D\mathcal{D}_{2-}\right. \nonumber \\ 
 &\left.+ \bar{\nu}''\mathcal{D}_{2+}\right){v} - \frac{1}{Re}\left(U'\mathcal{D}_{2+} + 2U'' D +  U'''\right)\frac{d\bar\nu}{dS} s,
\label{os_v2}
\end{align}
\begin{align}
(U-c){\eta} =~& -\beta U'{v}  - \frac{i}{k_x Re}\left(\bar{\nu}'D + \bar{\nu}\mathcal{D}_{2-}\right]{\eta} \nonumber \\
& + \frac{\beta}{Re} \left(U''  +  U'D\right)\frac{d\bar\nu}{dS} s,
\label{squire_v2}
\end{align}
\begin{equation}
(U-c)s = \frac{i}{k_x}S'{v}  - \frac{i}{k_x Pe}\mathcal{D}_{2-}s,
\label{species_v2}
\end{equation}
where $\eta=\partial u/\partial z-\partial w/\partial x$ is the perturbation vorticity in the $y$ direction, a prime and $D$ stand for a total and a partial derivative in the $y$-direction respectively, $\beta=k_z/k_x$ is a measure of the obliqueness of the perturbation with respect to the streamwise direction, $\mathcal{D}_{2-} \equiv D^2 - k^2 $ and $\mathcal{D}_{2+} \equiv D^2 + k^2$, with 
\begin{equation}
k^2=k_x^2 + k_z^2. 
\label{keqn}
\end{equation}
The equations \eqref{os_v2}-\eqref{species_v2} respectively are the modified Orr--Sommerfeld, modified Squire and the species balance equations. We notice that $v$ and $s$ are independent of $\eta$, but $\eta$ is slaved to $v$ and $s$, and so the system \eqref{os_v2} and \eqref{species_v2} provides the complete solution. These flows are often confined spatially, let us say between $y=-L$ and $L$. In the case of walls we have no slip and no penetration boundary conditions for the velocity, and the species concentration could be held fixed at the walls to $S(L)$ and $S(-L)$, so
\begin{equation}
v (\pm L) = D v (\pm L) = s(\pm L) = 0.
\end{equation}
The second condition arises from the continuity equation and represents the no-slip condition $u(\pm L)=0$. For a jet  wake or shear layer we would replace $L$ by $\infty$ and the equal sign equation by a $\to$. 

\citet{Squire_1933PRSA} (see also \citet{Drazin_Reid_2004book} and \citet{Schmid_Henningson_2001book}) proved that in a constant-viscosity, incompressible, unidirectional shear flow, every three-dimensional perturbation eigenmode that becomes unstable (i.e., exhibits exponential growth) is associated with a corresponding two-dimensional eigenmode that becomes unstable at a lower Reynolds number. The transformation between two and three dimensions is exact: by defining a reduced Reynolds number $Re_{2D} = k_x Re / k_{2D}$ (with $k_{2D}=k_x\sqrt{1+\beta^2}$), one maps the three-dimensional disturbance (with streamwise wavenumber $k_x$ and spanwise wavenumber $k_z=\beta k_x$) to a two-dimensional one of streamwise wavenumber $k_{2D}$. 
In the case of viscosity-stratified flows, Squire's theorem still applies, since the classical transformation can be extended by defining $k_{2D} \hat{s}_{2D} = k_x \hat{s}$, following which equations~\eqref{os_v2}, \eqref{squire_v2}, and \eqref{species_v2} for a 3D perturbation reduce to those for a 2D perturbation, with $Re_{2D} \le Re$. In other words, any three-dimensional modal instability has a counterpart that occurs at a lower Reynolds number in two dimensions. Most importantly, $Re_{cr}$, the smallest Reynolds number for instability, is always achieved by a two-dimensional mode. A stratification of density would render Squire's theorem invalid. Nonetheless, there too it provides a thumb rule, since the lowest Reynolds numbers are often for 2D perturbations. 
Squire’s theorem is further useful as it allows for the reverse process, i.e., the full three-dimensional spectrum can be obtained from solving just the two-dimensional problem. However, care must be taken: the species concentration eigenfunction gets rescaled when the transformation is applied, so the other eigenfunctions change as well, potentially affecting non-normal growth.

\begin{figure*}
\centering
\vskip-1.5in
\includegraphics[width=0.9\linewidth]{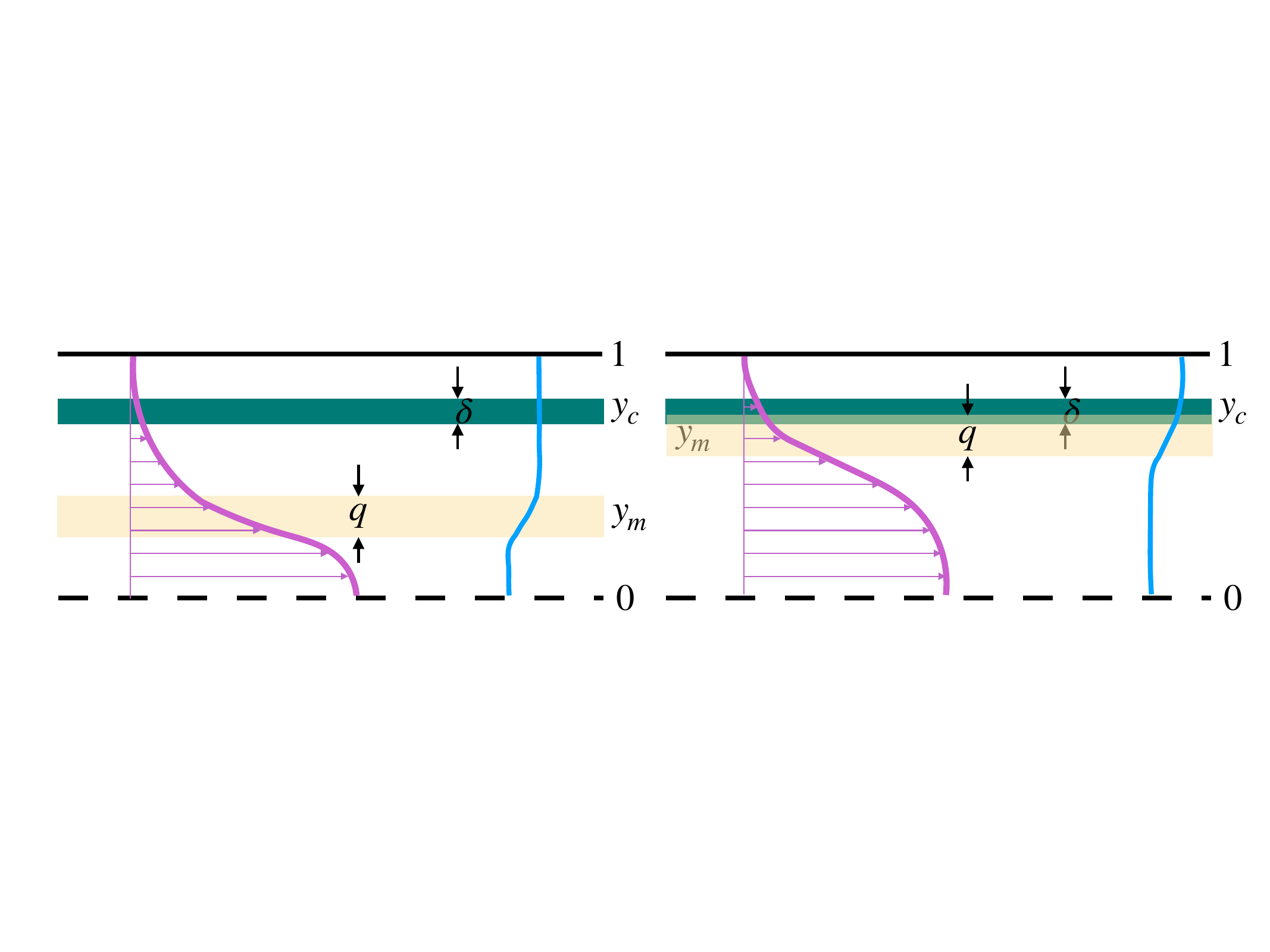}
\vskip-1.5in
\caption{Schematic of a half channel showing the critical layer for the species equation (in blue-green) and the layer where viscosity varies (in yellow). The momentum critical layer will be centred at $y_c$ as well, with a thickness of $\epsilon$. Profiles of velocity (in magenta) and viscosity (in blue) are sketched. Left: non-overlapping conditions, right: overlap conditions. Note that $q$ and $\delta$ are different in magnitude.}   \label{fig:schematic_overlap}
\end{figure*}

We now discuss the distinguished limits of the stability problem. We present material from
\citet{ranganathan2001stabilization} and \citet{Govindarajan_2004IJMF} in pedagogical form. We will be working in the limit of large Reynolds and large Péclet numbers.
The critical layer, shown schematically in figure \ref{fig:schematic_overlap}, is a layer centred around the location $y_c$ defined by $U(y_c)=c$, i.e., it is a layer within which the phase speed of the disturbance is comparable to the base flow velocity.  
However high be the values of $Re$ and $Pe$, viscous effects enter at $O(1)$ in this layer. Note that the left hand side of equations \eqref{os_v2} to \eqref{species_v2} are zero at $y=y_c$, which is a direct indication that viscous terms must be dominant in this neighbourhood. 

The momentum and the species perturbations $v$ and $s$ display inner solutions in the critical layer. Both their critical layers are centred at $y_c$ but differ in their thickness, and we term these $\epsilon \ll 1$ for the momentum critical layer and $\delta \ll 1$ for the species critical layer. The wall-normal variable is redefined as $Y \equiv (y-y_c)/\epsilon$ and ${\cal Y}\equiv (y-y_c)/\delta$, which are $O(1)$ within the respective critical layers. 
The base velocity and concentration can respectively be expanded in the neighbourhood of $y_c$ as 
\begin{equation}
U(Y) = c + \epsilon Y U'_c + O(\epsilon^2), \  S({\cal Y}) = S_c + \delta {\cal Y} {S}'_c + O(\delta^2),
\label{Uexpanded}
\end{equation}
where the subscript $c$ refers to quantities evaluated at $y=y_c$. By construction, $Y$ and $\cal Y$ are $O(1)$ within the critical layer. As seen in figure \ref{fig:schematic_overlap}, we have a mixed concentration layer of thickness $q$ centred around $y=y_m$ within which the viscosity is stratified. When the stratified layer is thin, i.e., $q \ll 1$, we may define a new variable $Y_m = (y-y_m)/q$, and expand the concentration within the stratified layer as
\begin{equation}
S(Y_m) = S_m + q {Y_m} {S}'_m + O(q^2),
\label{Sexpanded}
\end{equation} 
the subscript $m$ refers to the value at $y=y_m$. Note that the concentration $S$ attains different constant values on either side of the stratified layer. We wish to consider one special case: the overall changes in the concentration are small, but occur across a short width $q$, such that $S_m' \sim O(1)$ and also that $\bar\nu_m' \sim O(1)$. Under overlap conditions we have $S_c' \sim S_m'$, and when there is no overlap, $S_c' \simeq 0$.

It is clear by now that we have three length scales in the problem: $q$, $\epsilon$ and $\delta$, and the relative sizes of these, as well as where they are centred, will determine the dominant terms in the equation. Since we know how the concentration, and therefore viscosity, is stratified in the base flow, we know the location $y_m$, where stratification is centred, and $q$, the width of the stratified layer. 
But $\epsilon$ and $\delta$ are unknown and need to be determined. In equation \eqref{os_v2} it is the magnitude of viscosity variation rather than the concentration variation that determines the relative weights of the different terms, we shall work with $\nu$  as well as $s$ for the perturbations.
The unknown functions $v$ and $\nu$ may be expanded within the critical layer as
\begin{eqnarray}
[v(Y), s({\cal Y}),\nu({\cal Y})] = [\psi_0(Y),\Xi_0({\cal Y}),\chi_0({\cal Y})] +  \nonumber \\ \left[\epsilon \psi_1(Y),\delta  \Xi_1({\cal Y}), \delta  \chi_1({\cal Y})\right]  +  O(\epsilon, \delta)^2.
\label{chiexpanded}
\end{eqnarray}
To proceed, we must recognise the difference between situations corresponding to the left and right panels of figure \ref{fig:schematic_overlap}. On the left, the mixed and critical layers are well-separated while on the right they overlap. For non-overlap conditions we may derive a dominant balance at the lowest order from the species equation \eqref{species_v2} to be
\begin{equation}
\frac{d^2\chi_0}{d{\cal Y}^2} -i U_c' {\cal Y}  \chi_0  = 0,
\label{dom1}
\end{equation}
which we obtained by defining
\begin{equation}
\delta \equiv (k_x Pe)^{-1/3}.
\label{delta1}
\end{equation}
The dominant balance in the critical layer is unaffected by the viscosity stratification, and so we do not expect any dramatic changes to the stability character as compared to an unstratified flow. 
On the other hand, under overlap conditions, the lowest-order critical layer balance would be
\begin{equation}
\frac{d\ \Xi_0^2}{d{\cal Y}^2} -i S_c'{\psi_0}  = 0, \quad  {\rm yielding} \quad \delta \equiv \left[k_x Pe \right]^{-1/2}.
\label{dom2}
\end{equation}
Clearly, the critical layer balance is completely dominated by the mean gradient of concentration. Equation \eqref{dom2} sends out an important message. Up to now, we have been talking of the `overlap mechanism' as if it is a event that occurs when, by some external effort, care is taken to place a viscosity stratified layer within touching distance of the critical layer of whatever perturbation may be dominant. This is true in some instances, but there are instances when the placement of the stratified layer anywhere will trigger  perturbations {\em within that layer}, whose phase speed will then naturally match the velocity at the critical layer. High P\'eclet number flows are an excellent example. From equation \eqref{dom2} we see that where $S_c'$ is significant, i.e., near $y_m$, we will have high magnitudes of the eigenfunctions $\Xi_0$ and $\phi_0$, these quantities will be zero far away and thus support large gradients in the vicinity of $y_m$. In other words, the kinetic energy production layer naturally overlaps with the stratified layer, and we have achieved $y_c \sim y_m$.

For the perturbation momentum, we may similarly construct an equation for the balance in the critical layer at the lowest order as
\begin{eqnarray}
Y U_c'\frac{d^2 \psi_0}{dY^2}  = - \frac{i \bar{\nu}_c}{ \epsilon^3 k_x Re}\frac{d^4 \psi_0}{dY^4}
-  \epsilon k_x Sc U_c' \frac{d\bar \nu}{dS}\bigg|_c \frac{d^2\chi_0}{d{\cal Y}^2},
\label{compositeos}
\end{eqnarray}
where we have replaced $\delta$ by its value obtained in equation \eqref{dom2}.
The reader may convince themselves with some algebra that all the terms from the stability equations that we have omitted will appear only at higher order for the special case under consideration. Under non-overlap conditions, or if the Schmidt number $Sc \sim 1$, the third term above is obviously of higher order than the other two terms, and the dominant balance reduces just the traditional critical layer equation  derived by Lin in the absence of viscosity variation. However, if the two layers overlap and the Schmidt number is large, fundamental changes to the stability properties of the flow occur. For large Schmidt number where $Sc^3 \gg Re$ (a condition very easily met), a balance between the two terms on the right hand side ensues, and yields the rather unconventional scaling of
\begin{equation}
    \epsilon \equiv [k_x^2 Pe]^{-1/4} \qquad {\rm or} \quad \epsilon = \delta^{1/2}.
\end{equation}
We leave the finding of scalings and dominant balances in other situations as an exercise to the particularly interested reader. Note in particular the case of an overall viscosity contrast as big as or bigger than the viscosity itself. Here many of the effects of viscosity variation will be far stronger than in the case we discussed above. A simple experiment with water and glycerol flowing through a microchannel is such an example. Now, depending on the sign of the viscosity gradients, we may have high levels of stabilisation or destabilisation, and even new low Reynolds number instabilities. Profiles of velocity and viscosity corresponding to the latter situation are sketched in figure \ref{fig:schematic_overlap}, with the lower viscosity fluid in the core and the higher viscosity one in the annular region in this example. 

We may similarly define a wall variable $Y_w=y/\epsilon_w$, with $\epsilon_w \ll 1$. For the minor viscosity contrast we have considered, or for a case where the stratified layer and the wall layer are well-separated, we will get the classical balance
\begin{equation}
\bar\nu|_w \frac{ d^4 \psi_{0w}}{dY_w^4}  + i\frac{d^2 \psi_{0w}}{dY_w^2} = 0, 
\label{eq:wall}
\end{equation}
where $\epsilon_w \equiv (k_x Re)^{-1/2}$. It is only when there are extremely strong viscosity gradients near the wall that the dominant balance is interfered with. As noted above, the wall layer serves to dissipate perturbations, and the sign of such a strong viscosity gradient will decide whether stratification serves to increase or decrease dissipation.
If we were studying this question decades ago, we would have been solving the critical layer and wall layer equations separately at increasing orders in accuracy and conducting a matching with the outer solutions to obtain the complete solutions. But this elaborate procedure is unnecessary, since it is computationally quite inexpensive to solve the complete stability equations themselves as eigenvalue problems. The idea behind deriving the critical layer and wall layer equations was, as mentioned above, to show the physics of how even small viscosity variations can make a big difference to the stability of the flow. Moreover, the ability to derive highly reduced dominant balance equations proves extremely useful in a range of singular situations, by which the lowest-order effects can be isolated.

\section{Instabilities due to viscosity stratification in simple shear flows}\label{sec:instab_due_to_visc_strat}

The stage is now set to selectively examine the literature on linear instabilities, specifically, those involving exponential growth of perturbations, in viscosity-stratified shear flows. Our objective is not to duplicate the coverage of comprehensive reviews such as \citet{Govindarajan_Sahu_2014ARFM}, but rather to distil key physical mechanisms most relevant to our pedagogical goals and to highlight open questions. There is now ample evidence that viscosity stratification can significantly enhance or suppress instabilities, and in some cases even introduce entirely new instability mechanisms. However, the effects are often highly non-intuitive: both the direction (enhancing or suppressing) and magnitude of the influence can be difficult to predict a priori. This inherent unpredictability makes the stability and transition behavior of such flows a persistent and fascinating research area. We focus on parallel, wall-bounded laminar flows of miscible fluids, which serve as test-beds for instability analysis. In a few instances, we will touch upon immiscible fluid stability, for comparison. At the same time, we emphasize that extensions to more complex, spatially varying geometries are essential. Even modest geometric modifications can profoundly alter the stability landscape, sometimes introducing new singular perturbation structures, and thus represent a rich avenue for future investigation. In fact we may have to re-examine the adjective `canonical' that we bestow to parallel flows.

\subsection{Couette, Poiseuille and pipe flows}
\label{couette_pipe}
Couette flow refers to the motion between two parallel plates moving along themselves at different velocities, while plane Poiseuille flow (whose laminar profile was given in equation \eqref{plane_poise}) describes the pressure-driven motion between two stationary plates. In both cases, the laminar base state consists of a unidirectional shear flow parallel to the bounding plates. These flows seem so similar but differ in their stability behaviour due to the differing curvature of the base velocity profile. As mentioned in section \ref{sec:ShearFlowInstability}, plane Couette flow at constant viscosity is stable, in the traditional sense, to linear perturbations, i.e., all perturbation eigenvalues decay at any finite Reynolds number \citep{Romanov_1973FAIA}, whereas plane Poiseuille flow goes linearly unstable beyond a critical Reynolds number ($Re_{cr}$) of $5772.2$ (based on the maximum velocity and the channel half-width). In pioneering work on two immiscible fluids flowing in a Couette set-up, \citet{Yih_1967JFM} showed that the picture is quite the opposite from unstratified flow: when the thinner of the two layers is the more viscous, this flow can support exponentially growing longwave perturbations at any Reynolds number. Even when the thicker layer is more viscous, instability can arise at large viscosity contrast. The driving mechanism is the different velocity gradients on either side of the interface, which are the flow's response to the viscosity contrast. [We mention in passing that studies on immiscible interfaces have continued over the decades to throw up new instabilities, e.g. the low Reynolds number one of \citet{Mohammadi_Smits_2017JFM}.] 

Immiscible fluid interfaces have the additional feature of surface tension, which suppresses shortwave perturbations, while leaving long waves practically unaffected. Besides the absence of this effect, miscibility does not change the qualitative behaviour in Couette flow: the critical Reynolds number in the presence of thin mixed layers is fairly low and reduces as the Schmidt number increases. If we have three layers instead of two, the flow is unstable \citep{Jose_2024arXiv} when the outer layers are more viscous than the central layer, and modally stable otherwise. A viscosity increase towards the wall promotes an inflexional character in the mean velocity profile, and
produces nonlinear evolution via an `inviscid' instability \citep{Tendero_Ventanas_etal_2024arXiv}. 
We have seen in section \ref{sec:singular} that sharp viscosity gradients associated with two-fluid flow need special placement near the critical layer to change stability behaviour, whereas gentler viscosity gradients present everywhere, typically produced by temperature variations, always satisfy overlap conditions though the effect on the singularity is quantitatively weaker. Stratification in the latter scenario might just significantly modulate existing instabilities rather than create new ones, as seen in the early study of \citet{wall1996linear} on temperature-dependent viscosity. Viscosity gradients can stabilize or destabilize standard modes of instability (such as the Tollmien-Schlichting), depending on the sign of the gradient (wall heating/ cooling of gases and liquids will likely have opposite effects) and other parameters. Increase in the Schmidt or Prandtl number causes instability at lower Reynolds number, see e.g. \citet{Govindarajan_2004IJMF}.  

Pipe flow is important in industrial settings like petroleum extraction from natural reservoirs, and is interesting because, unlike in plane Poiseuille and Couette flow, Squire's theorem does not apply now. In fact the most unstable mode is often a helical mode. While the influence of viscosity stratification on linear instabilities has been well studied in this context, and reviews such as \citet{joseph1997core} have been available for a long time, the transition to turbulence of stratified flows remains an open question. But there are some things not yet solved in linear stability as well, e.g., a prediction of which kind of instability will be manifested under a given set of conditions. The limit of zero miscibility could be qualitatively different from finite but small miscibility. In immiscible core-annular flow through a pipe, \citet{usha2019interfacial} found answers which qualitatively differ from the miscible study of \citet{selvam2007stability}. In the immiscible case, when the annular fluid is the more viscous, the axisymmetric mode of instability is dominant, and when the core fluid is the more viscous, the corkscrew, or helical, mode is the most unstable. The dominance of the two modes is exchanged in miscible flow, underlining the fact that stability behaviour is not easy to predict.

Besides the convective instabilities we have discussed so far, where growing disturbances convect downstream from their source, some shear flows can display absolute instability, where perturbations continue to grow where they originate, while also infecting the upstream and downstream. Absolute instabilities are considered more conducive to attaining a turbulent state. For a detailed discussion in the context of constant-viscosity parallel shear flows, see \citet{drazin2002introduction,Drazin_Reid_2004book}.  It is known that viscosity-stratified pipe flow can be absolutely unstable for certain parameter ranges \citep{martin2009convective, selvam2009convective}, and therefore we may surmise that the route to turbulence will differ mechanistically from the constant viscosity case: where the pipe is always linearly stable and transition is triggered by algebraically growing modes. If such were to turn out to be the case, that would be new physics indeed. 

Shear flows of supercritical fluids have received some attention, but given their relevance to biochemical and energy engineering processes, warrant deeper investigation. Strong viscosity gradients are to be found in such flows, which can drive hydrodynamic instabilities. For instance, in Couette flow, \citet{bugeat2024instability} demonstrated that such gradients can generate vorticity waves through a delicate interplay between shear and vorticity. A particularly notable finding is that the mere presence of a local minimum in kinematic viscosity can render an otherwise stable stratified supercritical flow unstable. This mechanism closely resembles the instability reported by \citet{ren2019boundary} in the boundary layer over a flat plate immersed in supercritical $\text{CO}_2$.

While the preceding discussion focused on high-Reynolds-number flows in channels and pipes, these geometries also underpin a class of applications at much smaller length scales—namely, microfluidics. Viscosity-stratification-driven instabilities play a pivotal role in enhancing mixing between species, a key challenge due to the inherently laminar nature of microflows. Owing to the shared geometric configurations, we take a brief detour here from our scale-wise progression, from small particle scale dynamics discussed in section~\ref{sec:ParticleDynamics} to Earth-scale flows in section~\ref{sec:EarthFlows}, to focus on microfluidic channels and pipes for the remainder of this subsection. As we discuss at the end of this subsection, such small-scale flows also serve as valuable test-beds for understanding and designing larger-scale systems.

\subsubsection*{Microfluidic channels and pipes}

Microfluidic platforms, involving length scales of microns to millimetres, find applications in a wide range of fields \citep{stone2004engineering, squires2005microfluidics, whitesides2006origins, anna2016droplets, nunes2022introduction}. Controlled mixing in microfluidics devices is important in drug synthesis in microchannel reactors \citep{liu2021microfluidics}, in polymerase-chain-reactions to mix reagents \citep{zhang2006pcr}, in synthesis of nanoparticles using controlled diffusion \citep{wagner2004generation, marre2010synthesis, shang2017emerging}, in cellular assays \citep{el2006cells, huh2010reconstituting, sackmann2014present, zhang2018advances}, and in the food industry for toxin detection  \citep{neethirajan2011microfluidics} among other usages \citep{fani2013investigation, galletti2017unsteady, chan2019coupling, mariotti2021effect}. The Reynolds numbers range from $O(10^{-6})$ to $O(10)$, while Schmidt numbers in stratified fluids are typically of $O(10^3)$. Residence times within the flow geometries are small, but large enough for the continuum hypothesis to be valid. Notably, viscosity stratification effects are important at any Reynolds number, because they either contribute to singular behaviour when $Pe \gg 1$ which is frequently the case in microfluidic devices, or are $O(1)$ anyway when $Re \le O(1)$. And overlap effects could be important whenever P\'eclet numbers are high.

\begin{figure*}
\centering
\includegraphics[width=0.9\linewidth]{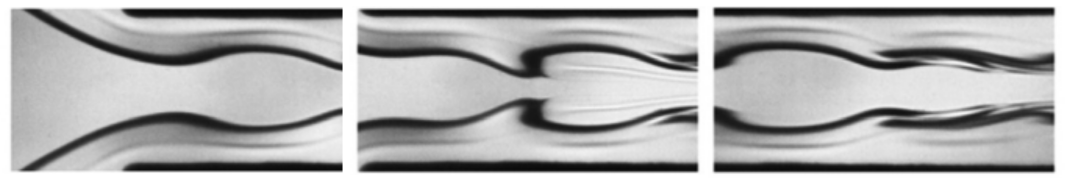}
\caption{Three snapshots of time evolution (progressively from left to right in $\mathcal{O}$(ms)) of ligament and vortex formation with two fluids in a microchannel with different viscosities. The flow is from left to right in each panel. The ratio of the viscosity of outer fluid to inner fluid is approximately 1000. Reprinted with permission from \citet{Hu_Cubaud_2016PRF}.}
\label{fig:cubaud_prf}
\end{figure*}
In situations where mixing is desired in a microchannel, diffusion alone is not sufficient when residence times are small. As a thought experiment, we consider a microchannel of $1$~mm in length, of square cross-section of $100 \mu$~m side, through which $2\mu l/$~s of pure water and sugar-water are introduced to flow side-by-side. The reader can calculate easily that diffusion times $(l^2/\kappa)$, where $\kappa$ is the diffusivity of sugar in solution, are far longer than residence times $(l/U)$. Therefore, a tracer will effectively be transported along the streamline it is originally placed upon. Other ways of mixing between reagents are needed, and instabilities can come in very handy in increasing passive mixing by disturbing the steady laminar flow. Viscosity contrast between reagents can generate such instabilities and induce mixing among the constituents. Most often in microchannels, the critical layer of the dominant perturbation naturally coincides with the stratified layer, producing an overlap effect, which can be exploited by precise control of relative flow rates and viscosity ratios \citep{lam2009micromixer, cubaud2014regimes, anna2016droplets, Cubaud_2020PRL}. In other words, a dominant balance similar to equation \eqref{dom2} ensures that production of perturbation energy is localised at the stratified layer. Sometimes, though, we see that waves produced by the instability undergo nonlinear saturation, and while we obtain a wavy interface between the two fluids, there is no mixing (see, e.g., \citet{carbonaro2025emergence}). Some wavy patterns give rise to partial or complete mixing, and future studies into the nonlinear regime with viscosity variations are warranted in microflows.

The experiment of \citet{Hu_Cubaud_2016PRF} consists of a low-viscosity fluid flowing in the core of a microchannel, surrounded by a miscible higher-viscosity fluid in the annular region (figure \ref{fig:cubaud_prf}). They explored a wide range of viscosity ratios: from $50$ to $5000$, across different Schmidt numbers. At low flow rates, diffusion dominates and leads to complete mixing, effectively eliminating the mixing layer. As the flow rate increases, the system transitions into a stable regime with a marked mixed layer, which remains so thin as to appear as a sharp interface, and eventually to an unstable regime at the highest flow rates. In this unstable regime, interfacial waves break up into ligaments and vortical structures, facilitating controlled mixing of the two fluids (figure~\ref{fig:cubaud_prf}).  
In a related setup investigated by \citet{Hu_Cubaud_2018PRL}, a lower-viscosity fluid overlays a higher-viscosity fluid in a microchannel. The qualitative behavior is similar: above a critical flow rate ratio, long-wavelength viscous waves form along the interface. These waves emit viscous ligaments or threads which provide high interfacial area, and this facilitates the entrainment of one fluid into the other. At larger Reynolds numbers, inertial-viscous waves appear, generating sharper wave crests and more vigorous ligaments that form recirculating vortices — providing more efficient mixing. \citet{Cubaud_2020PRL} developed a model to describe the diffusive behavior of miscible and partially miscible fluid pairs of differing viscosities in microchannels. Based on the formation and evolution of viscous threads as the two fluids are co-injected, they proposed a method to estimate the effective diffusivity. This diffusivity depends on a single fitting parameter, which in turn depends upon the viscosity ratio.

Viscosity stratification can also join hands with external forcing to significantly enhance mixing. In the two-layer miscible microflow configuration of \citet{dutta2019electric}, stratification alone is known to generate specific instability modes, and the addition of an electric field can pump up both the growth rates of these modes and the overall mixing levels. Other low-Reynolds-number instabilities may be generated by electrohydrodynamic forcing, for instance, due to contrasts in electrical properties of the reactants \citep{el2003electro, chang2007electrokinetic}, or by the use of deformable (soft) microchannel walls \citep{verma2013multifold}. Rotation provides another control parameter for mixing. Depending on the configuration, it can either suppress or enhance instability. The outcome is strongly influenced by the mixed layer thickness: spanwise rotation stabilizes flows with thinner mixed layers but destabilizes those with thicker ones \citep{sengupta2019coriolis}. This behavior is opposite to the non-rotating case, where thinner layers are more unstable \citep{Govindarajan_2004IJMF}. Interestingly, even in miscible fluid systems, transient interfacial tension can arise at early times \citep{carbonaro2025emergence}, which may lead to capillary instabilities and perhaps contribute to mixing.

Microfluidic flows can also serve as small-scale benchmarks for understanding fundamental fluid physics and informing the design of larger-scale systems. Conversely, experiments on larger scales can inform microfluidics, and we shall discuss an example in section \ref{sec:displacement}.

\subsection{Jets, mixing layers and films}
\label{sec:jets_films}

Jets and mixing layers are natural environments for the emergence of the overlap mode of instability. When one fluid jets into, or flows past, another, a mixed layer forms at the interface -- this layer is both velocity-sheared and viscosity-stratified. Often, the velocity profile supports a point of inflexion close to or within the stratified layer, and the critical layer too is located  within the same region.
In the numerical and experimental work of \citet{yang2024absolute} and \citet{Srinivasan_etal_2024PRF}, respectively, a jet of viscosity $\mu_1$ intrudes into a surrounding fluid of viscosity $\mu_2$, with the two fluids being miscible and of equal density. When $\mu_2/\mu_1 \gg 1$, the jet is absolutely unstable. At high viscosity contrast, helical instabilities are favoured \citep{li2009axisymmetric} over the axisymmetric modes which dominate at lower ambient-to-jet viscosity ratios. Viscosity stratification thus profoundly changes stability behaviour. Such flows can occur in hydrothermal vents, where they are accompanied by density gradients (Vinod Srinivasan, personal communication). A related phenomenon is seen in miscible Kelvin–Helmholtz layers: the instability is markedly enhanced when the slower-moving fluid is more viscous  \citep{Taguelmimt_etal_2016FTC}.  

Films, like jets, display instability at a rather low Reynolds number, where critical layers are not thin. Single or multilayer film deposition is of engineering importance in electronics and pharmaceuticals. A natural route for viscosity variation to arise across a film is the preferential evaporation of one component. Evaporation-induced increases in near-surface viscosity can significantly alter film stability. \citet{hong2024solutal} found that such stratification stabilizes a film over a substrate by suppressing Marangoni convection and rendering the dynamics more diffusive. A biological example of similar relevance is the tear film — a thin layer of fluid covering the cornea in our eyes — comprising oil, water, and a mucin-rich mucus layer. The variation of mucin concentration forms a layer of varying concentration that varies the viscosity, which in turn affects the stability and breaking properties of this protective layer. A model is developed for this problem by \citet{dey2019model}, and the tear film breakup time is found to be in better agreement with experimental results when continuous variations of viscosity across the film were prescribed, as compared to two-layer models of the tear film. With this model, it is seen that putting more viscous material near the bottom of the layer, i.e. the portion in contact with the eye, stabilises the film. 

Surface tension on the film surface typically suppresses shortwave instability modes and does not affect longwave modes, and this is the case with stratified flows as well \citep{choudhury2020enhanced}. But in the case of a film with a sharp viscosity gradient in its interior, we can have the two interfacial waves interacting with each other, with consequences that are unknown apriori. The reader is referred to studies of wave interaction, such as available in \citet{craik1988wave}. 

A film of liquid flowing down a solid plane inclined to the horizontal by an angle $\gamma$, is most unstable to long waves peaking at the surface, beyond a critical Reynolds number of $Re_{cr}=(5/4) \cot \gamma$ \citep{benjamin1957wave}, where the Reynolds number is based on the film thickness and surface velocity.  This means that a film falling down a vertical plate is always unstable. In the study of \citet{samanta2022role}, the effect of odd viscosity coefficients on the  stability of this system is examined. With odd viscosity, $Re_{cr}$ is non-zero for a vertically falling film, and the surface mode of instability is suppressed at any angle of inclination compared to the scalar viscosity case. However, with increase in $Re$ and at smaller inclination angles, the shear mode is dominant, and it is also stabilised analogous to the surface mode, in the presence of odd viscosity. With an increase in the odd to even viscosity ratio, the temporal growth rate of the shear mode decreases suggesting stability. The spatial and temporal variations of the stress tensor due to odd viscosity effects would bring in myriad possibilities for the dynamics, and would make for other future areas of study in these systems and others as well.

\subsection{Stability of particulate shear flows}\label{sec:ParticleShear}

We dealt with some aspects of particulate flows in section \ref{sec:ParticleDynamics}, and here we will briefly discuss large scale particulate flows with a mean shear and their stability characteristics. We advocate that this aspect receive attention in future. From volcanic eruptions to sediment transport in rivers and the ocean, large-scale particulate flows are ubiquitous. The local effective viscosity depends on the particle concentration, sizes and distribution. In the dilute limit, and for extremely small particles, we encountered the simplest form simplest form of this dependence in equation~\eqref{eq:SuspRheology}. In realistic geophysical and environmental settings, the particle concentration is usually heterogeneous, making the effective viscosity a function of space and time - even in the simple model of equation \eqref{eq:SuspRheology}.

\citet{harang2014kelvin} study instabilities and mixing in a viscosity and density-stratified flow with a free interface, that is representative of mud flow at the bottom of estuaries. They find some evidence for an overlap mode of instability. Notably, below a certain Reynolds number, viscosity contrast decreases the stability of  system. In a related study, \citet{ankush2023mixed} performed a linear stability analysis of vertical mixed convection in which viscosity depends on both temperature and solute concentration. They showed that the system can exhibit either stabilizing or destabilizing behavior, depending on the strength of the viscosity variations and the relative orientations of the temperature and solute gradients, with the Schmidt number playing a key role in modulating these effects. 

\begin{figure*}
\centering
\includegraphics[width=\linewidth]{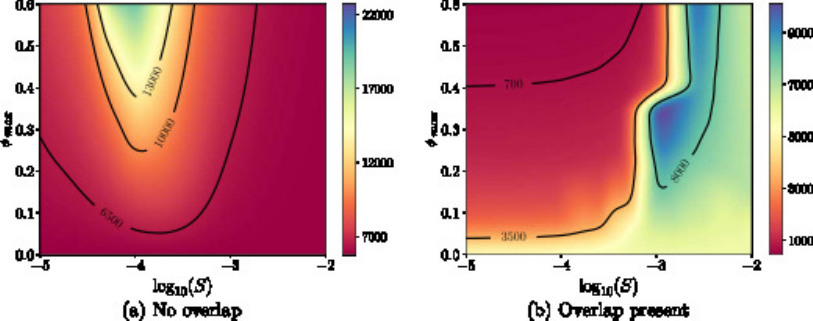}
\caption{Linear stability of a dusty channel flow. The critical Reynolds number for stability is shown in colour, as it varies with the Stokes number $St$ and with maximum particle concentration $\phi_{max}$. The particles are distributed non-uniformly in the flow and the particle layer overlaps with the critical layer in (b) whereas there is no such overlap in (a), showing significant alteration in stability regimes. A similar figure appears as Figure 6 of \citet{kumar2024mechanism}.} 
\label{fig_anup}
\end{figure*}

Even without an explicit change in the effective viscosity of the fluid, suspended inertial particles interact with the fluid in ways that can destabilize the flow. The particles experience a drag from the fluid and in turn exert reactive forces on it, giving rise to additional stresses. These particle-induced stresses — commonly represented by the particle stresslet \citep{batchelor1970stress} — can be interpreted as introducing a pseudo-viscosity. As a result, the overlap mechanism comes into operation in a particulate flow, and when the particle-laden layer overlaps with the critical layer, generates low Reynolds number instabilities \citep{kumar2024mechanism} even at constant effective viscosity. When the particles are in sufficient volume fraction to affect the effective viscosity as per equation \eqref{eq:SuspRheology}, it can further alter the growth rates of perturbations. A comprehensive understanding of how these effects interact during the transition to turbulence remains an open research question.
The linear stability mechanism in a dusty channel flow was recently demonstrated by \citet{kumar2024mechanism}, who considered particles concentrated into two thin layers symmetrically placed about the channel centerline. The critical Reynolds number for instability of the laminar flow, plotted in figure~\ref{fig_anup} as a function of the Stokes number $St$ and the maximum particle concentration $\phi_{max}$, reveals marked differences between non-overlap (left) and overlap (right) conditions. These findings underscore the destabilizing role of the overlap mechanism and its modulation by particle inertia and concentration. 

\subsection{Displacement flows} 
\label{sec:displacement}

When a lower viscosity fluid penetrates the higher viscosity one in a confined space, the interface becomes unstable at any flow speed and creates finger-like structures. This is the famous Saffman--Taylor \citep{saffman1958penetration} instability, driven entirely by viscosity contrast  [see also \citet{hill1952channeling,chuoke1959instability}, and the review of \citet{martinez2011transportation}]. Since this is a well-documented topic, we briefly summarise more recent developments. We restrict ourselves to the displacement of a miscible fluid by another in simple geometries with rigid boundaries.

Controlling the growth rate and dominant wavenumber of these fingers is of practical importance across applications — either to enhance mixing (e.g., in chemical reactors \cite{nijjer2018dynamics}) or suppress it (e.g., in oil recovery). The outcome depends sensitively on the viscosity ratio of the two fluids \citep{suekane2017three}. \citet{etrati2018two} and \citet{etrati2018viscosity} are representative theoretical and experimental studies respectively on displacement flow, that demonstrated that the interplay of density and viscosity contrasts, across a wide range of Reynolds and Froude numbers, where the Froude number is the ratio of the flow velocity to the gravity wave propagation speed, can be exploited to either enhance or suppress interfacial instabilities between displaced and displacing fluids. Works of \citet{taghavi2017buoyant, taghavi2018two} have developed semi-analytical models of miscible displacement flows where the viscosity ratio is an important parameter. More recently, through experiments on non-Newtonian fluids displacing a Newtonian fluid in a pipe, \citep{faramarzi2024buoyant} show that the difference in viscosity plays an important role in frontal dynamics and mixing.  

Chemical reactions where the product is of a different viscosity can be exploited in displacement flows. One example addresses the challenge of displacing a high-viscosity fluid from a pipe — a problem of considerable importance in oil and gas extraction from underground, where viscous drilling mud must be flushed out by a spacer fluid. If the spacer fluid is too viscous, excessive energy is required; if it is too dilute, it fails to efficiently displace the mud. \citet{burghelea2007novel} propose an elegant solution: a low-inertia, low-viscosity spacer fluid that undergoes an acid–base reaction upon contact with the high-viscosity fluid. The product of the reaction is a stiff gel, which is of higher viscosity than both reactants, and possesses non-Newtonian properties as well. In their experiments conducted in a half-centimetre pipe, \citet{burghelea2007novel} observe instability behaviour that can be attributed directly to the high levels of viscosity stratification provided at the miscible interface (between the reactants) by the product of the reaction, and in particular to a novel mixing mechanism driven by a non-monotonic viscosity profile. They gave evidence for this connection by linear stability analysis and by performing control experiments where the instability did not occur. As a result of their self-sustaining instability, the incoming spacer fluid engulfs the displaced viscous fluid, and the combined mixture then propagates rapidly through the channel. If scaled up and found to work, this would constitute a novel approach to efficiently oil recovery, whereas the standard approach is to introduce less viscous fluid along the walls and thence suppress instability. This mechanism also has broader implications beyond oil extraction. Another example of a chemistry-driven instability in displacement flows is in the study of \citet{maharana2021reaction}. Kelvin–Helmholtz-type billows can be supported at the interface of chemically reactive species, producing viscosity gradients through the formation of a product of a different viscosity. Both these studies are clearly encountering the overlap mode of instability, since perturbation kinetic energy is produced right at the interface where the product of the reaction introduces viscosity changes.

\subsection{Non-normal dynamics modified by viscosity stratification}\label{sec:NonNormalVVF}

As we discussed above, the transition to turbulence in shear flows is an open question, and just by modal stability analysis we fail to predict the transition Reynolds number, and miss the entire process leading up to turbulence. In such nonnormal systems, transient growth mechanisms associated with nonmodal algebraic growth rates can dominate the process of departure from the laminar state, even in the absence of exponentially growing modes as in pipe and Couette flows. Viscosity stratification alters the very nature of non-normality and provides alternative non-normal routes to disturbance growth. 

To examine non-normal effects, it is more convenient to employ the framework of an initial value problem; the key difference from the procedure for modal analysis is that the time dependence of the perturbation is not assumed a priori \citep{Schmid_Henningson_2001book,Schmid_2007ARFM}. As a result, the linearised equations, analogous to those of equations \eqref{os_v2} to \eqref{species_v2}, become: 
\begin{align}\label{nonnormal}
&\frac{\textrm{d}}{\textrm{d}t}\begin{bmatrix} v \\ \eta \\ s \end{bmatrix} = 
\mathcal{L}
\begin{bmatrix} v \\ \eta \\ s \end{bmatrix}= 
\begin{bmatrix}
{\mathbf L}_{OS} & 0 & {\mathbf L}_{vs} \\
{\mathbf L}_{\eta v}	 & {\mathbf L}_{SQ} & {\mathbf L}_{\eta s} \\
{\mathbf L}_{sv} & 0 & {\mathbf L}_{ss}
\end{bmatrix}
\begin{bmatrix} v \\ \eta \\ s \end{bmatrix}, \\
&\textrm{where} \nonumber\\ 
&{\mathbf L}_{OS} \equiv
D_{2-}^{-1} \bigg\{ik_x[U'' - U \mathcal{D}_{2-}]   + \frac{1}{Re}[\bar{\nu} \mathcal{D}_{2-}^2 \nonumber \\
&\quad\quad\quad+2\bar{\nu}'D\mathcal{D}_{2-} + \bar{\nu}''\mathcal{D}_{2+}] \bigg\}, \label{los} \\
&{\mathbf L}_{vs} \equiv	-\frac{ik_x D_{2-}^{-1}}{Re}	\left\{\left[U'\mathcal{D}_{2+} + 2U'' D +  U'''\right] \frac{d\bar\nu}{dS}\right\}, \label{lv_s} \\
&{\mathbf L}_{\eta v}	= -ik_z U'  \\
&{\mathbf L}_{SQ} =	-ik_x U + \frac{1}{Re}\left[\bar{\nu}'D + \bar{\nu}\mathcal{D}_{2-}\right], \\
&{\mathbf L}_{\eta s} = \frac{ik_z}{Re} \left[U''  +  U'D\right]\frac{d\bar\nu}{dS}, \\
&{\mathbf L}_{sv} = -S', \textrm{ and } {\mathbf L}_{ss} =	- ik_x U + \frac{1}{Pe} \mathcal{D}_{2-}.
\end{align}
In the above, ${\mathbf L}_{OS}$ and ${\mathbf L}_{SQ}$ respectively are the well-known Orr--Sommerfeld and Squire operators, modified here to account for viscosity variations. The equations above are supplemented by the initial conditions for the normal velocity $v$, normal vorticity $\eta$ and scalar $s$. 
The off-diagonal terms in the operator $\mathcal{L}$ are an indication of non-normality. For constant viscosity, the third row and third column would drop out. The truncated matrix is already non-normal, so significant algebraic growth is possible at Reynolds numbers below that at which exponential growth of instabilities is possible. In viscosity-varying flows, we see several extra contributions to non-normality, and thus scope for new kinds of algebraic growth, which can take the flow to turbulence, or other hitherto unknown states. This important aspect is largely unexplored, and we present it as one deserving immediate attention, where algebraic growth provides a competing mechanism to the overlap mechanism we have discussed at length in previous sections. 

We briefly review the well-known nonmodal, and hence algebraic, disturbance growth mechanisms in unstratified flows before discussing variable viscosity flows.
A common measure to estimate algebraic growth is the maximum possible gain over all initial conditions,
\begin{equation} G_{max}(t) = \max_{\forall \mathbf{u}(\mathbf{x},0)}\frac{||\mathbf{u}(\mathbf{x},t)||^2}{||\mathbf{u}(\mathbf{x},0)||^2}, \label{gmax} \end{equation} 
where $||\mathbf{u}(\mathbf{x},t)||^2$  is twice the perturbation kinetic energy at time $t$, and $\mathbf{u}(\mathbf{x},0)$ is the initial perturbation, subject to the constraint $\mathbf{u}(\mathbf{x},0) \neq 0$. The maximum possible gain in perturbation energy offers insight about the amplitude of the initial perturbation required to bring into play the nonlinear terms of the governing equations. Depending on the objective of the study, other options for the quantity we wish to optimise may be more suitable. For instance, a number of studies examine non-modal (and nonlinear) means to enhance mixing in the flow \citep{vermach2018optimal}.  

In the linearised setting, the optimal initial perturbation $\mathbf{u}(\mathbf{x},0)_{opt}$, which gives the largest energy growth, may be found by performing a singular value decomposition of the linearised operator \citep{Schmid_Henningson_2001book}. 
An alternative procedure for finding the optimal perturbation is direct-adjoint-looping \citep{cherubini2010rapid, pringle2010using}, an iterative method that involves the forward evolution of an initial condition followed by a backward evolution of the adjoint variable until convergence is achieved to a satisfactory level. At the end of every iteration, the guess for the optimal initial condition is updated by adopting a steepest descent algorithm, until the conditions of optimality are satisfied. 
Apart from doing so for the linearised system, we may also find the best growth afforded by the complete nonlinear system by applying adjoint-looping to it. The latter optimals depend on the initial amplitude specified, and could be quite different in structure from the linear optimals. 

In constant viscosity flows, the highest $G_{\max}$ from the linearised equations is frequently shown by linear combinations of streamwise independent perturbations, i.e., perturbations for which $k_x=0$. This growth is manifested in the form of long streamwise vortices interspersed by long streaks, with the latter moving at a different velocity from their surroundings. Streaks result from streamwise independent vortical perturbations displacing fluid elements between regions of high and low streamwise velocities resulting in velocity defects. They undergo secondary instabilities and, by the famous `self-sustaining process' of Fabian Waleffe \citep{Waleffe_1997PF}, the laminar flow is permanently abandoned for an unsteady and richer flow state in which streaks are repeatedly generated. The lift-up mechanism \citep{Landahl_1980JFM,Brandt_2014EJMBF}, which is central to most discussions on sub-critical transition in wall-bounded shear flows, continues to be the principal driver for algebraic growth during the linear stages. These streaks too undergo amplification through the lift-up mechanism that is followed by a nonlinear breakdown that regenerates vortical perturbations. This mechanism also plays a role in the nonlinear self-sustenance of turbulence \citep{Hamilton_etal_1995JFM,Jimenez_Pinelli_1999JFM,Schoppa_Hussain_2002JFM}.

Another nonmodal linear growth mechanism is the two-dimensional Orr mechanism, known for over a century \citep{Orr_1907PRIAA_alt, Roy_Govindarajan_2010BookChap}. Here, alternating positive and negative velocity structures arranged both along the streamwise and spanwise directions, initially tilted against the base shear, are reoriented by the background shear near the walls, resulting in transient energy amplification. The Orr mechanism is less talked about, since in parallel shear flows it is weak and short-lived compared to the lift-mechanism. However, in non-parallel flows the Orr mechanism can become extremely strong \cite{jotkar2019}, so it deserves further attention. 

How does this broad picture change upon introducing viscosity stratification? 
As a zeroeth-order effect of non-constant viscosity, there is a modulation of the mean flow. When the perturbations are streamwise independent, the extent to which they get amplified by the lift-up effect depends on the mean velocity gradient. This aspect of the self-sustaining process gets most prominently affected by viscosity variation. More intricate effects emerge as one accounts for the additional terms in the equations, including coupling terms with the species equations that bring in new nonlinearity. In an unstratified channel flow, symmetry ensures that the lift-up mechanism is equally prevalent at both walls. However, in an asymmetrically heated channel flow of a liquid with temperature-dependent viscosity, \citet{Thakur_etal_2021JFM} found that the lift-up mechanism is initially enhanced near the hot wall, where viscosity decreases toward the wall and velocity gradients are larger, and suppressed near the cold wall. Over time, the mixing of the temperature and hence viscosity, induced by velocity inflexions, weakens the viscosity gradient near the hot wall and strengthens it near the cold wall. Further details are provided as a schematic in \citet{Thakur_etal_2021JFM}. Consequently, although lift-up begins more strongly near the hot wall in liquid flows, it is sustained for longer near the cold wall. Similarly, for the Orr mechanism, the optimal initial perturbations are localized near the hot wall. This picture reverses in gaseous flows, where viscosity decreases with temperature: hence, the roles of the cold and hot wall are exchanged.

\begin{figure*}
\centering
\includegraphics[width=1.0\linewidth]{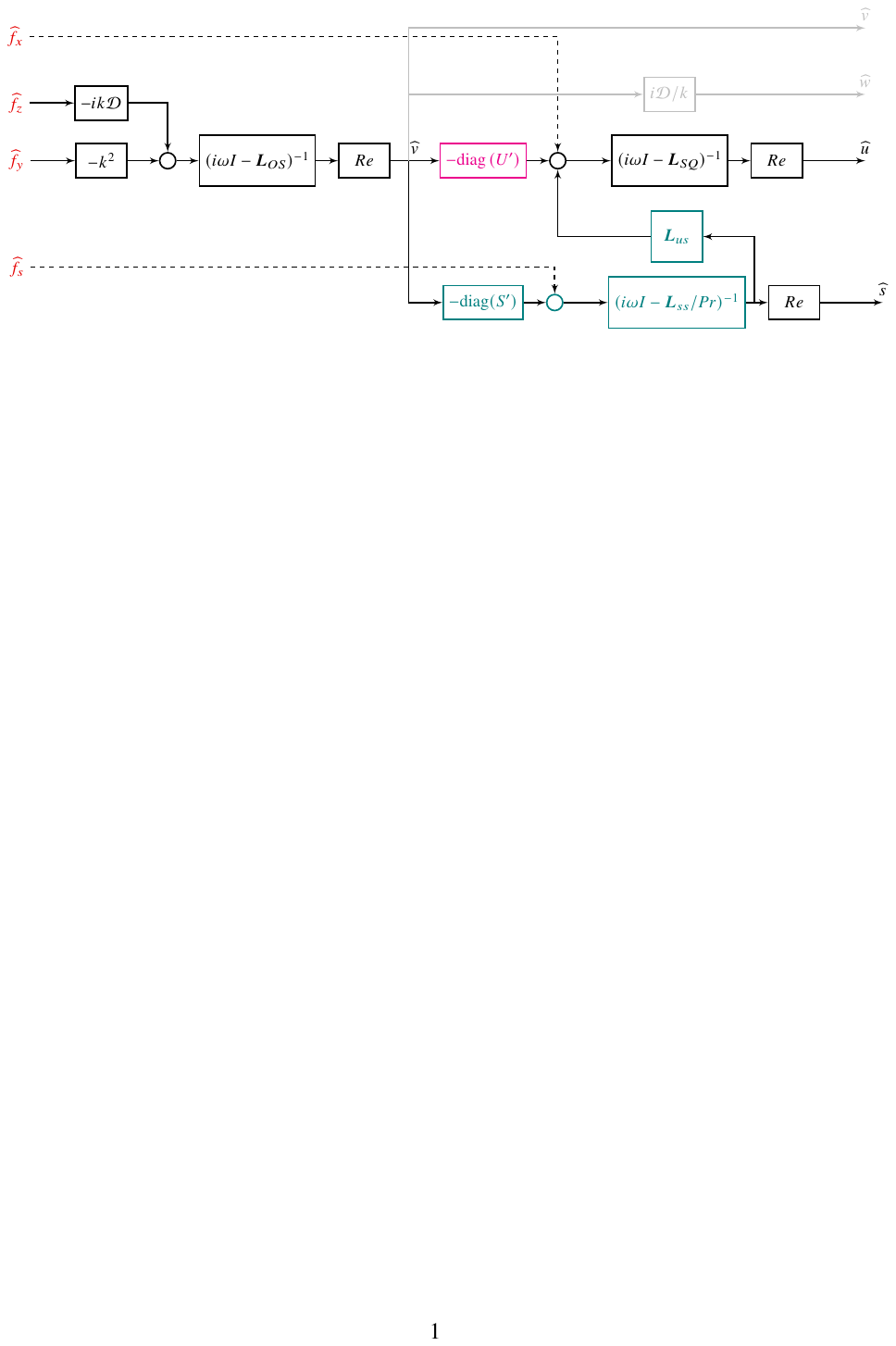}
\vspace{-6.2in}
\caption{Block diagram representing the amplification of the streak ($k_x = 0$) modes as a response to inputs $\widehat{f_x}$, $\widehat{f_y}$, $\widehat{f_z}$, and $\widehat{f_S}$. Diag(${U}'$) and diag($S'$) are diagonal matrices with elements $dU/dy$ and $dS/dy$. An amplification proportional to $Re^2$ is provided by $\widehat{f_y}$ and $\widehat{f_z}$ (solid lines), while the amplification is $O(Re)$ lower in the other two (dashed lines). Figure adapted from \citet{anagha_2025}.}
\label{fig:anagha}
\end{figure*}
Due to lift-up, streaks are ubiquitous in shear flows, including in fully developed turbulence. To get an idea of how streaks themselves are affected by viscosity variation, \citet{anagha_2025} study the response of the system to forcing. All nonlinear effects are clubbed into the forcing, so we may use the structure of equation \eqref{nonnormal} and set $k_x=0$ to select for streaks alone. Once this is done the terms in equation \eqref{nonnormal} simplify considerably. Moreover we now have $\eta = -i k_z u$, and the second equation may be written for $u$, with operators $\mathbf{L}_{\eta v}$ and $\mathbf{L}_{\eta s}$ replaced by proportional operators $\mathbf{L}_{u v}$ and $\mathbf{L}_{u s}$. We replace the time derivatives by their modal form, and add a forcing term to the equations. The block diagram representing the linearised process is shown in primitive variables in figure \ref{fig:anagha}. Since streaks are defined as regions of slower or faster streamwise velocity than the surroundings, the magnitude of $\hat u$ is direct a measure of the strength of streaks formed, and we focus on the routes that generate this quantity. The unimportant forcing directions (which scale by a single power of $Re$) are shown by dashed lines. We see two primary routes to streak formation (which scale by $Re^2$), where the black portions in the figure are common to both routes. Magenta represents the traditional route, modified by viscosity stratification since the profile $U'$ is now different. The other dominant route to streak formation (one via $\text{diag}(S')$) is completely new, and due to viscosity stratification. Interestingly, in channel flow, the two routes cooperate in one half to produce stronger streaks and compete in the other half to weaken them. Besides, the two routes behave differently in Couette and channel flow.

In shear flows involving a single fluid, we saw that two-dimensional disturbances show only modest levels of transient amplification by the Orr mechanism, and the entire process is short-lived. Beyond the initial consideration of the Orr mechanism by \citet{Thakur_etal_2021JFM,Jose_2024IJMF} showed that this century-old inviscid mechanism needs to be brought out from cold storage and used afresh in the case of viscosity-varying flows. Merely because the mean flow changes (its gradient $U'$ in particular), the Orr mechanism produces large algebraic perturbation growth, in situations where no exponential growth is possible. \citet{Jose_2024IJMF, Jose_2024arXiv} further show that viscosity stratified plane shear flows with two and three layers can support nonmodal perturbations that display significant amount of transient growth over a much longer period of time; note that the smooth three-layer flow is modally stable in the inviscid setting whenever the middle layer is not the least viscous of the three. Despite the relatively modest energy amplification, the Orr mechanism has been shown to be important even in self-sustaining processes in nonlinear turbulent shear flows. In this regard, it might not be unreasonable to anticipate these nonmodal processes to be relevant in the nonlinear dynamics of multi-layer plane shear flows.

As we have seen, viscosity stratification in laminar flows is now a well-researched area. And studies of fully-developed viscosity-stratified turbulence are increasingly appearing. But the route which takes a flow from laminar to turbulence, except for the first step, namely the departure from steady laminar flow into various unsteady states, is a wide open and most interesting question. In wall-bounded shear flow, we discussed two mechanisms which are at play: exponential and algebraic growth. In constant viscosity flow through straight pipes and channels, the latter wins, since $R_{AG} \ll R_{crit}$. But in viscosity-stratified flows exponential instability can appear at very low Reynolds number, and we could have $R_{crit} \sim R_{AG}$. In this case, either or a combination could be operative in driving the flow to turbulence. When both are in operation, exponential growth could overwhelm algebraic growth at long times. Further, as we increase the Reynolds number, how the flow will attain turbulence needs to be addressed. Moreover, at a particular fixed Reynolds number when the flow is neither steady laminar nor turbulent, the coherent structures formed can be different from those in constant viscosity flow, since the top-down symmetry is broken, and transitional states can display very different mixing and heat transfer characteristics from constant viscosity flow.

An important development over the past two decades in shear flow transition picture are edge boundaries and edge states (e.g., see \citet{Skufca_etal_2006PRL,Duguet_etal_2008PF,vollmer2009basin, schneider2010localized, Cherubini_etal_2008PF, de2012edge, kreilos2013edge, khapko2016edge, doohan2022state}), which emerged out of a careful dynamical systems analysis of nonlinear shear flows. In a highly non-trivial phase space, the edge states are nonlinear self-sustained solutions that lie on the stable manifold separating turbulent and relaminarising flow trajectories (also see the review of \citet{graham2021exact}). \citet{Rinaldi_etal_2018JFM} examined how the edge state has a lower (higher) fluctuation kinetic energy when the mean viscosity decreases (increases) away from the wall, which indirectly suggests how large a perturbation can be without the flow undergoing transition. A comprehensive understanding of the edge boundaries and the changes in the basins of attraction of laminar and turbulent flows in varying viscosity shear flows is however not yet available and this we believe is an academically rich area and has scientific and engineering applications like suppressing turbulence, delaying transition, or predicting turbulence using the knowledge of its state space (see e.g., \citet{suri2017forecasting, suri2024predictive}).

This section has highlighted the vast unknowns surrounding the transition to turbulence in viscosity-stratified shear flows. While understanding the complete route by which a laminar base state gives rise to turbulence is left to the future, we discuss the progress that has been made in characterizing the properties of fully developed turbulence in flows with spatially varying viscosity. These form the focus of the next section.

\section{Turbulent flows with variable viscosity}
\label{sec:turb}
\subsection{Wall-bounded flows}

The effect of viscosity stratification on turbulent plane channel flow was considered by \citet{Zonta_etal1_2012JFM}, who showed that the fluctuations about the mean are greater in the region with higher viscosity. The nonlinear flow allows for a greater mass flux through the less viscous region. To better appreciate this observation, it is useful to analyze the time evolution of asymmetric perturbations, which are precisely the ones obtained through optimisation \citep{Jose_etal_2020IJHFF,Thakur_etal_2021JFM}. As long as the perturbation amplitude is low, the fluctuations are stronger in the less viscous regions. As the flow transitions to a completely nonlinear state, the distribution of fluctuating vortical structures is consistent with what is observed in the turbulent state. Additionally, if the flow has stable density stratification, internal waves come into play \citep{Zonta_etal2_2012JFM}. The distribution of wall stresses and the effective heat transfer in the variable viscosity turbulent channel flow differ considerably from their constant-viscosity flow counterparts.  

\begin{figure}
\centering
\includegraphics[width=0.9\linewidth]{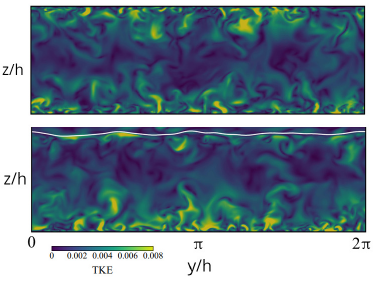}
\caption{Three dimensional direct numerical simulations of (top panel) a single fluid and (bottom panel) two immiscible fluid layers of equal densities but unequal viscosities, with the more viscous fluid ($m=2$) occupying the narrow region close to the top. The colour shows the turbulent kinetic energy at a certain streamwise location, where $y$ is the spanwise direction and the vertical direction is along the channel height from $-1$ to $+1$ (wall to wall). Adapted from \citet{roccon2021energy}.}
\label{fig:soldati}
\end{figure}

The quest for turbulent drag reduction is a problem that has attracted considerable attention for its importance in several practical applications. In this regard, \citet{roccon2021energy} study turbulent two-fluid flow through a channel, with one fluid occupying a narrow lubrication strip near the top wall, as seen in the lower panel of figure \ref{fig:soldati}. The white line denotes the interface between the two fluids. The turbulent kinetic energy in the single fluid (upper panel) indicates energetic structures near both walls, whereas a second immiscible fluid near one wall reduces the turbulent structures at that wall. This leads to significant drag reduction, which in this case is primarily achieved by surface tension, because the interface acts as an elastic layer and resists momentum transfer normal to it. In fact drag reduction is achieved even when the two viscosities are equal. But further increase in the viscosity of the lubricating layer can result in drag enhancement, since the gains made by surface tension are offset by the increased dissipation at high viscosity. 

Across a variety of incompressible wall-bounded turbulent shear flows, the profiles for the mean streamwise velocity are known to show a universal structure in the near-wall region upon suitable rescaling (using scales determined by the mean shear stress at the wall and the kinematic viscosity). The flow can be divided into regions on the basis of the dominance of viscous or inertial (Reynolds) stresses \citep{pope2001turbulent}. Organized in order of increasing distance from the wall are the viscous sublayer, the buffer layer, and the logarithmic layer. In the viscous sublayer, the flow remains practically laminar and momentum transfer is due to molecular diffusion. The buffer layer marks a transitional regime where both viscous and inertial effects are significant. In the logarithmic layer, inertial effects prevail and the mean velocity varies logarithmically with wall-normal distance. A vast literature exists on the structure of near-wall turbulence, including comprehensive reviews by \citet{jimenez2013near} and \citet{mckeon2017engine}. 

There have been attempts to describe turbulent near-wall profiles in the viscosity-stratified case in a similar manner, especially in the context of compressible flows. Some measure of success has been achieved in finding transformed mean velocity profiles that are suitable in the viscous sub-layer for turbulent channel flows \citep{Patel_etal_2015PF,Trettel_Larsson_2016PF} and the log-layer in boundary layer flows \citep{Zhang_etal_2012PRL}. Noting that these transformations do not work for all flows, \citet{Griffin_etal_2021PNAS} proposed an improved transformation performing which, compressible flow velocity profiles, where mean viscosity and density varied with distance away from the wall, collapsed onto the incompressible profile. The same transformation covered channel, pipe and boundary layer compressible flows. But this collapse requires different scalings in each layer (and is termed semi-local scaling). 

\subsection{Jets and free shear flows}
There have been a few numerical and experimental studies that examine the characteristics of a flow of a low viscosity turbulent jet into a more viscous ambient. Compared to a jet of identical viscosity injected into surrounding fluid, \citet{Talbot_etal_2013PS} show that there is a faster development of instabilities and enhanced kinetic energy dissipation in the near-field of a jet injected into higher viscosity fluid. The onset of instabilities leads to entrainment of high viscosity fluid into the jet which then contributes to the subsequent decay of the mean axial flow and a faster approach to isotropy in the downstream region. Fundamental changes to the large-scale structures too result when the two viscosities are different \citep{Voivenel_etal_2016PS}. And the jet is no longer able to maintain a self-similar profile.

\begin{figure}
\centering
\includegraphics[width=0.9\linewidth]{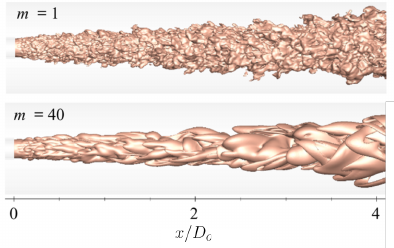}
\caption{Turbulent jet injected into a coflow with two viscosity ratios $m$; $m$ is the ratio of ambient and jet fluid viscosity and the fluids are miscible. $x$ is the streamwise direction and $D_o$ is the outer diameter of the pipe within which the jet is confined. The jet diameter is five times smaller than $D_o$. Adapted from \citet{Usta_etal_2023JFM}.}
\label{fig:usta_jet}
\end{figure}

The fact that the turbulence within a jet is significantly altered by the viscosity ratio $m$ between the surrounding and jet fluids was also demonstrated by
\citet{Usta_etal_2023JFM}. In their experiments and large-eddy simulations, a jet, maintained at a fixed Reynolds number based on inlet conditions, emerges into a pipe containing a second, miscible fluid in coflow within the pipe. Figure~\ref{fig:usta_jet} shows the cases of $m = 1$ and $m = 40$.
For higher surrounding viscosity ($m = 40$), the flow is dominated by large-scale structures within the jet and the surrounding fluid remains laminar. In contrast, at $m = 1$ (i.e., no viscosity stratification), both jet and surrounding fluid are turbulent. Noticeably, turbulent structures are of finer scale in the absence of viscosity contrast.

A key aspect of the dynamics for a jet flow is the mixing layer between the core flow and the ambient. In the planar setting, \citet{Taguelmimt_etal_2016FTC} considered the turbulent dynamics of two fluid streams of different viscosity. As in the case of a jet, there is an initial phase where the turbulent kinetic energy increases rapidly. Viscosity variation is shown to enhance the production of turbulent kinetic energy, with the production term becoming stronger due to a larger mean velocity gradient in the mixing layer. Later stage dynamics is similar to that of a jet in that the turbulent kinetic energy dies out faster than in an unstratified mixing layer. Another point we re-emphasize in this context is that large-scale quantities such as the mean velocity profile are considerably affected by viscosity variations \citep{Taguelmimt_etal_2016JT}. This is a significant point of departure from the traditional view that viscosity is only relevant at small scales in turbulence.

In homogeneous isotropic turbulence of two miscible fluids, the ultimate steady state would be one where the fluids are completely mixed and have no individual identity. But there are many situations such as Rayleigh--B\'enard turbulence where the top and bottom walls are maintained at different temperature for all time, so a viscosity difference is maintained on the large scale. And this difference can determine the structure of turbulence. The systems discussed above serve to highlight the main influences of variable viscosity. On the one hand, as shown by equation \eqref{eq:ReyOrrEnergy}, new terms appear in the equation for turbulent kinetic energy evolution directly on account of local fluctuations and mean gradients in viscosity. On the other hand, variable viscosity changes the mean velocity profile which in turn has a considerable effect on the production of turbulent kinetic energy. 

\subsection{Eddy viscosity in the ocean}\label{sec:EddyViscosity}
	
Earth's oceans are the most important sink of anthropogenically-generated heat and carbon. The ocean is turbulent in parts, and this, apart from wave mechanisms, helps mix and transport water mass, heat, and carbon around the globe. An understanding of the ocean's variable turbulent behaviour would serve to improve numerical global climate models, so that we may better address the most pressing question facing us: of climate change.
	
At extremely high Reynolds numbers, in the ocean or the atmosphere or even in large industrial processes, complete solutions of the Navier-Stokes equations are impossible, due to the large separation between the length scales at which the flow is forced and the scales at which most of the dissipation occurs. Also, often we are interested only in mean quantities and not in the details of the turbulence, and adopting the concept of eddy viscosity helps simplify our approach.

\begin{figure*}
\centering
\includegraphics[width=1\linewidth]{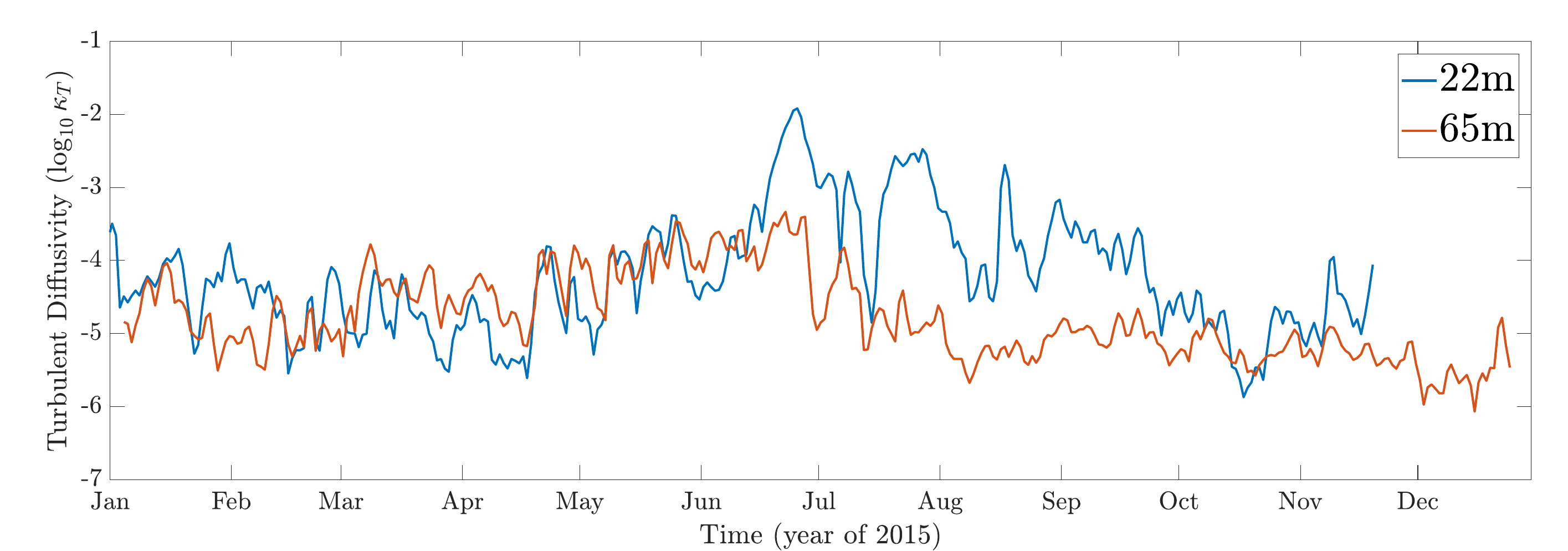}
\caption{Time series of daily-averaged eddy diffusivity $\kappa_T$ of the Bay of Bengal at two different depths (22m and 65m from the surface) calculated using mixing meters called $\chi$pods (\citep{moum2009mixing}). Note that the $y$ axis is on log scale, so differences are larger than they appear. Data taken from \citet{thakur2019seasonality}. }
\label{fig:seawater_visc}
\end{figure*}

\begin{figure*}
    \centering    \includegraphics[width=0.4\linewidth]{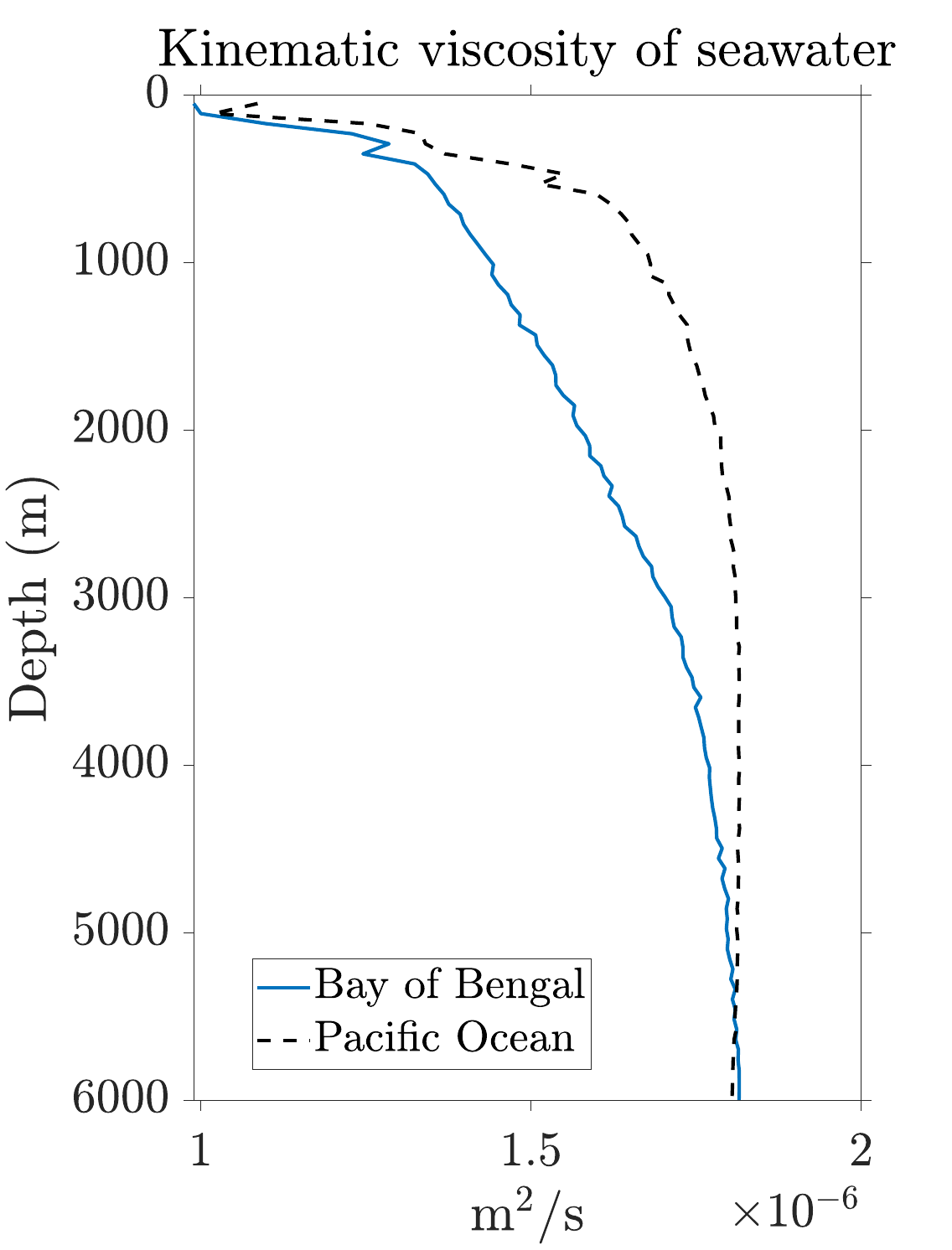}
\caption{Kinematic viscosity of seawater calculated for some representative temperature and salinity profiles for the Bay of Bengal and the Pacific Ocean using the empirical model of \citet{isdale1972physical} (also see \citet{sharqawy2010thermophysical}). The temperature and salinity are taken from the World Ocean Database \citep{boyer2018world}.}
    \label{fig:seawater_visc_only_nu}
\end{figure*}
To explain this, we perform a Reynolds decomposition, splitting the turbulent velocity and pressure fields into their mean (or expectation), denoted by uppercase, and a fluctuation, denoted by a hat, as
\begin{equation}
\begin{aligned}
\mathbf{u} (\mathbf{x},t)=\mathbf{U}(\mathbf{x},t) + \hat{\mathbf{u}}(\mathbf{x},t), \quad p(\mathbf{x},t)= P(\mathbf{x},t) + \hat p(\mathbf{x},t).
\end{aligned}
\label{reynolds_decomposition}
\end{equation}
Note that the `mean' velocity $\mathbf{U}$ is shown as a function of time. This is to allow for slow variations with time of this quantity whereas the perturbations are averaged out over shorter time intervals.
Substituting these into the Navier--Stokes equations \eqref{eq:ns}, taking an ensemble average across a distribution of perturbation quantities, and using the incompressibility condition in equation \eqref{eq:incomp}, we obtain the Reynolds equations 
\begin{eqnarray}
&\nabla\cdot \mathbf{U} = 0,\\
&\begin{split}
\frac{\partial U_j}{\partial t}  & + U_i \frac{\partial  U_j}{\partial x_i} = -\frac{1}{\rho} \frac{\partial P}{\partial x_j} + \\&+ \frac{\partial}{\partial x_i} \left[\nu \left(\frac{\partial U_i }{\partial x_j} + \frac{\partial  U_j}{\partial x_i}\right) - \langle \hat u_{i} \hat u_{j}\rangle\right],
\end{split}
\label{reyn_eqn}
\end{eqnarray}
where the angle bracket denotes mean values.
The covariances of velocity fluctuations, $ \langle \hat u_{i} \hat u_{j}\rangle$, are commonly called the Reynolds or turbulent stresses (whereas $\rho \langle \hat u_{i} \hat u_{j}\rangle$ are the actual stresses). So, we have four equations, i.e., three Reynolds equations and the mean continuity equation but have more variables because of the correlations of fluctuations: this is the classical closure problem in turbulence. 
Due to the observation that turbulence enhances mixing and momentum diffusion in a manner phenomenologically similar to molecular viscosity, the simplest turbulence models approximate the Reynolds stress tensor, $ \langle \hat u_{i} \hat u_{j}\rangle$, by relating it to the mean rate of strain and to local scalar quantities in the flow \citep{pope2001turbulent} -- much like how viscous stresses depend on strain via molecular viscosity.  
This is the eddy viscosity hypothesis and owing to Boussinesq, is given by 
\begin{equation}
-\langle{\hat u_i \hat u_j}\rangle = -\frac{2}{3}k\delta_{ij} + \nu_{T}(\boldsymbol{x} ,t) \left(\frac{\partial U_i }{\partial x_j} + \frac{\partial  U_j}{\partial x_i}\right),
\label{reynolds_stress}
\end{equation}
where $k = \langle{\hat u_i \hat u_j}\rangle/2$ is the turbulent kinetic energy and $\nu_{T}(\boldsymbol{x} ,t)$ is the eddy viscosity, which depends on the flow conditions and is a function of space and time. The isotropic part of the Reynolds stresses, $-2k\delta_{ij}/3$, can be absorbed into the mean pressure field, and only the remaining deviatoric (traceless) stress part is important for momentum transfer between mean and fluctuating components of the flow. The mean momentum equation now looks exactly like the Navier--Stokes equations for a laminar flow with the effective viscosity $\nu_T$ + $\nu$ and the modified mean pressure includes the term 2$\rho k$/3. This equation can be solved at a much coarser resolution than the complete turbulent flow, allowing a realistic simulation at Earth scale with reasonable accuracy in general circulation models. However, an appropriate form of the eddy viscosity $\nu_T$ needs to be prescribed as a function of space and time, and we do not delve into the multitude of ways expressions for $\nu_T$ are arrived at. This simplest of models performs quite well for most scenarios, but fails for specific systems like strongly swirling flows. A similar derivation as above for temperature lead us to the eddy diffusivity of temperature $\kappa_T$ that models the temperature fluxes in the mean scalar equation (see, e.g, chapter 4 of \cite{pope2001turbulent}). With an estimate of the turbulent Prandtl number, $Pr_T \equiv \nu_T/\kappa_T$, scalars provide good datasets to indirectly estimate the eddy viscosity and its variation with depth and time. However, estimating turbulent Prandtl number is not straightforward given that it depends on parameters of the flow like $Re$, $Pr$, $Ri$, turbulent Péclet number $Pe_T$, distance from the wall etc. \citep{kays1994turbulent,kapyla2022turbulent}. Unlike Schmidt and Prandtl numbers for molecular diffusivity, $Pr_T \sim O(1)$, since eddies transport and thus diffuse momentum, mass and heat in similar ways.

Turbulence in the ocean varies in space and across seasons. Wind and bottom topography stresses respectively drive turbulence in the uppermost and the lowermost layers. Internal-wave breaking contributes primarily to turbulence in the interior, but the interior can also be quiet for extended periods of time. Specialised turbulence measuring instruments like the Vertical Microstructure Profiler or $\chi$-pods that have been deployed in the different parts of the world's ocean measure temperature and velocity fluctuations to get estimates of the turbulent intensity in terms of eddy viscosities or temperature diffusivity. Measurements of $\kappa_T$, assuming $Pr_T$ to be unity, yield the eddy diffusivity $\nu_T$. Turbulent diffusivity is found to be dependent on surface forcing (wind stress, solar heat flux) as well as subsurface stratification \citep{cherian2020seasonal,thakur2019seasonality,warner2019feedback,moum2022deep,moum2021variations}. The upper ocean is more vigorously mixed, and turbulence typically decays with depth, but there is wide variability, including nonmonotonic variations. A time and space varying viscosity profile is therefore needed in hydrodynamic models. Time series of $\kappa_T$ are shown for a particular location in the Bay of Bengal in the Indian ocean in figure \ref{fig:seawater_visc}, at two different depths, of $22$ and $65$ metres below the sea surface. For simplicity, setting $Pr_T=1$, we make an estimate of $\nu_T$ as being equal to $\kappa_T$. However, $Pr_T$ is different for different kinds of stratified flows \citep{kays1994turbulent, venayagamoorthy2010turbulent}. The reader might be interested to take note of the variation with season and with depth. The upper ocean ($22 m$) diffusivity follows the surface forcing. Depending on the depth up to which surface effects are felt, i.e, the depth of the upper mixed layer,  the deeper ocean ($65m$) may follow surface forcing too, or show independent behaviour \citep{thakur2019seasonality}. Post the onset of the Indian summer monsoon (during July-September) winds are stronger and the eddy diffusivity at $22m$ is higher than normal. But these months are also the time when the ocean relatively quietens down at $65m$. Turbulence also varies at much smaller timescales than those visible in this figure. For example, nighttime cooling of the surface leads to convective mixing. Actual data point to the variation of $\nu_T$ with greater depth in the ocean. Below $O(100m)$ depth, the ocean becomes relatively quiet, and the sum of eddy and molecular viscosities is well approximated by the molecular viscosity itself, whose typical variation with depth is shown in figure \ref{fig:seawater_visc_only_nu}.

We illustrate the role of eddy viscosity and its variation by examining the Ekman spiral: flow which occurs just beneath the ocean surface, i.e., in the Ekman layer, in response to forcing by wind. Ekman layers are boundary layers found in the atmosphere, the ocean surface and the ocean bottom boundary, where the Coriolis force due to Earth's rotation balances viscous and Reynolds stresses. Since flows in these regimes are turbulent, the eddy viscosity replaces the molecular viscosity in relevance. The thickness of the Ekman layer $d_E \equiv (2 \nu_T/\mathcal{F})^{1/2}$, where $\mathcal{F}$ is the Coriolis frequency at the latitude of interest. It extends from the surface to a depth of a few tens to hundreds of metres, and as can be easily inferred, it grows in thickness as the strength of turbulence, and hence $\nu_T$, increases. The Ekman spiral was originally derived with a constant viscosity assumption by \citet{ekman1905influence}, as derived below.

In a coordinate system rotating about the $z$ (vertical) axis at a rate $\mathcal{F}$, we can represent the force balance within the Ekman layer upon neglecting nonlinear terms and molecular viscosity in the Navier-Stokes equation as
\begin{equation}
\mathcal{F} \hat{z} \times \mathbf{u_h} = \frac{-1}{\rho} \nabla p + \frac{\partial {\boldsymbol{\sigma_T}}}{\partial z}
\label{ekman}
\end{equation}
for the steady-state motion of a layer of fluid \citep{vallis2017atmospheric}, where $\mathbf{u_h} = (u,v)$ is the horizontal velocity, and the stress $\boldsymbol{\sigma_T} = \nu_T(z) [\partial \mathbf{u_h}/\partial z]$.	At the ocean free surface, continuity of stress provides
\begin{equation}
\sigma_{x}\bigg{|}_{z=0^+} = \left[\nu_T \frac{\partial u}{\partial z}\right]_{z=0^-} \  \mathrm{and} \  \sigma_{y}\bigg{|}_{z=0^+} = \left[\nu_T \frac{\partial v}{\partial z}\right]_{z=0^-},
\label{ekmanbc}
\end{equation}
where the stresses due to the air layer at $z=0^+$ are known. Here the superscripts + and - refer to approaching the surface from the atmosphere and within the ocean respectively. Deep down, as $z \rightarrow -\infty$, we have $u, v = (0, 0)$. Solving the above set of equations with a constant eddy viscosity gives the Ekman spiral:
\begin{equation}
u + i v = \frac{(\sigma_x + i \sigma_y)\bigg{|}_{z=0^+}}{\sqrt{\mathcal{F}\nu_T}}\exp\left[\frac{z}{d_E}\right]\exp\left\{i \left[\frac{z}{d_E} - \frac{\pi}{4}\right]\right\}.
\label{ekman_soln}
\end{equation}
The surface velocity in the ocean is at an angle of $\pi/4$ to the wind stress, and the spiral velocity field decays over the characteristic depth $d_E$.

Observational data often shows a departure from this theory, and several studies have accounted for these differences by including variations with depth in the eddy viscosity, and solving equations \eqref{ekman} and \eqref{ekmanbc}  with $\nu_T = \nu_T(z)$ (see, e.g. \citet{constantin2019atmospheric, constantin2021frictional,dritschel2020ekman,cronin2009near}. In particular, \citet{constantin2021frictional} find a departure from the classical deflection angle of 45$^{\circ}$, though the solution still remains a spiral and retains other features of the mass flow as before. As per our knowledge, these results have not been compared to actual measurements, probably because simultaneous measurements of ocean surface variables and eddy viscosity are not easy. Most large scale numerical models do not have the resolution to fully-resolve the Ekman layer and hence understanding the effect of changing turbulent viscosity within this layer is important for efficient parameterisation. 

The correct representation of eddy diffusivity is also important in a range of other situations. \citet{lentz1995sensitivity} study circulation in the inner ocean shelf. Upon varying the vertical structure of eddy viscosity they found that while the along-shelf circulation is not sensitive to the choice, the cross-shelf component is. Describing this circulation is needed for estimating how matter is transported away from the shelf. The influence of eddy viscosity on phytoplankton bloom rates \citep{siegel2002north}, phytoplankton settling, and the transport of larvae \citep{pineda2007larval,hale2020global} and sediments deserves attention. Micro- and nanoplastics in the ocean and atmosphere, and in sources of inland water used for irrigation and consumption are obviously deleterious to marine and human health \citep{carbery2018trophic,amaral2020ecology}. A particle smaller than the Kolmogorov scale effectively perceives laminar flow in its neighborhood, and its dynamics is determined by molecular viscosity, as we saw in section \ref{sec:ParticleDynamics}. To solve for large-scale flows containing an extremely large number of small particles, a judicious scale-dependent combination of the eddy and molecular viscosities must be designed. A simple addition of the two will not do. Such a description, which obviates having to solve the Navier--Stokes directly on large scale particle-laden flows, would be a boon. The local effective viscosity determines the effective particle Reynolds number and influences the morphology and distribution of organisms (and other solid material) with depth in the ocean \citep{vogel2020life}. Important progress has been made in numerical and observational studies that have looked into ocean debris and microplastics transport, like that of \citet{lebreton2012numerical, peng2021plastic}, and \citet{kooi2017ups} who use realistic profiles of seawater, including the variation of viscosity, in settling or oscillation of biofilm-covered microplastics. Including eddy viscosity variation in the proper manner in these calculations is likely to refine the estimates. The studies of \citet{thakur2022impact} and \citet{skitka2023epsilon}, that include comparison of model output to observational data, show the importance of improved eddy-viscosity parameterisation in better representing internal wave dynamics within high-resolution ocean hydrodynamic models. Deep-learning based improvements in vertical profiles of turbulent viscosity have shown advances in climate simulations, \citep{zhu2022physics} and in  estimating turbulent mixing values \citep{salehipour2019deep}. While gaps exist in the models in implementing better physics-derived viscosity parametrisation, the success so far validates the benefits of doing so, and also of using the synergy of observational data, numerical simulations, and machine learning techniques in the context of ocean dynamics and climate science.

We now turn to the influence of viscosity variations at the largest scales relevant to Earth, namely, flows in the mantle and associated magma and lava transport.

\section{Earth-scale consequences of viscosity stratification}
\label{sec:EarthFlows}

\subsection{Earth's Mantle}

The solid Earth is composed of metals and minerals with a metallic core covered by the silicate-rich mantle that accounts the bulk of the planet’s mass and volume \citep{birch1952elasticity}. While the mantle behaves as an elastic solid over short (less than years) timescales, supporting the propagation of seismic waves and response to external loading, it flows as a highly viscous fluid on geological timescales of thousands to millions of years. In the intermediate range, to account for both short and long timescale response, the viscous stress response of the mantle is generally modelled as that of a Maxwell material, with the relaxation time scale being a few hundred years \citep{peltier1974impulse, ribe2018theoretical}.

The mantle is in an unsteady dynamic state due to thermal convection, driven by an unstable temperature profile: Earth's outer core, just below the mantle is hot, and the near-crust region just above is far cooler. The temperature differences and the large length scales give moderate to large Rayleigh number. The dynamic mantle is what gives rise to the oceanic and continental crust, to volcanoes and earthquakes, and affects the magnitude and reorientation of Earth's magnetic field \citep{glatzmaiers1995three,schubert2001mantle,cathles2015viscosity}. The spatio-temporal response of the mantle at various timescales is a topic that is of great interest in Earth sciences \citep{korenaga2018crustal}. Crucially, the fluid-like mantle is far from uniform: temperature, pressure, and compositional gradients lead to spatial variations in viscosity of several orders of magnitude. Given the humongous magnitude of mantle viscosity, these variations may be small in a relative sense, but gradients can be high locally, and become dynamically important. Viscosity variations influence mantle convection which in turn affects plate tectonics and determines the long-term thermal and chemical evolution of Earth. In the following sub-sections, we explore what is known of how such viscosity stratification shapes mantle dynamics, governs the behavior of magma flows, and affects convective instabilities in Earth's interior. 

The mantle displays an extreme range of pressures and temperatures, which vary with space and time. Most studies which make broad estimates of mantle temperature agree on an overall increase in the adiabatic temperature with depth, and also on the presence of discontinuities in the temperature profile \citep{katsura2022revised}. This has consequences for physical properties, including the viscosity profile, and thence for the dynamics. Further, the mantle is heterogeneous in composition, and contains pockets of differing viscosity that can be a few tens of times higher or lower than the ambient. The high viscosity regions can survive much longer than the lower viscosity ones, leading to heterogeneity in the process of mixing \citep{gurnis1986stirring, kellogg1990mixing, van1999compositional}. Such regions can also aggregate to form large scale heterogeneities \citep{manga1996mixing}. The viscosity contrasts are dynamic, depending, beyond temperature, concentration and pressure, on whether the minerals are dry or wet, how their grain sizes are distributed, how the concentration of other elements like hydrogen vary \citep{steinberger2006models, ruh2022grain}, and also on the shear rate. Effective viscosity estimation is difficult, and further complicated by phase transitions, which occur under extreme pressure and temperature. Moreover, past deformations affect the present deformation. 

\begin{figure*}
\centering
\includegraphics[width=\textwidth]{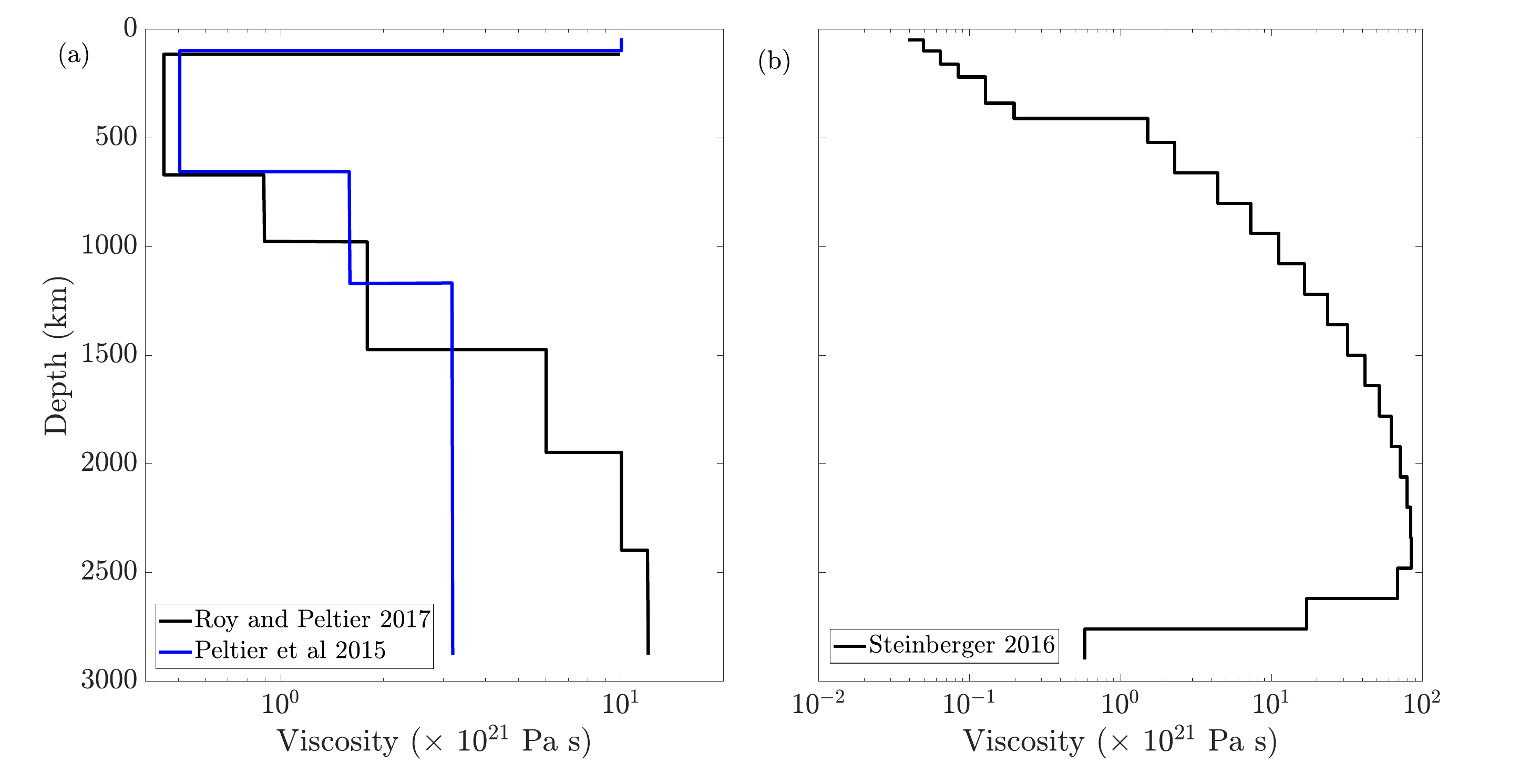}
\caption{(a) Sample profiles of mantle viscosity used in the models of \citet{roy2017space} (in blue) and \citet{peltier2015space} (in black) obtained using Earth's viscous response to ice sheet unloading. Adapted with permission from \citet{argus2021viscosity}, where further explanation is available. (b) Sample viscosity structure used in models to understand deeper-mantle dynamics. Contrast this to the profiles shown in (a), where the viscosity in the lowest part shows no decrease. This data appears in \citet{steinberger2016topography} and was used in modelling \citep{van2021record} (courtesy of Bernhard Steinberger and Douwe J J van Hinsbergen).}
\label{fig:diff_Mantle_visc_models}
\end{figure*}

Following the pioneering work of \citet{haskell1935motion} and later confirmed by many (e.g., \citet{forte2001deep, steinberger2006models}), the average viscosity of Earth's mantle has been estimated to be $\mathcal{O}$(10$^{21}$) -- $\mathcal{O}$(10$^{22}$) Pa s. We discuss a few estimates that have been made for how the profile of viscosity varies with depth, but we do not provide technical details of the methods used to arrive at them, nor are we able to evaluate them relative to each other. Though these estimates differ from each other, they agree to an increase on an average in viscosity with depth. While some studies, e.g. \citet{forte2001deep}, propose zigzag profiles for viscosity with one or more maxima at various depths, staircase structures, such as those in figure \ref{fig:diff_Mantle_visc_models} are often proposed as well. Later in this review, we will discuss these further. Noticeably, viscosity estimates can vary by a factor of about $100$ between studies. A factor of $100$ may appear as an extremely small percentage error in the viscosity, but note, from the definition $Ra_i \equiv g \Delta \rho_i L^3 /(\nu_{ref} \kappa_i)$, the Rayleigh number will be wrong by about this factor, which will make a big difference to the very nature of mantle convection. 
Early estimates of viscosity were made by solving an inverse problem, based on the rebound of Earth at different locations due to unloading of trillions of tons of surface ice during ice-age melting \citep{peltier1974impulse}. This rebound is called glacial isostatic adjustment. Used along with our knowledge of geoid anomalies (deviations of Earth's surface from the imaginary, irregular, and gravitationally equipotential surface), these provide bounds of viscosity at different depths, see e.g., \citet{peltier2015space}. Such inverse problems consider viscosity as the only free parameter, and that it only varies as a function of radius (depth). Estimates based on ice unloading are more sensitive to upper mantle viscosity, due to its lower viscosity and direct mechanical link to the lithosphere. This technique remains a standard up to today, and was recently used to infer the viscosity of Mars \citep{broquet2025glacial}. 
Figure \ref{fig:diff_Mantle_visc_models} (a) from \citet{argus2021viscosity} shows radial profiles of mantle viscosity obtained by inverse solution methods. The viscosity profiles are modified based on misfits between data and the computed values of crustal uplift and sea-level changes, and optimised until the best estimates are arrived at.

It is also important to gain knowledge of the viscosity of the lower mantle, which extends from 660 km to 2900 km below Earth's surface, for understanding the fate of the subducted oceanic lithosphere and hence the long-term changes to the sea level \citep{mao2018slab}. Also, the lower mantle is an important reservoir and pathway in the recycling of Earth minerals \citep{marquardt2015slab}. It is to be noted that the effective viscosity of the lower mantle is a topic of constant debate. For these viscosity estimates, comparing anomalies in geoid observations to numerical modelling, and including our knowledge of density anomalies as well as mineral physics constraints in setting up the simulations, has been proven to provide tighter bounds. Differences between observations and numerical geoid modelling have been reconciled by prescribing a radially varying viscosity in the models. And prescribing lateral variations in viscosity in certain regions like the core-mantle boundary and the upper mantle further improves the agreement between geoid modelling and observations. Many studies agree on a reduction of viscosity at the bottom of the mantle due to partial melting in the core--mantle boundary, see e.g. \citet{van2018atlas}. This reduction in the deepest part of the mantle, as seen in the viscosity profile in figure \ref{fig:diff_Mantle_visc_models} (b), which is derived from geoid measurements, seismic tomography, and constrained with mineral physics, would not be captured by models solely based on glacial isostatic adjustment, where sensitivity to the lower mantle viscosity is minimal (the method used to obtain profiles in figure \ref{fig:diff_Mantle_visc_models}(a)). Laboratory experiments on various mantle minerals generate bounds employed in viscosity estimates, most notably for the lower mantle. \citet{marquardt2015slab} conduct experiments on ferropericlase -- an oxide of Fe/Mg found in the lower mantle, to show an increase in the strength of this mineral at lower mantle pressures of $\mathcal{O}(100)$ GPa. Their estimate shows a progressive increase in viscosity up to a depth of 1500 km by more than two orders of magnitude. While more such experiments would be most enlightening, designing them poses challenges because of the extremely high pressures required and minuscule strain rates of such deformations.

To mention some zigzag viscosity profiles, \citet{rudolph2015viscosity} use a Bayesian inversion technique to determine mantle viscosity; and solutions using different values of the relevant free parameters in this inversion give similar radial profiles. They find viscosity maxima at 800--1200 km depth. \citet{forte2001deep} use surface geodynamic data to constrain the density and viscosity of Earth's mantle in a convective model and find two maxima in viscosity around $1000$ km and $2000$ km depth. On the other hand, a staircase structure is found by \citet{steinberger2016topography}, who uses topography details to constrain the viscosity and density structure.  Due to this viscosity structure in the model of \citet{forte2001deep}, the flow organises itself in the horizontal, and occurs primarily within layers. So deep mantle convection and mixing get suppressed. 


The $\mathcal{O}(10)^{21}$ Pa s viscosity makes the Reynolds numbers vanishingly small. For comparison, the dynamic viscosity of water at room temperature is $\mathcal{O}$(10$^{-3}$) Pa s. But mantle convection sets in because of the high Rayleigh number, up to $\mathcal{O}$(10$^{7}$). We estimate the Rayleigh number using estimates of the coefficient of thermal expansion of $\mathcal{O}$($10^{-5}$), thermal diffusivity of $\mathcal{O}$(10$^{-7}$), and density ($\approx$ 4000 kg/m$^3$) on geological timescales. Over long time scales, the mantle can be modelled as being in creeping flow (see e.g., \citep{Schubert_Turcotte_Olson_2001,ribe2018theoretical}). Hence the momentum equation \eqref{eq:ns} reduces to that in the Stokes limit, seen earlier as equation \eqref{eq:MassandMomentum}, where unsteady and nonlinear terms in the momentum are dropped, but the gravitational force which is a function of depth $z$ from Earth's surface: 
\begin{equation}
\frac{1}{Re}\nabla\cdot \bm{\sigma} + \frac{Ra(z)}{RePe}\delta T\mathbf{\mathbf{\underaccent{\tilde}{z}}} - \frac{1}{Ro}\mathbf{\mathbf{\underaccent{\tilde}{z}}} \times \mathbf{u} = 0,
\label{eq12}
\end{equation}
where $\delta T$ is the temperature variation from the hydrostatic value, $Ro \equiv U/\mathcal{F}L$ is the Rossby number and $\bm{\sigma}$ is given by equation \eqref{eq:StressDef}. The viscosity of the mantle, apart from being a function of pressure, temperature and concentration, is also a function of local strain rate. But to make computations feasible, it is standard to assume the mantle viscosity to be Newtonian, and to club all other dependencies into an empirical
explicit variation with depth, in terms of exponential functions, such as
\begin{equation}
\mu =  \exp(z), \ \  {\rm or} \ \ 
\mu = \frac{\mu_1}{\mu_0} + (1 - \frac{\mu_1}{\mu_0})\exp(-z),\label{exp2}
\end{equation}
where the viscosity is scaled by the surface dynamic viscosity $\mu_0$ (at $z = 0$) and $\mu_1$ in the second model is the asymptotic value of viscosity as $z \rightarrow \infty$, that represents the viscosity near the outer core. In the above equations, $z$ is scaled by the characteristic vertical scale $L$. The reader will immediately notice that smooth models like in equation \eqref{exp2} differ from the layered models of viscosity shown in Figure \ref{fig:diff_Mantle_visc_models}. The simpler forms are adopted as a first choice, due to the computational challenges presented by sudden jumps, and given our inability to infer the exact profile of mantle viscosity from surface measurements or seismic tomography. However, we have seen in section \ref{sec:singular} that the mathematical structure at sudden jumps can be special, and the resulting physics can be lost by such smoothing out. It would be important that mantle viscosity expressions be corrected as better knowledge is accrued. A dependence of mantle viscosity on temperature and pressure has been used, for example, by \citet{yuen1984stability} to study the stability of the oceanic lithosphere. Such dependencies affect shearing and mixing, create localised shear zones, and allow secondary instabilities to grow in the lower viscosity regions. In subduction zones, where a tectonic plate slides below another, the viscosity contrast with the ambient affects the mode of subduction \citep{ribe2010bending}. Theoretical and analytical efforts of subduction include slow viscous flow around a corner (e.g, \citet{ribe2007analytical, ribe2018theoretical}). Besides these studies, which, for better tractability, assume Newtonian rheology, there are studies allowing for non-Newtonian rheology for the upper mantle, while the lower mantle is still prescribed to be Newtonian (e.g., see the review of \citet{faccenna2014mantle}). The upper mantle viscosity now is modelled as a power law 
\begin{equation}
\mu = K[T,c,P]({\mathbf x})(\epsilon_{ij}\epsilon_{ij})^{1/2n-1/2},  
\end{equation}
which brings in further nonlinearity into the equations. In the above, $\epsilon_{ij}$ is $(i,j)^{\textrm{th}}$ element of the rate of strain tensor $\epsilon_{ij} = (\partial_iu_j + \partial_ju_i)/2$, $K$ is a function of temperature, concentration and pressure. The exponent $n$ is often prescribed a value of 3.5 (see \citet{schellart2024subduction}).  

We now discuss mantle convection in some detail. Textbooks like that of \citet{gerya2019introduction} are likely to be helpful to readers interested in developing or improving models of mantle convection and other geodynamic processes. Convection is crucial for transporting heat from Earth's core into the mantle, transporting heat within the mantle, and for bringing about mixing and redistribution of Earth's materials. Besides large-scale motion, mantle plumes, also resulting from buoyancy, are a smaller scale phenomenon that can nevertheless be powerful. The role of viscosity stratification in modifying mantle convection remains an interesting area of research \citep{mao2019dynamics}. The interplay of density and viscosity gradients can also give rise to interesting mixing dynamics and instabilities. A series of numerical simulations related to mantle convection were conducted using temperature-dependent viscosity with many orders of magnitude variation, accounting for nonlinear rheology, phase transitions, and even compressibility effects \citep{tackley1996effects, tackley2000self, tackley2008modelling}. \citet{zhong2000role} simulated 3D mantle convection with layered (piecewise constant) viscosity profiles, with temperature-dependent viscosity profiles, as well with combinations of the two. Both types of viscosity profiles  show the emergence of long-wavelength structures, and the variation of viscosity within the mantle in both cases serves to prevent cold downwellings from breaking up into plumes at deeper depths, unlike in a constant viscosity model. Viscosity stratification was shown to enable these downwelling flows to reach the core-mantle boundary, a qualitative departure from constant viscosity findings. These results complement similar effects on downwelling structures obtained by \citet{zhang1995influences} who use a viscosity profile similar to that in figure \ref{fig:diff_Mantle_visc_models} (b). While this study does not bring out major qualitative differences between discontinuous and smooth viscosity profiles, the former lies open to further study. To faithfully simulate abrupt and large jumps in viscosity, specialised numerical methods are needed, and recent works have focused on developing efficient numerical schemes faithful to such physics. Inclusion of mineral physics data \citep{stemmer2006new,furuichi2011development,kronbichler2012high,heister2017high} makes for more realistic viscosity variations. High viscosity contrasts create narrower plumes in these models, suppress vertical motion and lead to higher mixing within horizontally organised layers. Instability studies and analyses of simulation results are needed to check whether the behaviour accrues from an overlap instability. Another important question is to understand how perfectly or imperfectly mixed the mantle is, following convection. To further understand mantle dynamics, more fundamental studies on smaller-scale temporal and spatial viscosity variations, and their comparison to large-scale realistic simulations, will add value, and some model problems discussed in section \ref{conv_w_visc} touch upon this briefly.

\subsection{Magma and Lava Flows}

Besides large-scale motions in the mantle as a whole, a range of flows at smaller scales are taking place, probably all the time. Various dynamics and processes within the mantle and Earth's crust produce local melting, leading to what are termed `magma chambers' made up of molten or partially molten rocks. This molten material is at a higher temperature than the material above and so, under favourable conditions, pushes its way upwards through porous solid or a semi-molten matrix, and finally erupts at Earth's surface through fracture points, giving rise to volcanoes and/ or new crust \citep{katz2022dynamics, katz2022physics}. The molten rocky material that makes its way out to Earth's surface is called lava. 

Magma is composed of silicate melts and other oxides along with their crystals, and its composition changes as it rises upwards \cite{cooper2014rapid}. The change in chemical composition means that it is not possible to ignore chemistry in these problems if we wish to obtain an accurate description of the underlying mechanisms. The large variations in the mineral concentration, crystal size, and gas bubble density give rise to corresponding variations in viscosity. Temperature variations too are important here, and the viscosity of molten magma can vary by about fifteen orders of magnitude: in the range of $10^{-1}-10^{14}$ Pa s \citep{caricchi2007non,gonnermann2007fluid,giordano2008viscosity}. This molten material has to move through surrounding highly-viscous non-eruptible mushy layers composed of crystals, other solids and melt \citep{cashman2017vertically}. The mantle has a porosity of $\approx$ 0.1\% but this number varies in space and time \citep{katz2022physics}. 

As the molten magma travels through regions of varying physical and chemical composition, it interacts with the surrounding matrix \citep{khodakovskii1995melt}. When chemically-reactive magma rise through porous matrix with a solubility gradient, \citet{aharonov1995channeling} show that this creates high-porosity passageways for the melt to rise through. As a result the mechanical stability of the matrix could be affected. The channel length scales in such a flow are chosen based on the physical parameters, mostly based on the Damk\"ohler (ratio of chemical reaction time to transport time) and Péclet numbers \citep{spiegelman2001causes}. This is a reaction-infiltration instability in which a combination of reaction, advection and solid compaction team up: the magma melt dissolves the surrounding matrix, and thus creating additional localised pathways to penetrate the matrix. The increased flow cross-section creates greater advection which in turn speeds up matrix dissolution in a runaway process. The excess fluid pressure gives rise to compaction of the solid matrix, which enhances the instability and sets a preferred length scale. More recent works like that of \citet{jones2018reaction} expand on this to show the effect of advection and diffusion too in setting the length scale of the instability. This instability is easily achieved at smaller compaction length [a length scale that is derived from the mobility and the bulk viscosity of the semi-molten matrix] when the reaction rates are higher. However, the instability, and hence transport, can be affected by variable viscosity, and as shown by \citet{hewitt2010modelling}, the variability of the matrix or bulk viscosity can lead to suppression of the instability. The laboratory experiments of \citet{whitehead1991instability}, mimicking hot geophysical flows in an elastic chamber, show fingering instability in a temperature-dependent fluid [a fluid whose viscosity and other properties vary significantly with temperature] that gets cooled as it rises upwards through a slot. The fingering instability is caused by the cooling, and increase in viscosity of the fluid as it rises. Some early studies of \citet{ockendon1977variable,pearson1977variable} show the effect of wall temperature in modifying velocity fields in a high Prandtl number temperature-dependent fluid flowing through a long narrow channel. Along similar lines, a recent study of a variable viscosity low-Reynolds-number flow through a slender pipe, in the limit of high Peclet number, shows that the low viscosity fluid near the walls has lower advection rates than the higher viscosity core \citep{louis2023effect}. These studies point out that in flow through narrow conduits or through mush, viscosity variation affects the flow field and stability. Viscosity and its variation, within the magma and in the surroundings, thus determine the magma's flow speed and behaviour, and thence its eruption dynamics \citep{giordano2008viscosity,pec2017reaction}. This is a multiphase problem that includes the solid matrix, or mush, and the molten magma flow \citep{mckenzie1984generation}. We submit that magma flows call out for further investigation: the many parameters in this flow, including the geometrical properties of the magma chamber and the passage of exit, could yield new modes of instability. Also, the interaction of molten magma with ground water is important for magma mixing that generates violent eruptions \citep{gonnermann2007fluid}. Analytical models of interactions of such flows with chemical changes are directions of work that will contribute significantly to the understanding of these flows.

\begin{figure*}
\centering
\begin{subfigure}{0.42\textwidth}
\includegraphics[width=\textwidth]{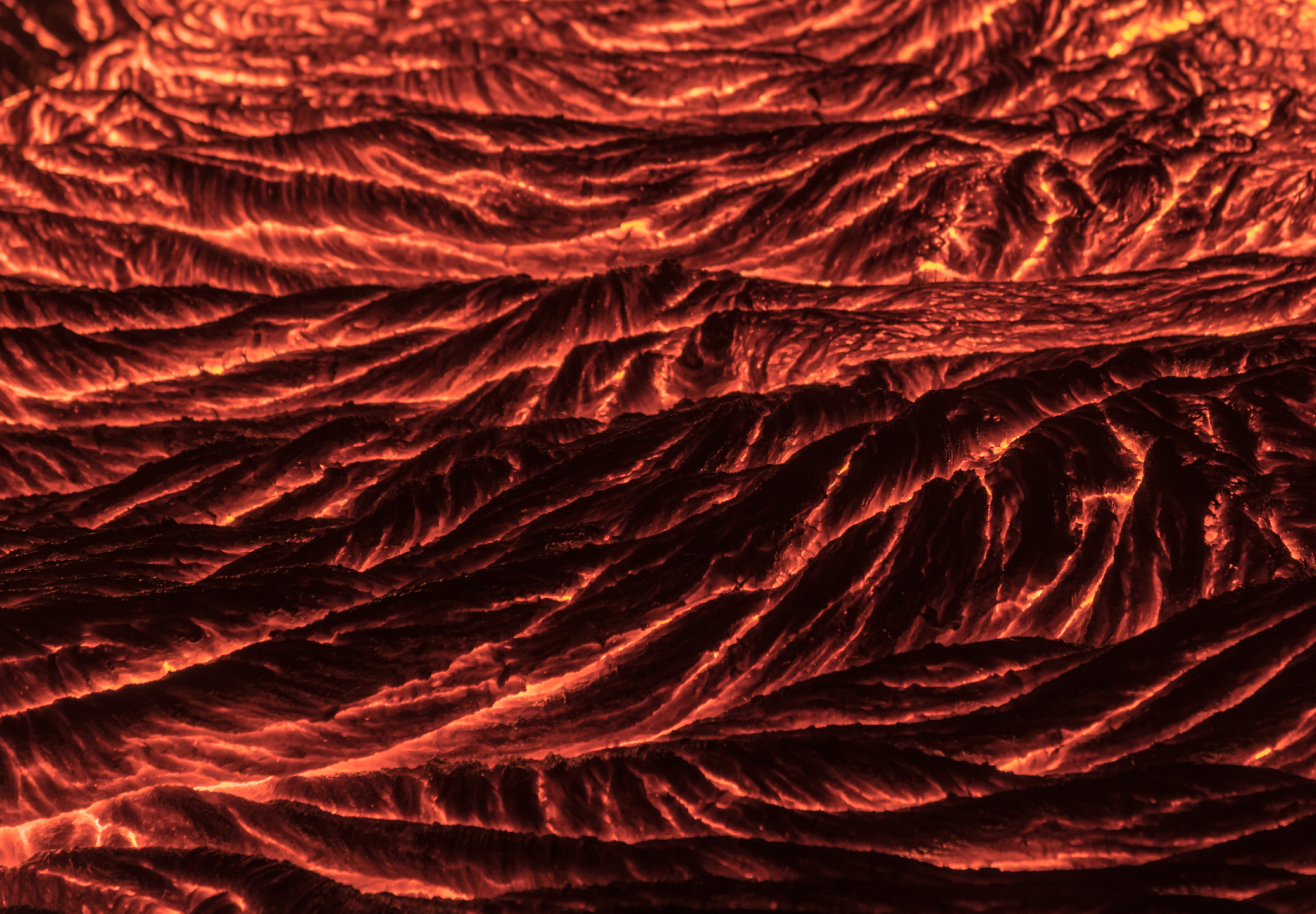}
\caption{The ropey structure of Pahoehoe lava from K\={i}lauea volcano}
\label{fig:first}
\end{subfigure}
\hfill
\begin{subfigure}{0.45\textwidth}
\includegraphics[width=\textwidth]{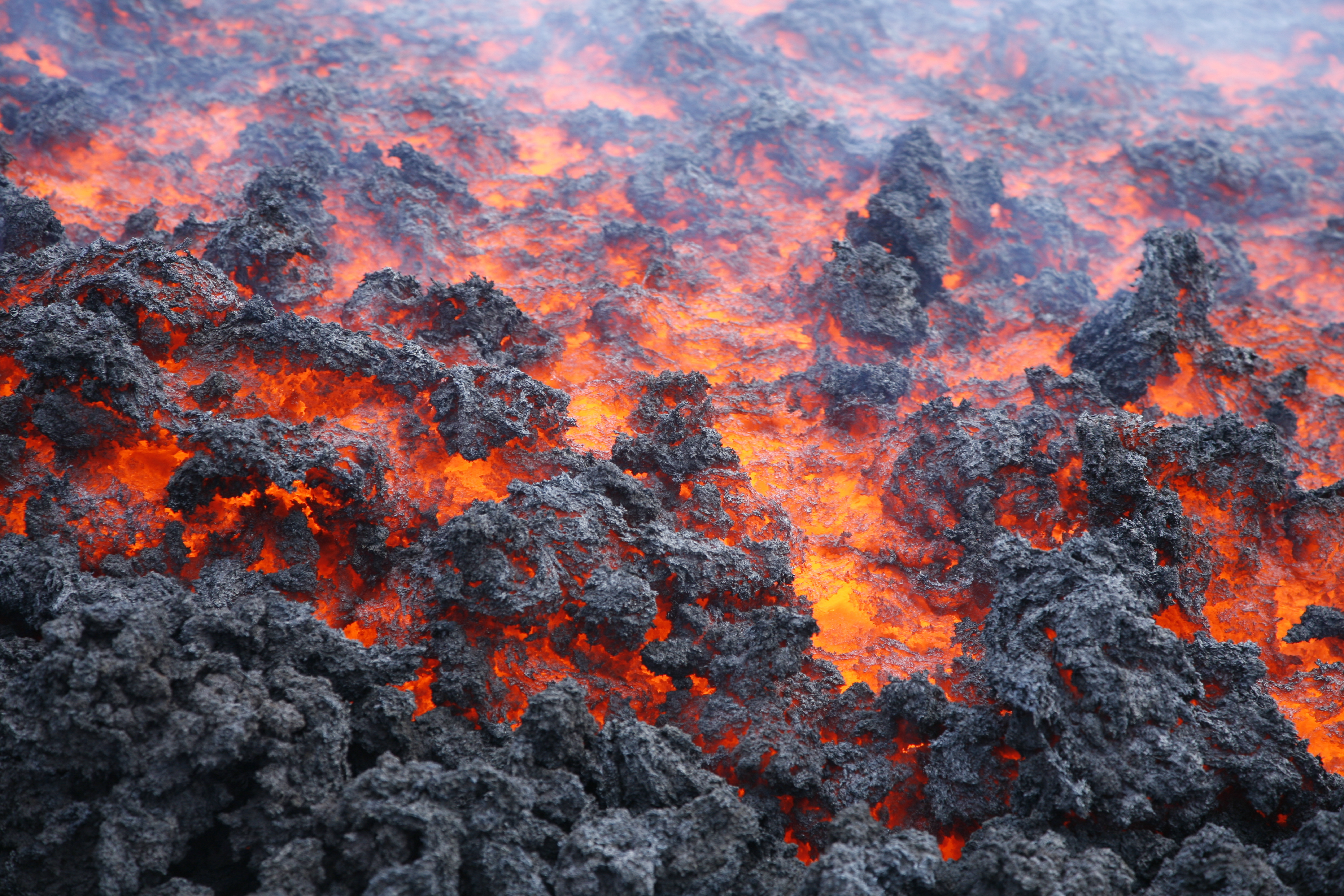}
\caption{The granular structure of `a`a lava from Etna volcano}
\label{fig:second}
\end{subfigure}
\caption{Images showing the difference in the structure of Pahoehoe (left) and `a`a (right) lava flows. `a`a lava is much more viscous than Pahoehoe. From Tom Pfeiffer (www.volcanodiscovery.com)}
\label{fig:lava_fig}
\end{figure*} 

Upon eruption at Earth's surface, magma undergoes abrupt cooling that initiates crystallisation. This increases the effective viscosity and, going by the laboratory experiments of \citet{himo2021chaos}, endows the effective viscosity with a host of possibilities for time and space variations. Pahoehoe is a primitive form of lava that typically emerges from the volcanic vent. It flows easily due to its lower viscosity as a result of its higher temperature. Under cooling and upon encountering steep slopes, pahoehoe can transform into a'a' lava \citep{cashman1999cooling}. These two types of lava flows differ in their structure with one being more liquid-like and the other more granular (see Figure \ref{fig:lava_fig}). \citet{culha2023yih} proposes that the Yih instability we discussed earlier \citep{Yih_1967JFM} increases the viscosity of the lava via mechanical stirring that furthers the rate of crystallisation. They conduct field experiments on Pahoehoe lava and suggest that this lava may have internal layers with varying speeds and viscosity that might initiate mixing and crystallisation. In their model, they have a more viscous and less dense layer flowing above a less viscous and more dense lower layer, and this combination lies in the parameter range for an instability. They argue that such instability can lead to mixing between layers (see their Figure 2). Model analogues of such flows, which lend themselves to theoretical treatment, include films of varying viscosity flowing down an incline. The linear stability of such a flow \citep{usha2013linear} reveals that the surface instability mode of \citet{benjamin1957wave} can be complemented by an overlap mode at viscosity interfaces. Film breakup is a nonlinear feature that could be studied by theoretical approaches. The contributions due to phase-change thermodynamics may be nontrivial, since latent heat provides a source of buoyancy.

\subsubsection{Convection with varying viscosity}\label{conv_w_visc}

Mantle convection and lava flows can be informed by fundamental studies on model flows. Some works, not directly relatable to realistic mantle dynamics, like the linear stability of three layer vertical Poiseuille flow \citep{renardy1987viscosity} find that the relative position of the heavy layer with respect to the more viscous layer determines which mode of instability (varicose or snake in their work) or both are unstable. The experiments on density and viscosity stratified layered gravity currents of \citet{amy2005density} show that viscosity contrast control the relative position of the more dense layer with respect to the nose of the current. And the lock exchange problem of \citet{allen2022mixing} show that unless viscosity of the two fluids are near equal, the difference in viscosity leads to a reduction in large scale mixing. 

A discussion of mantle convection also naturally leads to a discussion on convection and rotating convection of fluids in general. Given that the Reynolds number of mantle flows is most often infinitesimally small, the steady slow viscous flow approximation is valid. By the Taylor Proudman theorem, an ideal (inviscid) rotating fluid in the absence of the nonlinear convective term in the rotating Navier-Stokes equations with adverse (heated from below) thermal gradient cannot have any non-uniform motion along its axis of rotation \citep{chandrasekhar2013hydrodynamic}. This system can only be rendered thermally unstable with the introduction of viscosity (with nonlinearity still neglected) -- another classical case of viscosity making the flow unstable. \citet{chandrasekhar2013hydrodynamic} considers linear hydrodynamic stability of an internally-heated spherical shell, including rotation and gravity variation with the radius, and derives the critical Rayleigh number for the onset of convection. Inspired by the need to understand onset of convection in rotating spherical systems, \cite{dormy2004onset,ardes1997thermal} worked on this question and obtained possible patterns of convection in spherical shells. Some of the earliest numerical simulations of heated convection ignored the variations in viscosity even though they were designed to address mantle convection, but they emphasised the importance of incorporation of proper rheology \citep{mckenzie1974convection,trompert1998mantle}. Whereas convection due to bottom heating is extensively covered in other reviews like \citet{ahlers2009heat,chilla2012new}, we discuss below those studies where the effect of viscosity variation has been quantified. As with the other flows discussed in this review, viscosity variations play an important role in convection dynamics as well.

The effect of viscosity variation with temperature and depth (pressure) in the limit of infinite Prandtl number (a good approximation for mantle $Pr$) was addressed in one of the earliest such studies by \citet{torrance1971thermal} where they provide a scaling for heat transfer with the Rayleigh number for a given mantle viscosity model. Prescribing an exponential dependence of viscosity on temperature, \citet{stengel1982onset} show that the onset of convection is determined by ratio of viscosities of the top layer and bottom layers (with gravity acting downwards). The system is stabilised up to a viscosity ratio of $\mathcal{O}(1000)$, beyond which it starts to destabilise, and display a lower critical $Ra$. \citet{trompert1998mantle} performed  numerical simulations of Stokes flow in a bottom heated box with an open top. With a prescribed viscosity contrast of $10^5$ across the vertical extent of the box, they observe several features of mantle convection, like plate tectonics, downwellings and horizontal divergences of the flow. \citet{moresi1995numerical} use a mind-boggling temperature-dependent viscosity contrast of $10^{14}$ and find convection to be qualitatively different between Rayleigh numbers of $\mathcal{O}{(10^7)}$ and $\mathcal{O}{(10^8)}$, with the latter being more chaotic. With increase in viscosity contrast at a constant Rayleigh number, convection became akin to that in a stagnant-lid situation, no longer capturing plate-tectonics-like behaviour such as happens on Earth. With increase in viscosity contrast, the heat transfer rate decreased, the temperature drop across the  bottom boundary later decreased, and thickness of the bottom thermal boundary layer increased. Physical experiments of \citet{davaille1993transient} had a viscosity contrast up to $\mathcal{O}(10^6)$ with $Ra$ in the range $\mathcal{O}(10^6) -\mathcal{O}(10^8)$. The upper lid of their experimental tank was cooled. With large viscosity contrasts, they found  the fluid in the upper part of the tank to be quieter, and the fluid in the lower part of the tank being more well mixed. 

With an increase in the temperature difference between the top and bottom walls to reach higher Ra, non-Oberbeck-Boussinesq (NOB) effects, due to the dependence of properties like viscosity and thermal expansion with temperature, kick in. These include a deviation in the centre temperature from the algebraic mean of the two boundary plates. NOB effects have been shown to bring in asymmetry of the top and the bottom boundary layers. \citet{ahlers2006non} studied the NOB effects for convection in water where the effect of viscosity variation with temperature was included in laminar boundary layer theory, similar to \citet{wall1997linear}, that provided an agreement with their numerical simulations for the correction needed in the centre temperature and temperature drop across the thermal boundary layers due to NOB effects. They extended Prandtl boundary layer theory for temperature-dependent viscosity and thermal diffusivity. The degree of NOB effects introduced in a convective system has to be material dependent, as different fluids respond differently to changes in temperature \citep{chilla2012new}.

\citet{yoshida2006low} finds that only with temperature and depth-dependent viscosity in a spherical shell and with convection simulations of Rayleigh number $\mathcal{O}(10^7)$, the inferred patterns of convection in non-Earth planets is represented. \citet{zhang2023effect} consider convection simulations of lunar mantle with pressure-dependent effective viscosity with an Arrhenius profile that increased with depth and found reduction in plume size in the mantle. In related areas, studies of heat and material fluxes in such environments that take into account chemical changes are topics of wide interest. Also, understanding the length and time scales of mantle plumes and how they affect local changes in viscosity remain uncertain parameters in modelling sea level changes and understanding ocean circulation patterns \citep{steinberger1998advection, koppers2021mantle}.

Some studies have looked at turbulent Rayleigh-B\'enard convection and much work remains to be done in terms of understanding the effect of viscosity stratification in the early stages of instability development as well as how it couples in the presence of rotation and other external fields. As discussed in section \ref{sec:NonNormalVVF}, preliminary efforts of \citet{Thakur_etal_2021JFM}, not directly related to convection, study the stabilising role of viscosity variations in a heated channel, and find that the structures of highest nonmodal growth are located towards the walls with higher temperature (and consequently lower viscosity). The role of these structures in the onset and sustenance of full-fledged turbulence is left as an open question. 


\section{Connections across scales, future directions and conclusions}
\label{sec:conclusions}

Microorganism dynamics, industrial and geophysical flows can inform each other on the physics of viscosity variation. Microscale viscosity variations affect individual particle behaviour on the small scale which will determine their collective behaviour at large scales. Secondly, similar behaviour occurs in Earth's mantle, where objects made of solid crystalline material are interspersed with flowing matter. Though the sizes of the solid objects are large in the mantle, the system is in creeping flow, which enables a direct transfer of knowledge from one length scale to the other. Shear flow extends across all scales, and though we have given more attention to high Reynolds number flows, the approaches we describe can be used to understand other flows and other ranges of nondimensional numbers. The introduction of singularities, inflexional profiles, extra sources of nonlinearity and the breaking of symmetry by viscosity variations is universal across scales and problems. Further connections have been made elsewhere in this review.

A host of future directions for research have been discussed across the sections, and the reader is encouraged to return to those for context. Here we do not repeat all of them, but indicate some broad directions.

The dynamics of collections of particles in constant-viscosity turbulent and laminar flows is being investigated through many lenses: active particles and their collective motion, and snow avalanches, to name two. The motion of a single particle modifies that of neighbouring particles, and so flow behaviour on the scale of interparticle separation will determine large-scale features. And variations in viscosity, even if small, on these scales, can determine interparticle interactions, which in turn determine clustering, clumping, coalescence, chaotic mixing or its suppression, and the feedback forcing from particles to the turbulence. Understanding these processes in varying viscosity situations will involve the microscale approaches we outlined. Single particle dynamics in viscosity-varying flow is beginning to be studied, but learning about the dynamics of collections of particles is wide open. The interaction between two spheres that we discussed is a pointer that viscosity stratification will bring in surprises.

Finding modal laminar instabilities in viscosity-varying flows is now a mature topic, but features like chemical reactions, and aims like achieving efficient and predictable mixing in microchannels, continue to be of interest. Non-modal growth due to the new non-normal structure of viscosity-stratified flows is by no means a solved problem: it poses mathematical challenges and enjoys high applicability at all length scales, including the microscopic. To mention just one open question: we have very little idea how viscosity variation will impact the laminar-turbulent separatrix in phase space, and the edge state. That bring us to transition to turbulence: an extremely important process which is poorly understood except in special situations. A range of flows are transitional, i.e., neither laminar nor fully turbulent. There is an array of routes to turbulence that different flows take, and asking how stratification affects these routes, and whether it generates hitherto unknown routes, is most appealing. We discussed low Reynolds number laminar instabilities due to stratification, but we do not know whether they will trigger turbulence.

Canonical turbulent flows on the intermediate scale, such as jets, plumes, mixing layers, boundary layers, channels and pipes, need to be studied in varying-viscosity situations. These canonical flows form important components of processes on the large scale in the ocean, mantle and the atmosphere, and are relevant in describing them. Questions would include what the dominant coherent structures are, how the basic length scales such as jet width would evolve with distance or time and how mixing or chemical reactions will be affected. Further advances in imaginative and physics-informed modelling and measurements of eddy viscosity and its variations are called for, to improve understanding of mixing and a variety of other process in the atmosphere and the ocean. Complementarily, better knowledge of molecular viscosities within the mantle will aid in better descriptions of convection within it.

Several assumptions have been made in studies on viscosity variation for the sake of simplicity. This is a natural approach when nothing is known about a question. But once some inroads have been made in understanding it, it is proper to re-examine the validity of the assumptions. Releasing them would constitute natural and important extensions. The use of constant species diffusivities is one such, since in reality, these are functions of temperature, pressure and concentration. Their variations introduce new couplings between the species and Navier-Stokes equations, with the promise of new richness. Almost no shear flow is parallel, and it is known that releasing the parallel-flow assumptions changes flow behaviour qualitatively. The combination of varying geometry and viscosity provides a palette of possibilities. Viscosity-varying flows in a rotating frame of reference is largely unexplored, and relevant on the large scale, e.g., to mantle dynamics.

The kinetics of phase transition can lead to mushy layers where solid and liquid phases co-exist. Local viscosity here is space- and time-dependent, and this presents another relatively clean canvas to work on. The findings in this context will illuminate dynamics within phase change processes at every scale.

One interesting approach to research, which applies to us too, is to work backwards from the final product to ask the `why' questions. The kitchen is a great place to begin, where fundamental studies on the production of sugar syrup, cake (or idli) batter or chocolate might reveal ways to improve efficiency and taste, and to design better industrial scale food processes.

We hope the reader is convinced that studies on viscosity-stratified flows are exciting, reveal unexpected physics, and connect across disparate length and time-scales. Working on them can keep scientists busy and happy for years to come. 

\section{Acknowledgements}

We are grateful to Dylan Reynolds, Attreyi Ghosh and Srikanth Sastry for discussions on odd viscosity, the mantle, and solidification, respectively. Thanks to Ganga Prasath, Samriddhi Sankar Ray, Sriram Ramaswamy, Anubhab Roy, and Brato Chakrabarti for reading the draft paper and providing important input, Anugraha A for making Figure \ref{fig:schematic} and Anagha Madhusudhanan for Figure \ref{fig:anagha}. The work of RG and SKJ at ICTS-TIFR was supported by the Department of Atomic Energy, Government of India, under Project No. RTI4001. The work of RT was supported by the seed grant of Indian Institute of Technology Delhi and the Prime Minister Early Career Research Grant (Project No. RP05041G\_SN) of the Anusandhan National Research Foundation (ANRF) of India. The work of AS was supported by the U.S. Department of Energy, Office of Science, Advanced Scientific Computing Research (ASCR) Early Career Research Program. This paper describes objective technical results and analysis. Any subjective views or opinions that might be expressed in the paper do not necessarily represent the views of the U.S. Department of Energy or the United States Government. Sandia National Laboratories is a multimission laboratory managed and operated by National Technology and Engineering Solutions of Sandia, LLC, a wholly owned subsidiary of Honeywell International Inc., for the U.S. Department of Energy’s National Nuclear Security Administration under contract DE-NA0003525.

\bibliography{references.bib}  

\end{document}